\documentclass[sigconf]{acmart}

\usepackage{tikz}
\usepackage{booktabs}
\usepackage{pgfplots}
\usepackage{tabularx}
\pgfplotsset{compat=1.17}
\usepackage{pgfplotstable}

\definecolor{customPink}{HTML}{f5d5e0}
\definecolor{customPurple}{HTML}{7b337d}

\newcommand{\h}[2]{%
  \textcolor{customPurple!#1!gray}{#2}%
}

\usepackage{balance}
\usepackage{multirow}

\usetikzlibrary{shapes, arrows, positioning}
\AtBeginDocument{%
  }

\copyrightyear{2026}
\acmYear{2026}
\setcopyright{cc}
\setcctype{by}
\acmConference[AHs 2026]{The Augmented Humans International Conference 2026}{March 16--19, 2026}{Okinawa, Japan}
\acmBooktitle{The Augmented Humans International Conference 2026 (AHs 2026), March 16--19, 2026, Okinawa, Japan}
\acmDOI{10.1145/3795011.3795019}
\acmISBN{979-8-4007-2351-3/2026/03}

\begin{document}

\title[SoK: Pedagogical, Health, Ethical, and Privacy Challenges of XR in ECE]{SoK: Understanding the Pedagogical, Health, Ethical, and Privacy Challenges of Extended Reality in Early Childhood Education}

\author{Supriya Khadka}
\affiliation{%
  \institution{Coventry University}
  \city{Coventry}
  \country{UK}}
\email{khadkas25@uni.coventry.ac.uk}

\author{Sanchari Das}
\affiliation{%
  \institution{George Mason University}
  \city{Fairfax, Virginia}
  \country{USA}}
\email{sdas35@gmu.edu}

\renewcommand{\shortauthors}{Khadka and Das}

\begin{abstract}
Extended Reality (XR) combines dense sensing, real-time rendering, and close-range interaction, making its use in early childhood education both promising and high risk. To investigate this, we conduct a Systematization of Knowledge (SoK) of $111$ peer-reviewed studies with children aged $3-8$, quantifying how technical, pedagogical, health, privacy, and equity challenges arise in practice. We found that AR dominates the landscape ($73\%$), focusing primarily on tablets or phones, while VR remains uncommon and typically relies on head mounted displays (HMDs). We integrate these quantitative patterns into a joint risk and attention matrix and an~\emph{Augmented Human Development (AHD)} model that link XR pipeline properties to cognitive load, sensory conflict, and access inequity. Finally, implementing a seven dimension coding scheme on a $0 - 2$ scale, we obtain mean scholarly attention scores of $1.56$ for pedagogy, $1.04$ for privacy (primarily procedural consent), $0.96$ for technical reliability, $0.92$ for accessibility in low resource contexts, $0.81$ for medical and health issues, $0.52$ for accessibility for disabilities, and $0.14$ for data security practices. This indicates that pedagogy receives the most systematic scrutiny, while data access practices is largely overlooked. We conclude by offering a roadmap for \emph{Child-Centered XR} that helps HCI researchers and educators move beyond novelty to design systems that are developmentally aligned, secure by default, and accessible to diverse learners.
\end{abstract}


\maketitle

\section{Introduction}
\label{sec:introduction}

Early childhood is one of the most formative periods of human development, during which children rapidly acquire cognitive, emotional, social, and motor competencies that shape later learning and behavior~\cite{britto2017nurturing, unesco_ecce_need_know_2025}. Although often defined as birth to 8 years~\cite{UNICEF_2025_ECD}, we operationalize \emph{early childhood} as ages \emph{$3-8$}, the stage in which children begin structured learning while still relying on play, social interaction, and multisensory experiences~\cite{KennedyEtAl2012_Literacy3to8}. 

In this context, advances in \textit{Extended Reality (XR)} have intensified interest in immersive learning for early childhood education (ECE). XR encompasses \textit{Augmented Reality (AR)}, \textit{Virtual Reality (VR)}, and \textit{Mixed Reality (MR)}~\cite{fast2018testing}, where AR overlays virtual elements onto the physical world~\cite{azuma1997survey, duezguen2020towards, noah2025pins}, VR replaces the surrounding environment through head-mounted displays (HMDs)~\cite{steuer1992defining}, and MR supports bidirectional interaction through spatial mapping and projection~\cite{milgram1994taxonomy}. XR can increase engagement, support embodied interaction, and improve conceptual understanding~\cite{villena2022effects, noah2021exploring, wangImpactsAugmentedReality2024}, with demonstrated benefits in language and literacy~\cite{zhang2025analyzing,chen2022effects,haoming2024systematic}, STEM~\cite{liu2024augmented,mansour2025embodied}, socio-emotional and environmental learning~\cite{balchaImpactAugmentedRealityassisted2025,zhang2023using}, and accessibility support~\cite{lin2016augmented,chițu2023exploring,zhang2022virtual}. At the same time, XR design for young children presents distinct challenges: developmental limits on visual perception, balance, motor control, and executive functions mean that systems calibrated for adults can impose excessive cognitive load or create sensory conflict~\cite{bexson2024safety}. Concerns regarding VR increase because manufacturers warn against extended HMD use for young children~\cite{bexson2024safety,pawelczyk2025understanding}, which leads many studies to adopt projection-based or non-wearable systems~\cite{hsiaoUsingGestureInteractive2016a, baehoikyoungVerificationVRPlay2025a, besevliMaRTDesigningProjectionbased2019}. XR systems also collect extensive behavioral and biometric data, including gaze, gestures, head pose, speech, and physiological signals~\cite{masai2020eye, sun2022augmented, gnacek2024avdos, noah2022security}, which, when involving minors, raise concerns about privacy, data minimization, secondary use, and compliance with COPPA~\cite{kishnani2024dual,adhikari2025natural} and GDPR-K~\cite{jawalkar2024ethical,kaimara2022could,tazi2024we}. Transparency gaps further complicate parental and educator oversight~\cite{liInnovativeApplicationAI2025a, bexson2024safety}, and additional ethical issues include content appropriateness, commercialization and gamification risks, and blurred boundaries between real and fictional experiences~\cite{yuanExperimentalStudyEfficacy2022, yilmazAreAugmentedReality2017, fridbergThematicTeachingAugmented2024, bailey2019virtual}. 

To address these gaps, we present a Systematization of Knowledge (SoK) that synthesizes the technical, pedagogical, developmental, data protection, and equity challenges of XR in early childhood education. We distinguish this work as an SoK rather than a traditional systematic review because it moves beyond aggregating literature to construct a novel interpretative framework (\textit{Augmented Human Development (AHD)}) that structurally maps these disparate risks. Drawing from $111$ peer-reviewed publications retrieved from major HCI, education, and psychology databases (e.g., ACM, IEEE, Scopus, ERIC), we analyze how XR systems are designed, deployed, and studied for ECE. Our investigation follows four research questions:

\begin{itemize}
    \item \textbf{RQ1:} How do technical and pedagogical factors constrain the design, implementation, and classroom integration of XR systems for early childhood education (ECE)?
    \item \textbf{RQ2:} What health-related risks and constraints arise when children aged $3-8$ interact with immersive XR systems (e.g., visual strain, cybersickness, postural and motor effects), and how are they measured or reported?
    \item \textbf{RQ3:} In what ways do XR deployments in ECE surface accessibility and equity challenges (e.g., disability support, inclusive interaction design, and suitability for low-resource environments)?
    \item \textbf{RQ4:} How do existing XR-for-children studies handle privacy-sensitive data practices (e.g., data collection, biometric tracking, parental consent, and regulatory compliance), and which privacy risks remain under-addressed?
\end{itemize}

\noindent\textbf{Contributions:}
\begin{itemize}
    \item We conduct the first Systematization of Knowledge (SoK) that characterizes the XR-in-ECE design and deployment space, deriving a structured taxonomy of technical limitations, developmental and health constraints, privacy and data security risks, and accessibility and equity barriers.
    \item We introduce the Augmented Human Development (AHD) framework, which models XR child–system interaction as a function of cognitive, sensory, environmental, and developmental parameters for children aged $3-8$, and use it to analyze how specific XR design choices propagate into developmental risk.
    \item We distill mitigations and design patterns reported across $111$ empirical studies into an actionable framework that specifies concrete configuration, interaction, and data-governance guidelines for XR researchers, system designers, and ECE practitioners.
\end{itemize}

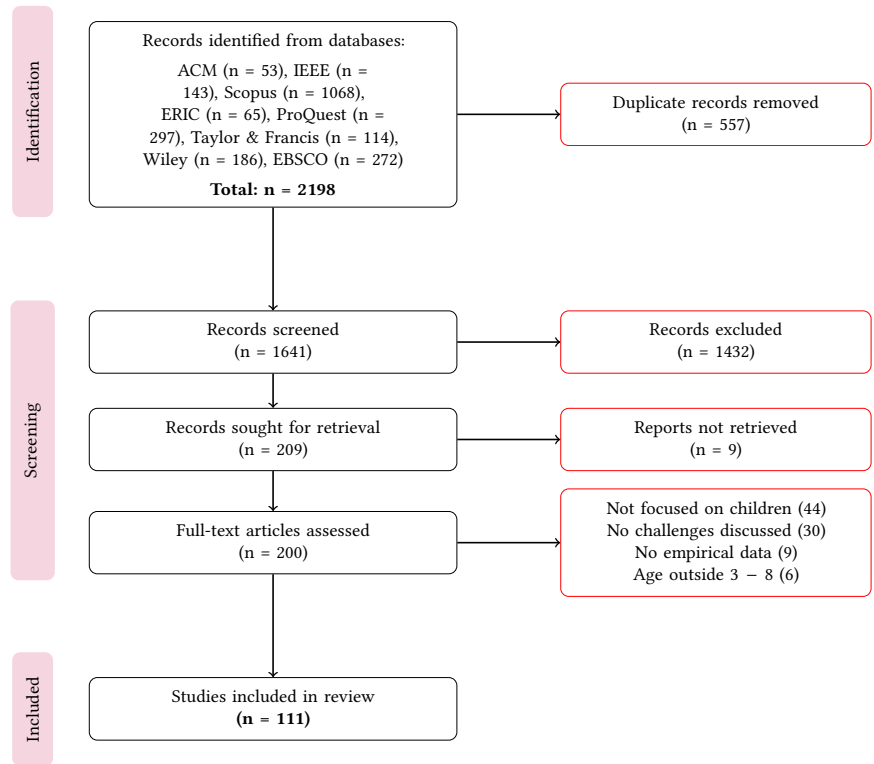
\begin{figure*}[ht]
\centering
\resizebox{0.65\linewidth}{!}{
\begin{tikzpicture}[
  node distance=1cm,
  every node/.style={font=\normalsize, align=center},
  process/.style={rectangle, draw, text width=6cm, minimum height=1cm, rounded corners, inner sep=6pt},
  processR/.style={rectangle, draw, text width=5cm, minimum height=1cm, rounded corners, inner sep=6pt, color=red, text=black},
  arrow/.style={->, thick}
]

\node (id) [process] {Records identified from databases:\\[4pt]
ACM (n = 53), IEEE (n = 143), Scopus (n = 1068),\\
ERIC (n = 65), ProQuest (n = 297), Taylor \& Francis (n = 114),\\
Wiley (n = 186), EBSCO (n = 272)\\[4pt]
\textbf{Total: n = 2198}};

\node (excluded0) [processR, right=1.8cm of id] {Duplicate records removed\\(n = 557)};
\draw [arrow] (id.east) -- (excluded0.west);

\node [rotate=90, fill=customPink, rounded corners, text width=3.2cm, inner sep=7pt] at (-4.2,.05) {Identification};

\node (dup) [process, below=1.8cm of id] {Records screened\\(n = 1641)};
\draw [arrow] (id) -- (dup);

\node (excludedn) [processR, right=1.8cm of dup] {Records excluded\\(n = 1432)};
\draw [arrow] (dup.east) -- (excludedn.west);

\node (screen) [process, below=0.6cm of dup] {Records sought for retrieval\\(n = 209)};
\draw [arrow] (dup) -- (screen);

\node (excluded1) [processR, right=1.8cm of screen] {Reports not retrieved\\(n = 9)};
\draw [arrow] (screen.east) -- (excluded1.west);

\node [rotate=90, fill=customPink, rounded corners, text width=4.4cm, inner sep=7pt] at (-4.2,-5.7) {Screening};

\node (eligible) [process, below=0.7cm of screen] {Full-text articles assessed\\(n = 200)};
\draw [arrow] (screen) -- (eligible);

\node (excluded2) [processR, right=1.8cm of eligible] {
Not focused on children (44)\\
No challenges discussed (30)\\
No empirical data (9)\\
Age outside $3-8$ (6)\\
};
\draw [arrow] (eligible.east) -- (excluded2.west);

\node (included) [process, below=1.8cm of eligible] {Studies included in review\\\textbf{(n = 111)}};
\draw [arrow] (eligible) -- (included);

\node [rotate=90, fill=customPink, rounded corners, text width=1.5cm, inner sep=7pt] at (-4.2,-10.4) {Included};

\end{tikzpicture}
}
\caption{PRISMA Flow Diagram Illustrating the Study Identification and Selection Process.}
\label{fig:prisma}
\Description[A PRISMA flow diagram showing the selection process of 111 studies from an initial pool of 2198 records.]{The diagram illustrates the four-stage systematic review process. 
1. Identification: A total of 2198 records were identified from eight databases (ACM, IEEE, Scopus, ERIC, ProQuest, Taylor & Francis, Wiley, EBSCO). 557 duplicate records were removed.
2. Screening: 1641 records were screened, and 1432 were excluded, leaving 209 records sought for retrieval. 
3. Eligibility: 9 reports were not retrieved. 200 full-text articles were assessed for eligibility. From these, 89 were excluded for reasons including: not focusing on children (44), no challenges discussed (30), no empirical data (9), or age falling outside the 3-8 range (6).
4. Included: The final set comprises 111 studies included in the qualitative synthesis.}
\end{figure*}

\section{Method}
\label{sec:method}
We conducted a systematic literature review following the 2020 Preferred Reporting Items for Systematic Reviews and Meta-Analyses (PRISMA) guidelines~\cite{moher2009preferred, alzahrani2025systematic}, a widely adopted standard for the transparent identification, screening, and synthesis of empirical evidence, as summarized in Figure~\ref{fig:prisma}. Additionally, we adopted the methodological design established in prior works by the last author of this paper~\cite{huang2025systemization,shrestha2022sok,duzgun2022sok,tazi2022sok,das2022sok,zezulak2023sok,tazi2023sok,tazi2022sok,grover2025sok,tazi2024sok,saka2025sok,podapati2025sok,majumdar2021sok,agarwal2025systematic,kishnani2023blockchain,jones2021literature,noah2021exploring,das2019all,shrestha2022exploring}. 

\subsection{Search Strategy}
We searched eight major academic databases commonly used in human–computer interaction, educational technology, and learning sciences research~\cite{newman2020systematic, valente2022analysis}: ACM Digital Library, IEEE Xplore, Scopus, ERIC, ProQuest, Taylor \& Francis, Wiley Online Library, and EBSCOhost. We limited coverage to the period $2010-2025$ to capture the modern wave of XR research in education, which coincides with the wider availability of consumer-grade AR/VR hardware and software platforms~\cite{avila2014virtual}. We focused on studies whose \textbf{title}, \textbf{abstract}, or \textbf{author keywords} referenced three conceptual areas: XR technologies, educational practice, and early childhood. To maximize recall while maintaining conceptual precision~\cite{keele2007guidelines}, we used multiple keyword variants. The final search string was: \textit{("augmented reality" OR "virtual reality" OR "mixed reality" OR "extended reality" OR AR OR VR OR MR OR XR OR "head-mounted display")} \textbf{AND} \textit{(educat* OR learn* OR teach* OR training OR pedagog* OR "instructional design" OR "educational technolog*" OR "digital learn*" OR "immersive learn*" OR gamif* OR "serious game*")} \textbf{AND} \textit{("early childhood" OR "early years" OR preschool* OR kindergarten OR nursery OR "young children")}. 

\subsection{Screening and Selection Process}
All records were imported into~\textit{Zotero}~\cite{zotero} for reference management. Duplicate records were identified using Zotero's built-in duplicate detection and manually verified, resulting in $557$ duplicates. The remaining $1641$ unique records were screened in two stages: (1) title and abstract screening and (2) full-text eligibility assessment. Both stages were guided by the inclusion and exclusion criteria specified below and informed by established practices for systematic reviews in HCI and educational technology~\cite{keele2007guidelines}. After full-text screening, $111$ studies were retained for synthesis (Figure~\ref{fig:prisma}).

\textbf{Inclusion criteria:}
(1) investigates AR/VR/MR/XR use in early-childhood learning;  
(2) reports empirical data from human participants (experiments with pre/post measures, user studies, case studies with $N>1$, observations, or surveys; pilot studies were included when results were reported~\cite{montoya-rodriguezEducationalInterventionTheory2025, liangExploitationMultiplayerInteraction2017, liangHandGesturebasedInteractive2017, liInnovativeApplicationAI2025a, dalimTeachARInteractiveAugmented2016, korosidouEffectsAugmentedReality2024});  
(3) includes participants aged $3-8$ years;  
(4) discusses at least one empirical or theoretical XR-related challenge;  
(5) is an English, peer-reviewed journal article or full conference paper; 
(6) published between 2010 and 2025.

\textbf{Exclusion criteria:}
(1) studies not meeting the above criteria;  
(2) duplicate or redundant publications;  
(3) single-participant case reports;  
(4) papers without empirical data;  
(5) studies that do not report any XR-related challenges;  
(6) work on non-technological interactive learning activities.

\subsection{Data Extraction and Management}
For each eligible study, we extracted bibliographic metadata (title, authors, venue, year) and key descriptive variables, including \textit{participant age}, \textit{XR modality (AR/VR/MR)}, \textit{sample size}, \textit{geographic context}, \textit{learning domain}, and \textit{session duration}. We also collected feasibility-related details such as \textit{device type}, \textit{teacher training requirements}, and \textit{classroom or deployment constraints}, which are critical for understanding real-world adoption in early childhood settings~\cite{blackwell2014factors}. To analyze challenges, we extracted qualitative descriptions of \textit{privacy}, \textit{data security}, \textit{accessibility and equity}, \textit{health concerns}, \textit{technical limitations}, \textit{pedagogical issues}, and \textit{reported mitigation strategies}. This extraction followed principles of qualitative content analysis, with iterative refinement of coding categories~\cite{elo2008qualitative}. All extracted data were compiled into structured tables to support systematic comparison and synthesis.

\subsection{Analytical Framework}
To move beyond descriptive synthesis and answer our research questions in a theoretically grounded manner, we employed a single analytical framework rooted in the \emph{Augmented Human Development (AHD)}, which has two layers:

\begin{enumerate}
    \item a \textbf{conceptual layer} (AHD), which characterizes child and system's interaction in terms of cognitive, sensory, environmental, and developmental factors; and
    \item two \textbf{operational dimensions}, which quantify (a) the level of scholarly attention and (b) the estimated real-world risk associated with each challenge category.
\end{enumerate}

\begin{figure*}[ht]
\centering
\includegraphics[width=0.7\linewidth]{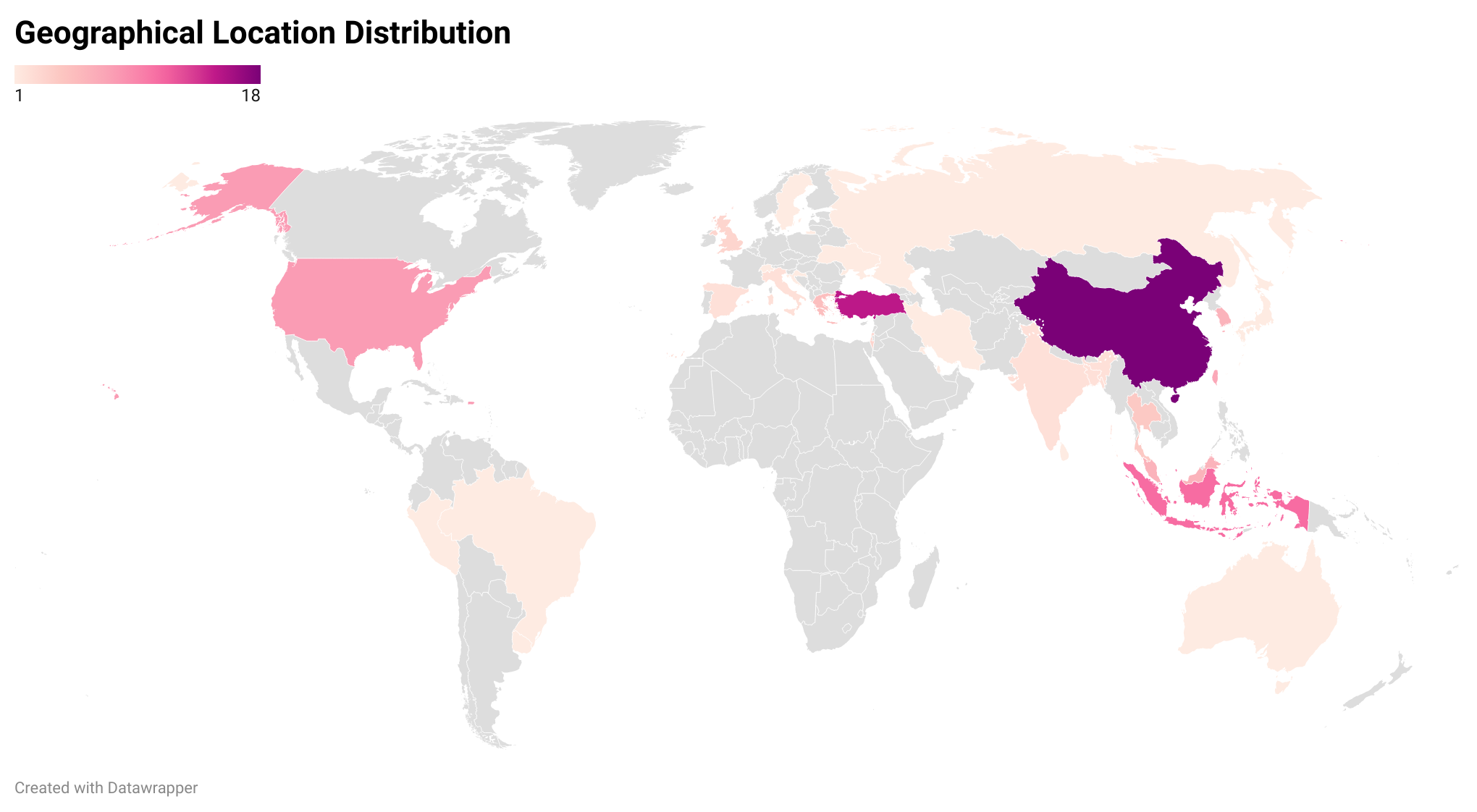} 
\caption{Geographic Distribution of the \textbf{111} Included Studies.}
\label{fig:map}
\vspace{-3mm}
\Description[A world map highlighting the geographic distribution of the 111 included studies.]{A world map visualizing the countries of origin for the 111 analyzed studies. The distribution shows high concentrations of research output from China, Turkey, and Indonesia. Other regions are shaded to indicate lower frequencies of study origin.}
\end{figure*}

\subsubsection{Augmented Human Development}
\label{sec:ahd}
To interpret the diverse challenges of XR in ECE, we introduce the \emph{Augmented Human Development (AHD)} framework. AHD conceptualizes how XR technologies interact with children’s developing cognitive, sensory, social, and motor systems. Rather than treating XR as a neutral delivery medium, AHD emphasizes that immersive technologies modulate developmental processes in dynamic, time-dependent ways. This makes AHD particularly relevant for children aged $3-8$, whose perceptual, motor, attentional, and socio-emotional systems are undergoing rapid change. At any moment \(t\), we characterize the interaction between a child and an XR system as:
\begin{equation}
AHD(t) = f\big(C(t), S(t), E(t), D(t)\big),
\end{equation}
where:
\begin{itemize}
    \item \(C(t)\) denotes \textbf{cognitive load}: the demands an XR activity places on attention, memory, and executive functions.
    \item \(S(t)\) captures \textbf{sensory stimulation and integration}, including motion cues, visual complexity, and sound, and embodiment.
    \item \(E(t)\) reflects the \textbf{environmental and social context}, such as teacher mediation, availability of guidance, ambient noise, and opportunities for co-located collaboration.
    \item \(D(t)\) represents the child's \textbf{developmental profile}, including age, motor abilities, language proficiency, sensory sensitivity, and prior technological experience.
\end{itemize}

This decomposition does not attempt to fully specify the function \(f(\cdot)\). Instead, AHD serves as a conceptual scaffold that explains why XR challenges manifest differently in early childhood than in older learners or adults. e.g.,:
\begin{itemize}
    \item Tracking errors, jitter, and visually unstable augmentations primarily affect \(S(t)\), which can cascade into elevated \(C(t)\) or disorientation.
    \item Pre-literacy barriers and complex multimodal instructions overload \(C(t)\), especially in tasks requiring fine motor actions or dual-task coordination.
    \item Classroom orchestration, noise, and turn-taking constraints reshape \(E(t)\), influencing how children interpret and engage with XR experiences.
    \item Motor immaturity, short attention spans, and varying sensory thresholds, all part of \(D(t)\), set strict boundaries on acceptable interaction workloads.
\end{itemize}

A key component here is that, we highlight how XR risks and mitigations operate across multiple human–system dimensions simultaneously. Many challenges reported in the literature, such as cognitive overload, novelty-driven distraction, VR-induced fatigue, or inequitable access, are not isolated failures but coupled effects that emerge from misalignment among \(C(t)\), \(S(t)\), \(E(t)\), and \(D(t)\). 

\subsubsection{Operational Dimensions}
\noindent{\textbf{Dimension 1: Scholarly Attention}}
In addition to the AHD perspective, we score how strongly each challenge is addressed in the literature: every study is rated $0-2$ per category. \textbf{0} = Not addressed; \textbf{1} = Mentioned briefly or partially addressed; \textbf{2} = Substantively discussed or investigated.

The coding categories are: \textbf{(1) Privacy} : consent procedures, data transparency, anonymization, regulatory compliance (e.g., COPPA, GDPR-K). \textbf{(2) Data Security} : encryption, access control, data storage, and system integrity. \textbf{(3) Accessibility \& Equity - Disability Support} : multimodal access, adaptive interfaces, and accommodations for children with special needs. \textbf{(4) Accessibility \& Equity - Low-Resource Settings} : suitability for low-cost devices, offline use, bandwidth constraints, and teacher support limitations. \textbf{(5) Medical / Health} : eye strain, cybersickness, ergonomics, posture, overstimulation, and photosensitivity. \textbf{(6) Technical / Hardware} : latency, tracking failures, resolution constraints, software bugs, and network or hardware limitations. \textbf{(7) Pedagogical} : curriculum alignment, age appropriateness, cognitive load, classroom management, and teacher mediation. These map naturally onto the AHD components: for instance, Medical/Health and many Technical issues influence \(S(t)\) and \(C(t)\), while Pedagogical and Accessibility/Equity concerns are tightly coupled to \(E(t)\) and \(D(t)\). Coding was performed by two authors, with disagreements resolved through discussion to achieve a shared interpretation of the categories~\cite{elo2008qualitative}.

\noindent{\textbf{{Dimension 2: Real-World Risk}}
\label{sec:risk-framework}
The second dimension of the framework estimates the \emph{real-world risk} of each challenge category, independent of how often it appears in the literature. Rather than relying on the 111 reviewed studies, we estimate real-world risk using technical documents and external standards. This helps identify areas where potential harm to children or institutions may be high even when scholarly attention is low~\cite{liInnovativeApplicationAI2025a}. We adopt a 3$\times$3 matrix combining:

\textbf{Likelihood} of occurrence in typical early childhood XR deployments:
(1) \textbf{Low} (requires rare or exceptional circumstances),  
(2) \textbf{Medium} (plausible in common deployments),  
(3) \textbf{High} (likely or recurrent in most deployments).

\textbf{Impact} on the child, educator, or institution if the risk occurs:
(1) \textbf{Low} (minor discomfort or brief disruption to learning),  
(2) \textbf{Medium} (reversible health effects, moderate data or pedagogical issues),  
(3) \textbf{High} (lasting physical or psychological harm, major privacy breaches such as GDPR-K/COPPA violations, or long-term negative educational or social consequences). The resulting risk level is computed as:
\[
\text{Risk Level} = \text{Likelihood} \times \text{Impact}.
\]

Conceptually, Likelihood and Impact are influenced by the AHD components. e.g., a design that chronically overloads \(C(t)\) or overstimulates \(S(t)\) in typical classroom conditions is likely to receive high Likelihood and Impact scores. In our Discussion in section~\ref{sec:discussion}, we juxtapose these risk scores with the scholarly attention scores to reveal gaps where high-risk challenges receive comparatively limited attention in the existing XR-for-early-childhood studies.

\begin{figure*}[ht]
\centering
\begin{minipage}{\textwidth} 
    \centering
    \caption{Prevalence of Technical Challenges by Device Type.}
    \label{fig:tech-heatmap}
    \renewcommand{\arraystretch}{1.5}
    \setlength{\tabcolsep}{4pt}
    \small

    \providecommand{\h}[2]{\textcolor{customPurple!#1!black}{\textbf{#2}}}

    \begin{tabular}{|l| >{\centering\arraybackslash}p{2cm} | >{\centering\arraybackslash}p{2cm} | >{\centering\arraybackslash}p{2cm} | >{\centering\arraybackslash}p{2cm} |}
    \hline
    \textbf{Challenge Category} & \textbf{\shortstack{Mobile/Tablets}} & \textbf{HMDs} & \textbf{\shortstack{PC/Laptops}} & \textbf{\shortstack{Specialized}} \\
    \hline
    \textbf{Target \& Tracking Loss} & \h{100}{Pervasive} & \h{40}{Occasional} & \h{70}{Common} & \h{40}{Occasional} \\ 
    \hline
    \textbf{HMD Misalignment} & \h{1}{None} & \h{40}{Occasional} & \h{1}{None} & \h{1}{None} \\ 
    \hline
    \textbf{Hardware Incompatibility} & \h{20}{Rare} & \h{1}{None} & \h{1}{None} & \h{40}{Occasional} \\ 
    \hline
    \textbf{Calibration \& Setup} & \h{20}{Rare} & \h{1}{None} & \h{1}{None} & \h{1}{None} \\ 
    \hline
    \textbf{Speech Recog. Failure} & \h{20}{Rare} & \h{1}{None} & \h{100}{Pervasive*} & \h{1}{None} \\ 
    \hline
    \end{tabular}

    \vspace{2mm}
    \makebox[\linewidth][c]{%
    \parbox{0.9\textwidth}{ 
    \centering
        \footnotesize
        \emph{Note:} Terms indicate reporting frequency: Pervasive ($>40\%$), Common ($25$--$40\%$), Occasional ($10$--$25\%$), Rare ($<10\%$). \\ 
        *High failure rate for PC/Laptops must be interpreted with caution due to the small sample size in this category.
    }
}
\end{minipage}

\Description[A table illustrating the prevalence of technical challenges across different device types.]{A heatmap table categorizing technical challenges by prevalence (Pervasive, Common, Occasional, Rare) across four hardware categories. Key findings include: Target and Tracking Loss is 'Pervasive' for Mobile/Tablets; HMD Misalignment is 'Occasional' for HMDs; Speech Recognition Failure is 'Pervasive' for PC/Laptops.}
\end{figure*}

\section{Results}
\label{sec:results}

The final cohort comprises 111 studies published between 2010 and 2025. As visualized in Figure~\ref{fig:map}, the geographic distribution was diverse, with notably high concentrations of research originating from China (n=18), Turkey (n=15), and Indonesia (n=11).

Technologically, the field is heavily skewed towards AR (73\%, n=81), followed by VR (20\%, n=21) and MR (4.5\%, n=5), with 3 studies combining AR and VR. This correlates strongly with the hardware platforms: consumer Mobile/Tablets dominated (63\%, n=70), while HMDs (9.9\%, n=11) and PC/Laptops (3.6\%, n=4) were rare. Despite the prevalence of consumer hardware, a significant cluster utilized Specialized Devices (13.5\%, n=15). This experimental category includes MR markers \cite{burlesonActiveLearningEnvironments2018a} and custom setups featuring pico-projectors, depth cameras, or motion sensors \cite{besevliMaRTDesigningProjectionbased2019, panIntroducingAugmentedReality2021, hsiaoUsingGestureInteractive2016a, baehoikyoungVerificationVRPlay2025a}. A further 9.9\% (n=11) used a combination of multiple device types. 

In terms of pedagogical aims, Language \& Literacy (37\%, n=41) and STEM (29\%, n=32) accounted for over two-thirds of applications. The remaining studies addressed Health \& Social Sciences (19\%, n=21), Humanities \& Arts (8\%, n=9), with a small minority targeting \textit{combined} domains (7\%, n=8). The most frequently studied demographic was the $5-6$ age group (n=22). Applying the `Scholarly Attention' framework (Section~\ref{sec:method}) to this cohort reveals a distinct hierarchy in research depth. 

\textbf{Pedagogical Challenges} (avg=1.56) received the most substantial discussion, followed by \textbf{Privacy} (avg=1.04). A middle tier emerged comprising \textbf{Technical Issues} (0.96), \textbf{Low-Resource Accessibility} (0.92), and \textbf{Medical/Health Concerns} (0.81). Conversely, critical gaps remain: \textbf{Disability Support} (0.52) was largely overlooked, while \textbf{Data Security} (0.14) was almost entirely ignored.

\subsection{Technical Limitations and Solutions (RQ1a)}
Technical challenges represented a moderately cited barrier (avg score=0.96). However, our analysis reveals that these failures are not uniform; rather, as shown in Figure~\ref{fig:tech-heatmap}, they stem from specific frictions between \textit{System Stability Requirements ($S(t)$)} and \textit{Developmental Constraints ($D(t)$)}.

\subsubsection{Device-Specific Failure Modes: The Bio-Technical Mismatch}

\textbf{Mobile and Tablet devices} ($n=70$) exhibited a strong correlation with \textbf{Target Recognition and Tracking Failures}, the most frequently reported issue in the cohort~\cite{kulasekaraGameCentricElearning2025, calle-bustosAugmentedRealityGame2017, wangARcampPTangibleProgramming2024, syahidiARChildAnalysisEvaluation2019a, aladinARTOKIDSpeechenabledAugmented2020a, prekaAugmentedRealityQR2019a, ayuagungmasaristamyAugmentedRealityCultural2024a, hossainAugmentedRealityBasedElementary2021a, lewis-presserDesigningAugmentedReality2025, joDevelopmentUtilizationProjectorRobot2011a, puDevelopmentSituationalInteraction2018, kisnoDigitalStorytellingEarly2022, paliwalEnhancingEducationAugmented2024}. This failure mode is bio-physical rather than purely technical: the stability required for monocular camera tracking ($S(t)$) conflicts with the developing fine motor skills ($D(t)$) of children aged $3-8$, resulting in jitter and drift. Uncontrolled classroom lighting ($E(t)$) frequently exacerbated this, causing devices to lose target lock.

\textbf{HMDs} presented an inverted risk profile. While tracking issues were lower~\cite{liangExploitationMultiplayerInteraction2017, liangHandGesturebasedInteractive2017, panResearchApplicationImmersive2021, shimadaVRHandHygiene2017}, failures were dominated by \textbf{Misalignment and Obstruction}~\cite{passigSolvingConceptualPerceptual2014a, montoya-rodriguezEducationalInterventionTheory2025}. This represents a hardware-anthropometry conflict: standard headsets are not calibrated for early childhood cranial dimensions ($D(t)$), creating a physical barrier to the senses ($S(t)$).

\textbf{Hardware Incompatibility} was observed in low-end smartphones unable to run high-specification apps \cite{calle-bustosAugmentedRealityGame2017, syahidiARChildAnalysisEvaluation2019a, ayuagungmasaristamyAugmentedRealityCultural2024a, lawAugmentedRealityTechnology2025, kusumaningsihUserExperienceMeasurement2021}, and in complex specialized setups involving robots or projection \cite{burlesonActiveLearningEnvironments2018a, joDevelopmentUtilizationProjectorRobot2011a, besevliMaRTDesigningProjectionbased2019}. Meanwhile, \textbf{Speech Recognition Failure} was clustered in PC/Laptop setups \cite{dalimTeachARInteractiveAugmented2016, hanExaminingYoungChildrens2015a} and one smartphone study \cite{aladinARTOKIDSpeechenabledAugmented2020a}, suggesting a vulnerability in legacy microphone arrays compared to modern mobile hardware.

\subsubsection{Infrastructural Barriers}
Beyond device-specific issues, \textbf{Network and Synchronization} errors (latency, connection loss, multi-device de-synchronization) were hardware-agnostic challenges reported across all platforms~\cite{tanApplicationArtbasedKnowledge2025, liangExploitationMultiplayerInteraction2017, liInnovativeApplicationAI2025a, fajrieAugmentedRealityMedia2022, hoInteractiveMultisensoryVolumetric2023, besevliMaRTDesigningProjectionbased2019, dengARCatTangibleProgramming2019, stearneAugmentedRealityPlaygrounds2025}. These represent a friction between System requirements ($S(t)$) and Environmental infrastructure ($E(t)$). Distinct architecture-specific conflicts, such as distributed system resolutions for Leap Motion~\cite{liangExploitationMultiplayerInteraction2017} and dual-camera issues~\cite{kimVirtualStorytellerUsing2011}, remained significant hurdles, with several errors unresolved in the literature~\cite{gweonMABLEMediatingYoung2018, alcornDiscrepanciesVirtualLearning2013, zakraouiStudyChildrenEngagement2021}

\subsubsection{Strategies for Mitigation}
The literature addressed these technical barriers through three distinct approaches:

\begin{enumerate}
    \item \textbf{Environmental Scaffolding ($E(t)$):} To counter tracking and lighting failures, researchers enforced strict operational controls. These included ensuring high-intensity illumination~\cite{calle-bustosAugmentedRealityGame2017, aladinARTOKIDSpeechenabledAugmented2020a, paliwalEnhancingEducationAugmented2024, besevliMaRTDesigningProjectionbased2019} and maintaining optimal camera distance~\cite{calle-bustosAugmentedRealityGame2017, syahidiARChildAnalysisEvaluation2019a, fengEffectsARLearning2022, isikarslanogluThinkTogetherDesign2024, huangUsingAugmentedReality2016a, raduComparingChildrensCrosshair2016} and tilt~\cite{lewis-presserDesigningAugmentedReality2025}. For speech recognition, mitigations focused on ensuring a quiet setting~\cite{aladinARTOKIDSpeechenabledAugmented2020a} and placing microphones closer to participants~\cite{dalimTeachARInteractiveAugmented2016, aladinARTOKIDSpeechenabledAugmented2020a}.
    \item \textbf{Human-in-the-Loop Mediation:} For hardware issues that lacked software patches (specifically HMD misalignment), studies relied on active adult intervention ($E(t)$). Teachers or researchers were required to manually adjust headsets or guide the child during the session~\cite{montoya-rodriguezEducationalInterventionTheory2025, passigSolvingConceptualPerceptual2014a}.

    \item \textbf{Architectural Optimization ($S(t)$):} To address hardware performance and network lag, developers employed asset optimization techniques (e.g., ``low-poly modeling'', Level of Detail (LOD) management)~\cite{tanApplicationArtbasedKnowledge2025}. Operational tweaks included using splash screens to conserve battery~\cite{lawAugmentedRealityTechnology2025} and upgrading hardware to higher processing capacities~\cite{besevliMaRTDesigningProjectionbased2019}. For connectivity issues, offline architectures were prioritized~\cite{tanApplicationArtbasedKnowledge2025}, while complex multi-sensor conflicts were resolved using distributed architectures~\cite{liangExploitationMultiplayerInteraction2017} and edge computing nodes~\cite{liInnovativeApplicationAI2025a}.
\end{enumerate}

\subsection{Pedagogical Integration Issues (RQ1b)}
Pedagogical challenges represented the most frequently cited barrier in the cohort. Unlike technical failures, these challenges stem from the complexity of integrating immersive tools into the delicate ecology of early childhood classrooms. Our analysis reveals that these challenges are not monolithic; rather, they form a ``pedagogical triad'' of friction between the Child ($D(t)$), the Teacher ($E(t)$), and the Design ($S(t)$):


\subsubsection{The Child: Cognitive and Developmental Mismatches}
The primary friction at the learner level is the tension between the technology's cognitive demands ($C(t)$) and the child's developmental capacity ($D(t)$).

\textbf{The Engagement-Distraction Paradox:}
A central pattern across the literature is a tension between engagement and distraction. In AHD terms, this manifests when high-intensity \textit{Sensory Stimulation ($S(t)$)} exceeds the child's available \textit{Cognitive Capacity ($C(t)$)}. 

Most studies (107 of 111) document positive learning outcomes, but our analysis indicates these gains are often driven by a Novelty Effect, where the ``wow'' factor of XR spikes sensory input but fails to deepen conceptual understanding~\cite{montoya-rodriguezEducationalInterventionTheory2025, topuExaminationPreSchoolChildrens2023a, yuanExperimentalStudyEfficacy2022, tanApplicationArtbasedKnowledge2025, gecu-parmaksizAugmentedRealityBasedVirtual2019, wuComparingEffectsAR2025, hermanIntegratingSocialLearning2025, gecu-parmaksizEffectAugmentedReality2020a, alcornDiscrepanciesVirtualLearning2013, stearneAugmentedRealityPlaygrounds2025}. Only a few studies attempted to control for this effect using multi-phase or `surprise' designs \cite{yuanExperimentalStudyEfficacy2022, alcornDiscrepanciesVirtualLearning2013}, leaving the role of genuine learning versus novelty largely unresolved.

Four studies reported no significant differences between XR and traditional instruction \cite{caiCaseStudyUsing2023, yilmazEducationalMagicToys2016a, chenUsingAugmentedReality2019a}, and one found that XR users performed worse \cite{dallolioImpactFantasyYoung2024}. These mixed findings point to an underlying mechanism that links high engagement and underperformance: \textbf{cognitive overload}.

Cognitive overload was explicitly reported in 40 studies and typically attributed either to XR’s high-intensity sensory input or to complex task structures~\cite{kulasekaraGameCentricElearning2025, burlesonActiveLearningEnvironments2018a, yilmazExaminationVocabularyLearning2022, vasilyevaASDAutisticSpectrum2025a, lewis-presserDesigningAugmentedReality2025, idrisDevelopmentPatternLearning2021a}. This overload produced a \textbf{cognitive tunnel effect}, where children focused so heavily on interface mechanics ($S(t)$) that they lost sight of the educational goal ($C(t)$)~\cite{fengEffectsARLearning2022}. Distraction, therefore, is not a behavioral failure but a symptom of a system design that ignores the executive function limits of the developmental profile ($D(t)$)~\cite{tanApplicationArtbasedKnowledge2025, yilmazAreAugmentedReality2017, antoniaArtfulThinkingAugmented2018, fajrieAugmentedRealityMedia2022, lawAugmentedRealityTechnology2025, jamiatEffectsAugmentedReality2020}.

The literature converges strongly on mitigation strategies, treating distraction not as a behavioral problem but as a symptom of overload. Key design principles include reducing multimedia clutter \cite{simsekEffectAugmentedReality2024}, simplifying game mechanics \cite{kulasekaraGameCentricElearning2025, lewis-presserDesigningAugmentedReality2025, puDevelopmentSituationalInteraction2018, lorussoSemiimmersiveVirtualReality2020, vate-u-lanAugmentedReality3D2012}, and favoring active interaction \cite{liangHandGesturebasedInteractive2017, khairulanuardiIncreaseReadingHabit2022, gavishAugmentedVirtualitySystems2022}. Procedural supports ($E(t)$) were equally vital: habituation time was used to let children process the novelty before learning began~\cite{liangExploitationMultiplayerInteraction2017, poobrasertAugmentedLearningEnvironments2023, simsekEffectAugmentedReality2024}, and structured ``distraction tasks'' were introduced to manage and redirect off-task behavior~\cite{yuanExperimentalStudyEfficacy2022, zakraouiStudyChildrenEngagement2021}.

\textbf{Developmental Constraints:} 
A major source of mismatch arises from developmental differences that XR designs often fail to account for ($D(t)$ vs $S(t)$). Many studies implicitly adopt a ``one size fits all'' approach \cite{weiDevelopingCuboidRecognitionbased2023, borovanskaEngagingChildrenUsing2020a}, deploying the same interface across wide age ranges despite large variations in cognitive capacity. Multimedia elements that 5-year-olds found engaging were overwhelming for 3-year-olds \cite{weiDevelopingCuboidRecognitionbased2023}, and these disparities become even more pronounced with greater age gaps \cite{borovanskaEngagingChildrenUsing2020a}. Such mismatches increase the likelihood of overload, confusion, and disengagement.

A closely related developmental barrier is \textbf{pre-literacy}. Many XR systems rely on text-based instructions ($S(t)$) that pre-readers cannot decode ($D(t)$)~\cite{yuanExperimentalStudyEfficacy2022, wangARcampPTangibleProgramming2024, lewis-presserDesigningAugmentedReality2025, puDevelopmentSituationalInteraction2018, jamiatEffectsAugmentedReality2020, tuliEvaluatingUsabilityMobileBased2021a, panResearchApplicationImmersive2021, lorussoSemiimmersiveVirtualReality2020, weiEffectsARbasedVirtual2023}. This resulted in task errors, slowed interaction, or dependence on adults. Additionally, some younger children experienced \textbf{reality confusion}, indicating a friction where the System's visual fidelity ($S(t)$) blurs the boundaries of the Child's developing reality distinction ($D(t)$)~\cite{fridbergThematicTeachingAugmented2024}.

To address these constraints, studies introduced image-based or icon-driven interfaces \cite{yuanExperimentalStudyEfficacy2022, wangARcampPTangibleProgramming2024, puDevelopmentSituationalInteraction2018, tuliEvaluatingUsabilityMobileBased2021a}, recorded audio prompts \cite{jamiatEffectsAugmentedReality2020, weiEffectsARbasedVirtual2023}, and teacher mediation where adults read instructions aloud \cite{weiEffectsARbasedVirtual2023}. These adaptations reduced cognitive burden and improved accessibility for younger children.

\subsubsection{The Teacher: Instructional Readiness and Ecology}
The second friction point concerns the teacher’s capacity to orchestrate the Environmental context ($E(t)$) to accommodate the immersive System ($S(t)$).

\textbf{Teacher Readiness:} Across the literature, a critical environmental deficit is the limited pedagogical confidence teachers feel when adopting XR tools~\cite{aydogduAugmentedRealityPreschool2022, dasilvaCuboKidsProposal2020, sariDevelopingFinancialLiteracy2022, joDevelopmentUtilizationProjectorRobot2011a, kisnoDigitalStorytellingEarly2022, raptiEnrichingTraditionalLearning2023, liInnovativeApplicationAI2025a, safarEffectivenessUsingAugmented2017, balchaImpactAugmentedRealityassisted2025, vatamaniukRoleInteractiveMethods2024, poobrasertAugmentedLearningEnvironments2023}. Many teachers struggled to align the system's technical demands with existing pedagogical routines. Evidence indicates that typical 8-week training programs are inadequate; meaningful competence i.e. the ability to effectively modulate $E(t)$ around $S(t)$, requires extended, ongoing support over an academic year or longer~\cite{lewis-presserEnhancingPreschoolSpatial2025}.

To address these readiness gaps, studies highlighted comprehensive professional development to increase environmental capacity. These included targeted training on specific XR applications~\cite{aydogduAugmentedRealityPreschool2022, dasilvaCuboKidsProposal2020, sariDevelopingFinancialLiteracy2022, joDevelopmentUtilizationProjectorRobot2011a, kisnoDigitalStorytellingEarly2022, raptiEnrichingTraditionalLearning2023, hermanIntegratingSocialLearning2025, safarEffectivenessUsingAugmented2017, balchaImpactAugmentedRealityassisted2025, chenUsingAugmentedReality2019a}, the formation of XR-focused communities of practice~\cite{paliwalEnhancingEducationAugmented2024}, and long-term implementation cycles embedded in the school year~\cite{lewis-presserEnhancingPreschoolSpatial2025}. Some studies proposed hybrid models pairing teachers with AI or XR experts for continuous guidance~\cite{liInnovativeApplicationAI2025a}.


\begin{figure*}[ht]
\centering
\begin{minipage}{\textwidth}
    \centering
    \caption{Prevalence of Reported Health and Safety Risks by Device Type.}
    \label{fig:health-heatmap}
    \renewcommand{\arraystretch}{1.5}
    \setlength{\tabcolsep}{4pt}
    \small
    
    \providecommand{\h}[2]{\textcolor{customPurple!#1!black}{\textbf{#2}}}
    
    \begin{tabular}{|l| >{\centering\arraybackslash}p{2cm} | >{\centering\arraybackslash}p{2cm} | >{\centering\arraybackslash}p{2cm} | >{\centering\arraybackslash}p{2cm} |}
    \hline
    \textbf{Risk Category} & \textbf{\shortstack{Mobile/Tablets}} & \textbf{HMDs} & \textbf{\shortstack{PC/Laptops}} & \textbf{\shortstack{Specialized}} \\
    \hline
    \textbf{Cybersickness} & \h{1}{None} & \h{100}{Pervasive} & \h{1}{None} & \h{20}{Rare} \\ 
    \hline
    \textbf{Ergonomic Strain*} & \h{40}{Occasional} & \h{100}{Pervasive} & \h{70}{Common} & \h{70}{Common} \\ 
    \hline
    \textbf{Visual Fatigue} & \h{30}{Rare} & \h{40}{Occasional} & \h{1}{None} & \h{1}{None} \\ 
    \hline
    \textbf{Sedentary Behavior} & \h{70}{Common} & \h{30}{Rare} & \h{70}{Common} & \h{40}{Occasional} \\ 
    \hline
    \end{tabular}

    \vspace{2mm}
    \makebox[\linewidth][c]{%
        \parbox{0.9\textwidth}{ 
            \centering 
            \footnotesize
            \emph{Note:} Values reflect reporting frequency (not clinical severity): Pervasive ($>40\%$), Common ($25$--$40\%$), Occasional ($10$--$25\%$), Rare ($<10\%$). \\ 
            *Ergonomic strain denotes arm/neck fatigue (Mobiles/PCs) or cranial load (HMDs).
        }
    }
\end{minipage}

\Description[A table illustrating the prevalence of health and safety risks across different device types.]{A heatmap table categorizing health risks by prevalence (Pervasive, Common, Occasional, Rare) across four hardware categories. Key findings include: Cybersickness is 'Pervasive' for HMDs; Ergonomic Strain is 'Pervasive' for HMDs and 'Common' for PC/Laptops; Sedentary Behavior is 'Common' for Mobile/Tablets but 'Rare' for HMDs.}
\end{figure*}

\textbf{Classroom Management and Social Ecology:} XR also introduced frictions at the Social Ecology level ($E(t)$). Although often designed to promote collaboration, the System ($S(t)$) frequently isolated learners, disrupting the social environment: children became possessive of devices, resisted taking turns, or engaged in conflict rather than cooperative play~\cite{topuExaminationPreSchoolChildrens2023a, prekaAugmentedRealityQR2019a, liuApplicationTabletbasedAR2023a, hossainAugmentedRealityBasedElementary2021a, raptiInvestigatingEducatorsStudents2025}. Teachers reported difficulty managing the resulting noise and overstimulation~\cite{demirdagInvestigationEffectivenessAugmented2025a, panIntroducingAugmentedReality2021, duzyolInvestigationEffectAugmented2022a, korosidouEffectsAugmentedReality2024, wangImpactsAugmentedReality2024}. The immersive nature of XR also weakened teacher authority by shifting children’s attention to the device. Further complicating the ecology, parents in the MABLE study questioned the pedagogical value of using screens to teach self-regulation, perceiving it as contradictory to reducing screen ($S(t)$) dependency~\cite{gweonMABLEMediatingYoung2018}. To re-engineer the environment ($E(t)$), researchers recommend conducting activities individually or in small groups to reduce classroom chaos~\cite{topuExaminationPreSchoolChildrens2023a, demirdagInvestigationEffectivenessAugmented2025a, duzyolInvestigationEffectAugmented2022a, topuEffectsUsingAugmented2024a, chenUsingAugmentedReality2019a}. Additional mitigation strategies included using sound amplifiers to manage ambient noise~\cite{demirdagInvestigationEffectivenessAugmented2025a} and clearly defining supportive teacher roles during AR activities~\cite{demirdagInvestigationEffectivenessAugmented2025a, fengEffectsARLearning2022, chenUsingAugmentedReality2019a, drljevicInvestigatingDifferentFacets2022}. At the same time, to strengthen children’s peer interaction, several studies designed activities that used the full physical room beyond the device~\cite{burlesonActiveLearningEnvironments2018a} and incorporated structured group tasks~\cite{topuExaminationPreSchoolChildrens2023a, prekaAugmentedRealityQR2019a, mamani-calapujaLearningEnglishEarly2023}.

\subsubsection{The Design: Methodological and Content Blindspots}
The final friction stems from weaknesses in designing the System ($S(t)$) to match the Child ($D(t)$), and in the methodological capacity to measure this interaction.

\textbf{Content Suitability and Ceiling Effects:} Several studies exhibited a misalignment between the System's content ($S(t)$) and the Child's developmental baseline ($D(t)$). An AR intervention for teaching ``circles'', e.g., showed no measurable benefit because children had already mastered the concept with real objects~\cite{gecu-parmaksizAugmentedRealityBasedVirtual2019}, illustrating a clear \textbf{ceiling effect}~\cite{shahabUtilizingSocialVirtual2022} where the technology adds sensory complexity without pedagogical value. Other tasks, such as ``Musical Puzzles,'' were too abstract for this age group~\cite{lorussoSemiimmersiveVirtualReality2020}, reflecting an excessive \textbf{Cognitive Load ($C(t)$)} imposed by highly symbolic content. In some cases, overly vivid animations ($S(t)$) made children passive observers rather than active meaning-makers, leading to poorer outcomes than traditional print materials~\cite{wuComparingEffectsAR2025, luoEffectDifferentCombinations2023a, huangUsingAugmentedReality2016a}. These findings highlight a recurring issue: identifying content ($S(t)$) that is genuinely developmentally appropriate ($D(t)$)

\textbf{Methodological Flaws:} Most implementations were short-term  ($<60$ mins), making it impossible to distinguish between a temporary response to high Sensory Stimulation ($S(t)$) (Novelty) and genuine Developmental Gain ($D(t)$)~\cite{lorussoSemiimmersiveVirtualReality2020}. Furthermore, Environmental constraints ($E(t)$) regarding privacy often reduced data fidelity; e.g., one study avoided audio recordings and instead captured interview responses on paper~\cite{lawAugmentedRealityTechnology2025}, limiting the ability to analyze the richness of the child-system interaction. The field also still lacks clear design principles tailored to early childhood~\cite{simsekEffectAugmentedReality2024}. To address these design frictions, researchers proposed mitigation strategies that bridge the physical and digital worlds. One approach is ``Scaffolding Reality'', where children begin with physical creation ($D(t)$) and then scan their work into AR ($S(t)$), bridging concrete and digital representations~\cite{fridbergThematicTeachingAugmented2024}. Another strategy is the ``flipped classroom'' model, reserving AR for home exploration while keeping the classroom Environment ($E(t)$) focused on hands-on, imaginative activities to maintain curricular balance~\cite{idrisDevelopmentPatternLearning2021a}.

\subsection{Health, Safety, and Ergonomics (RQ2)}
This section examines the physiological and behavioral impact on the child. Our analysis reveals that health risks follow a causal link where the hardware choice ($S(t)$) dictates the complexity of the physiological strain on the child ($D(t)$).

We identified three primary domains of risk: \textbf{Neuro-vestibular} (Motion Sickness), \textbf{Musculoskeletal} (Ergonomics), and \textbf{Behavioral} (Sedentary Habits). Figure~\ref{fig:health-heatmap} visualizes the safety profile of each hardware category.

\subsubsection{Physiological Risks: The System-Body Conflict}
The literature indicates a distinct trade-off where the immersion level of the System ($S(t)$) creates specific strains on the Child's Physiology ($D(t)$).

\textbf{Neuro-vestibular Conflict (Cybersickness):}
This risk is predominantly associated with \textbf{VR HMDs}. It arises from a sensory conflict where the visual inputs from the System ($S(t)$) mismatch the vestibular inputs from the Child's physical body ($D(t)$). Studies warned about nausea, dizziness, and fatigue caused by this disconnect~\cite{vasilyevaASDAutisticSpectrum2025a, chuEffectsNonwearableDigital2023, liangExploitationMultiplayerInteraction2017, raptiInvestigatingEducatorsStudents2025, anTeachersPerceptionsEarly2023, dallolioImpactFantasyYoung2024, baehoikyoungVerificationVRPlay2025a, shimadaVRHandHygiene2017}. Unlike other devices, this risk is intrinsic to the medium of VR; even high-end hardware cannot fully eliminate the disconnect for young children with developing vestibular systems. To mitigate this, researchers heavily relied on Environmental controls ($E(t)$), reducing session duration~\cite{liangExploitationMultiplayerInteraction2017}, ensuring mandatory breaks~\cite{liangHandGesturebasedInteractive2017}, allowing habituation time~\cite{raptiInvestigatingEducatorsStudents2025}, and relying on active adult intervention to resolve immediate physical distress~\cite{anTeachersPerceptionsEarly2023, vasilyevaASDAutisticSpectrum2025a, baehoikyoungVerificationVRPlay2025a}.

\textbf{Musculoskeletal Strain:}
Ergonomic risks are serious concerns when it comes to XR usage. These risks bifurcated based on the device interaction model are:
\begin{itemize}
    \item \textbf{HMDs (Cranial Load):} The most critical ergonomic failure was the physical weight of headsets ($S(t)$). Standard devices are rarely designed for early childhood anthropometry ($D(t)$), leading to heavy front-loading on the neck and discomfort~\cite{raptiInvestigatingEducatorsStudents2025, anTeachersPerceptionsEarly2023, liangExploitationMultiplayerInteraction2017, shimadaVRHandHygiene2017}. Since hardware modification is difficult, mitigation relied almost exclusively on adult intervention ($E(t)$) to manually adjust the fit and support the device.
    \item \textbf{Mobiles/Tablets (Postural Load):} While lighter, handheld devices introduced strains and arm fatigue. This manifests as a friction where the \textbf{System's requirement for rigid camera angles ($S(t)$)} exceeds the \textbf{Child's static muscle endurance ($D(t)$)}~\cite{yilmazExaminationVocabularyLearning2022, lewis-presserDesigningAugmentedReality2025, tuliEvaluatingUsabilityMobileBased2021a, duzyolInvestigationEffectAugmented2022a, isikarslanogluThinkTogetherDesign2024}. To reduce this load, researchers employed Environmental supports ($E(t)$) and design adaptations. Tablet holders were a frequent solution to prevent strain~\cite{demirdagInvestigationEffectivenessAugmented2025a, wangARcampPTangibleProgramming2024, lawAugmentedRealityTechnology2025, borovanskaEngagingChildrenUsing2020a, tuliEvaluatingUsabilityMobileBased2021a, shimadaVRHandHygiene2017}. Interaction design choices included simplifying touch inputs ($S(t)$)~\cite{caiCaseStudyUsing2023} or using tangible materials that required no clicking~\cite{burlesonActiveLearningEnvironments2018a}. Other strategies involved providing rest periods~\cite{lewis-presserDesigningAugmentedReality2025, montoya-rodriguezEducationalInterventionTheory2025}, utilizing devices with wider fields of view to accommodate children's height~\cite{dalimTeachARInteractiveAugmented2016}, and encouraging peer collaboration so children could assist one another with ergonomic difficulties~\cite{kimOTCObjectCamera2019}.
\end{itemize}

\textbf{Visual Health:} 
Concerns regarding visual fatigue, eye strain, and headaches were pervasive across all screen-based interventions regardless of immersion level~\cite{wuComparingEffectsAR2025, mamani-calapujaLearningEnglishEarly2023, muangmoolDevelopmentSocialInteraction2023a, zhouUseAugmentedReality2020, huangUsingAugmentedReality2016a, linUsingHomebasedAugmented2025a}. The consensus mitigation was strict time-gating ($E(t)$); sessions were rarely permitted to exceed 20 minutes~\cite{muangmoolDevelopmentSocialInteraction2023a}, with some protocols enforcing limits as strict as 10 minutes to protect visual health~\cite{wuComparingEffectsAR2025}.

\subsubsection{Behavioral Risks: Displacement and Addiction}
Beyond immediate physical pain, the literature raised broader concerns about the long-term behavioral footprint of XR ($D(t)$ vs $S(t)$).

\textbf{Sedentary Lifestyle:} A significant discourse emerged regarding the risk of the System ($S(t)$) displacing essential physical activity ($D(t)$)~\cite{prekaAugmentedRealityQR2019a, abrarAugmentedRealityEducation2019a, fajrieAugmentedRealityMedia2022, wuComparingEffectsAR2025, dilekeryigitImpactAugmentedReality2025, panIntroducingAugmentedReality2021, mamani-calapujaLearningEnglishEarly2023, lorussoSemiimmersiveVirtualReality2020, balchaImpactAugmentedRealityassisted2025, dallolioImpactFantasyYoung2024, linUsingHomebasedAugmented2025a, stearneAugmentedRealityPlaygrounds2025}. Notably, several studies explicitly designed the XR experience to require physical activity, countering sedentary behavior by design~\cite{mamani-calapujaLearningEnglishEarly2023, balchaImpactAugmentedRealityassisted2025, stearneAugmentedRealityPlaygrounds2025}.

\textbf{Psychosocial Effects:} Finally, concerns regarding social isolation, reduced self-motivation, and compulsive use were noted~\cite{yilmazAreAugmentedReality2017, lawAugmentedRealityTechnology2025, hanExaminingYoungChildrens2015a, gweonMABLEMediatingYoung2018, balchaImpactAugmentedRealityassisted2025, huangUsingAugmentedReality2016a}. Mitigation in this domain relied on ``Human-in-the-Loop'' supervision ($E(t)$), with parents and teachers acting as active regulators of screen time rather than passive observers~\cite{fajrieAugmentedRealityMedia2022, linUsingHomebasedAugmented2025a}.

\subsection{Accessibility and Equity Barriers (RQ3)}
RQ3 addresses the dual challenges of accessibility and equity. In the AHD framework, these represent critical failures where the \textbf{System ($S(t)$)} is designed for a narrowly defined ``standard'' \textbf{Child ($D(t)$)} and a resource-rich \textbf{Environment ($E(t)$)}, systematically excluding diverse populations.

\subsubsection{Inclusion for Children with Disabilities: The System - Development Gap}
While numerous studies incidentally suggest that XR could support children with special needs~\cite{caiCaseStudyUsing2023, aguirregoitiaExperienceApplicationAugmented2017, demirdagInvestigationEffectivenessAugmented2025a, yilmazAreAugmentedReality2017, raptiEnrichingTraditionalLearning2023, tuliEvaluatingUsabilityMobileBased2021a, liangExploitationMultiplayerInteraction2017, shoshaniVirtualProsocialReality2023a, liangHandGesturebasedInteractive2017, dalimTeachARInteractiveAugmented2016}, this was rarely the primary design focus. We identified a specific subset of 13 papers designed explicitly for children with disabilities, highlighting a gap where standard XR interfaces ($S(t)$) fail to accommodate atypical developmental profiles ($D(t)$). Researchers justified this focus by noting that existing interventions are typically designed for older participants~\cite{vasilyevaASDAutisticSpectrum2025a, shahabUtilizingSocialVirtual2022, lorussoGiokAlienARbased2018, lorussoSemiimmersiveVirtualReality2020}, or by highlighting a critical scarcity of apps for autism~\cite{shahabUtilizingSocialVirtual2022, alcornDiscrepanciesVirtualLearning2013} and learning disabilities~\cite{poobrasertAugmentedLearningEnvironments2023}.

Within this subset, the largest group ($n=6$) focused on children with \textbf{Autism Spectrum Disorder (ASD)}~\cite{vasilyevaASDAutisticSpectrum2025a, chuEffectsNonwearableDigital2023, lorussoGiokAlienARbased2018, linUsingHomebasedAugmented2025a, alcornDiscrepanciesVirtualLearning2013, shahabUtilizingSocialVirtual2022}. Notably, the technological choice here deviated from the general trend: VR was the preferred medium ($n=4$)~\cite{vasilyevaASDAutisticSpectrum2025a, chuEffectsNonwearableDigital2023, alcornDiscrepanciesVirtualLearning2013, shahabUtilizingSocialVirtual2022}, likely because VR allows for total control of sensory stimulation ($S(t)$), reducing environmental distractions ($E(t)$) that trigger distress in neurodiverse children. A further five studies targeted \textbf{learning disabilities}~\cite{lawAugmentedRealityTechnology2025, jamiatEffectsAugmentedReality2020, paliwalEnhancingEducationAugmented2024, lorussoSemiimmersiveVirtualReality2020, poobrasertAugmentedLearningEnvironments2023}, relying almost exclusively on AR. The remaining two studies addressed physical disabilities~\cite{chien-yuAugmentedRealitybasedAssistive2010, baehoikyoungVerificationVRPlay2025a}. Overall, this confirms that inclusive design remains an edge case rather than a core component of the XR pipeline. In contrast to the ASD group, studies targeting \textbf{learning disabilities} relied almost exclusively on AR ($S(t)$), with only one exception using semi-immersive VR~\cite{lorussoSemiimmersiveVirtualReality2020}. The remaining two studies addressed physical disabilities~\cite{chien-yuAugmentedRealitybasedAssistive2010, baehoikyoungVerificationVRPlay2025a}, split evenly between platforms.

Overall, this technological divergence offers a critical AHD insight. While the broader cohort is heavily skewed towards AR (73\%), the accessibility literature demonstrates a significant shift toward VR. This suggests that for specific neurodiverse profiles ($D(t)$), the complete sensory control ($S(t)$) of VR is a functional necessity rather than a luxury, whereas general learning goals are adequately served by the lighter sensory overlays of AR. However, consistent with the wider field, the VR solutions employed were predominantly non-immersive to ensure safety and usability for the early-childhood group.

\subsubsection{Constraints in Low-Resource Settings: The System-Environment Conflict}

A significant majority of the reviewed literature ($n=73$) explicitly addressed the friction between high-tech system requirements ($S(t)$) and resource-constrained environments ($E(t)$). These studies consistently identified high cost and resource inaccessibility as primary challenges. This concern strongly correlated with hardware selection: handheld devices were the device of choice in 63\% of the studies, justified by the argument that ubiquitous hardware lowers the barrier to entry ($E(t)$). However, our analysis reveals a critical lack of consensus regarding what constitutes ``low-resource'' technology. First, a dominant group argues that VR HMDs are prohibitively expensive ($S(t)$) for these Environments ($E(t)$)~\cite{chien-yuAugmentedRealitybasedAssistive2010, tuliEvaluatingUsabilityMobileBased2021a, hermanIntegratingSocialLearning2025, anTeachersPerceptionsEarly2023, baehoikyoungVerificationVRPlay2025a}. Consequently, these studies present AR on standard mobile phones as the solution to equity~\cite{kulasekaraGameCentricElearning2025, wangARcampPTangibleProgramming2024, syahidiARChildAnalysisEvaluation2019a, yilmazAreAugmentedReality2017, antoniaArtfulThinkingAugmented2018, kurniawanARtraceAugmentedReality2019, prekaAugmentedRealityQR2019a, ayuagungmasaristamyAugmentedRealityCultural2024a, abrarAugmentedRealityEducation2019a, hossainAugmentedRealityBasedElementary2021a, lewis-presserDesigningAugmentedReality2025, sariDevelopingFinancialLiteracy2022, paliwalEnhancingEducationAugmented2024, lewis-presserEnhancingPreschoolSpatial2025, hoInteractiveMultisensoryVolumetric2023, duzyolInvestigationEffectAugmented2022a, bubpamasDevelopmentEarlyChildhood2024a, isikarslanogluThinkTogetherDesign2024, khairulanuardiIncreaseReadingHabit2022, kusumaningsihUserExperienceMeasurement2021}.

In direct contrast, a second group considers tablets and mobile phones themselves to be ``luxury items'' ($S(t)$) that act as an obstacle to access in under-funded contexts ($E(t)$)~\cite{fajrieAugmentedRealityMedia2022, lawAugmentedRealityTechnology2025, borovanskaEngagingChildrenUsing2020a, dilekeryigitImpactAugmentedReality2025, topuEffectsUsingAugmented2024a, huangUsingAugmentedReality2016a}. Adding a third dimension, several studies challenge the notion that VR is inherently exclusive, presenting low-cost HMDs (e.g., cardboard viewers) as a viable solution~\cite{vasilyevaASDAutisticSpectrum2025a, shoshaniVirtualProsocialReality2023a, raptiInvestigatingEducatorsStudents2025, muangmoolDevelopmentSocialInteraction2023a, hsiaoUsingGestureInteractive2016a, kusumaVirtualRealityLearning2018, shimadaVRHandHygiene2017}. Thus, while there is universal agreement that cost reduction is essential, the specific technology that achieves this remains relative to the local economic context. Beyond hardware, infrastructural constraints ($E(t)$) such as unstable internet connectivity were cited as a barrier to cloud-dependent Systems ($S(t)$)~\cite{tanApplicationArtbasedKnowledge2025, aladinARTOKIDSpeechenabledAugmented2020a, lawAugmentedRealityTechnology2025, dilekeryigitImpactAugmentedReality2025, liInnovativeApplicationAI2025a, safarEffectivenessUsingAugmented2017, topuEffectsUsingAugmented2024a}. Finally, software licensing costs were noted as a restriction, with studies proposing open-access models to align the tech with the economic reality of public education~\cite{dasilvaCuboKidsProposal2020, caiCaseStudyUsing2023, lewis-presserDesigningAugmentedReality2025, gecu-parmaksizEffectAugmentedReality2020a}.

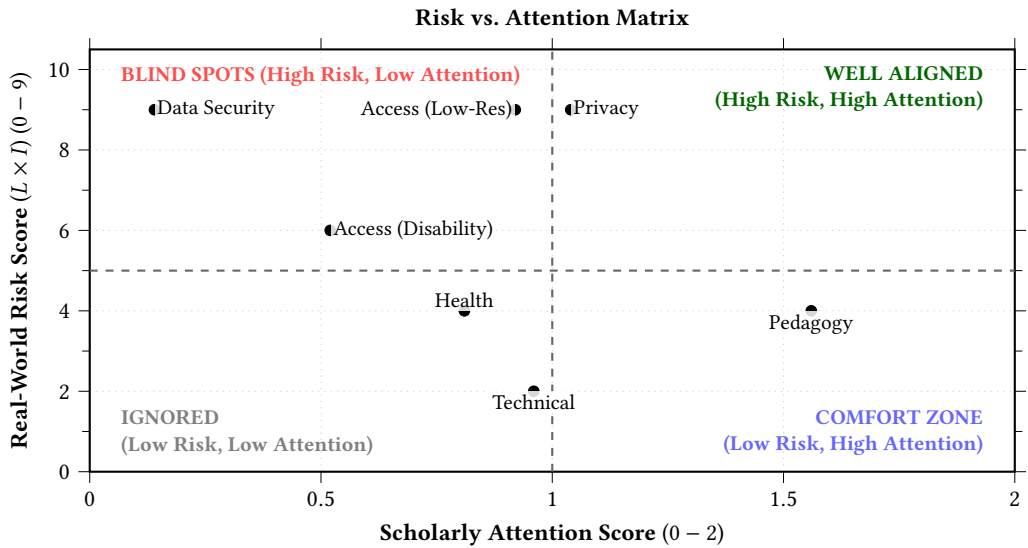
\begin{figure*}[htbp]
\centering
    \begin{tikzpicture}
        \begin{axis}[
            width=0.78\linewidth,
            height=7.2cm,
            xlabel={\textbf{Scholarly Attention Score} ($0-2$)},
            ylabel={\textbf{Real-World Risk Score} ($L \times I$) ($0-9$)},
            xmin=0, xmax=2,
            ymin=0, ymax=10.5,
            xtick={0, 0.5, 1, 1.5, 2},
            ytick={0, 2, 4, 6, 8, 10},
            minor y tick num=1,
            axis line style={draw=black, thick},
            tick style={draw=black},
            tick align=outside,
            grid=major,
            grid style={dotted, black!20},
            title={\textbf{Risk vs. Attention Matrix}},
            clip=false,
            scatter/classes={
                a={mark=*, blue!60, draw=black, mark size=3pt},
                b={mark=*, red!70, draw=black, mark size=3pt},
                c={mark=square*, orange, draw=black, mark size=3pt}
            }
        ]
            \draw[black!60, dashed, thick] (axis cs:1, 0) -- (axis cs:1, 10.5);
            \draw[black!60, dashed, thick] (axis cs:0, 5) -- (axis cs:2, 5);

            \node[anchor=north west, align=left, font=\bfseries\small, color=red!70] 
                at (axis cs:0.05, 10.3) {BLIND SPOTS (High Risk, Low Attention)};
            
            \node[anchor=north east, align=right, font=\bfseries\small, color=green!40!black] 
                at (axis cs:1.95, 10.3) {WELL ALIGNED\\(High Risk, High Attention)};
            
            \node[anchor=south east, align=right, font=\bfseries\small, color=blue!60] 
                at (axis cs:1.95, 0.2) {COMFORT ZONE\\(Low Risk, High Attention)};
            
            \node[anchor=south west, align=left, font=\bfseries\small, color=gray] 
                at (axis cs:0.05, 0.2) {IGNORED\\(Low Risk, Low Attention)};

            \addplot[
                scatter,
                only marks,
                scatter src=explicit symbolic,
                visualization depends on={value \thisrow{label}\as\mylabel},
                visualization depends on={value \thisrow{anchor}\as\myanchor},
                nodes near coords={\mylabel},
                every node near coord/.append style={anchor=\myanchor, font=\small, color=black, fill=white, inner sep=1pt, opacity=0.85, text opacity=1}
            ] table [meta=class] {
                x      y    class   label                 anchor
                0.14   9    b       {Data Security}       west
                0.52   6    b       {Access (Disability)} west
                1.04   9    c       Privacy               west
                0.92   9    c       {Access (Low-Res)}    east
                1.56   4    a       Pedagogy              north
                0.96   2    a       Technical             north
                0.81   4    a       Health                south
            };
        \end{axis}
    \end{tikzpicture}
    \caption{`Risk Assessment Matrix' Mapping Scholarly Attention against Calculated Real-World Risk.}
    \label{fig:risk-matrix}
    \Description[A scatter plot mapping seven challenge domains based on Scholarly Attention versus Real-World Risk.]{A scatter plot titled 'Risk vs. Attention Matrix' divided into four quadrants. The X-axis represents 'Scholarly Attention Score' (0 to 2), and the Y-axis represents 'Real-World Risk Score' (0 to 9).
- The 'Blind Spots' quadrant (High Risk, Low Attention) contains 'Data Security' (Risk 9, Attention 0.14) and 'Access (Disability)' (Risk 6, Attention 0.52).
- The 'Well Aligned' quadrant (High Risk, High Attention) contains 'Privacy' (Risk 9, Attention 1.04) and 'Access (Low-Res)' (Risk 9, Attention 0.92).
- The 'Comfort Zone' quadrant (Low Risk, High Attention) contains 'Pedagogy' (Risk 4, Attention 1.56) and 'Technical' (Risk 2, Attention 0.96).
- 'Health' (Risk 4, Attention 0.81) falls into the 'Ignored' quadrant (Low Risk, Low Attention).}
\end{figure*}

\subsection{Privacy and Data Practices Landscape (RQ4)}
Our analysis exposes a critical gap in the \textbf{Environmental Governance ($E(t)$)} of XR data. While the \textbf{System ($S(t)$)} relies on continuous, high-fidelity sensing of the physical world and learner actions, specifically through camera feeds, microphone input, and motion sensors, the literature focuses almost exclusively on procedural consent rather than technical data protection. Despite the sensitive nature of the population (children aged $3-8$) and these data-intensive capabilities, only a small fraction of studies ($n=7$) discussed technical security risks or architectural mitigations~\cite{vasilyevaASDAutisticSpectrum2025a, lorussoGiokAlienARbased2018, liInnovativeApplicationAI2025a, gweonMABLEMediatingYoung2018, wuComparingEffectsAR2025, lewis-presserDesigningAugmentedReality2025, lewis-presserEnhancingPreschoolSpatial2025}. Instead, our analysis reveals that the literature predominantly conflates ``privacy'' with procedural consent. This represents a failure of the research environment ($E(t)$) to match the sensing capabilities of the tools ($S(t)$) being deployed; while VR studies occasionally flagged biometric risks~\cite{liInnovativeApplicationAI2025a}, the video and audio capture inherent to the dominant AR interventions ($73\%$) remained largely unscrutinized from a data security perspective.

This governance gap manifests as a distinct disparity between the type of data collected ($S(t)$) and the measures used to protect it ($E(t)$). While 61 studies addressed privacy, they did so almost exclusively through the lens of research ethics (e.g., obtaining parental consent) or basic \textbf{data anonymization} techniques, such as blurring faces in multimedia and random coding~\cite{topuExaminationPreSchoolChildrens2023a, tanApplicationArtbasedKnowledge2025, yilmazAreAugmentedReality2017, lawAugmentedRealityTechnology2025, chien-yuAugmentedRealitybasedAssistive2010, wuComparingEffectsAR2025, lewis-presserDesigningAugmentedReality2025, liInnovativeApplicationAI2025a, gweonMABLEMediatingYoung2018, wangImpactsAugmentedReality2024, fridbergThematicTeachingAugmented2024, isikarslanogluThinkTogetherDesign2024, linUsingHomebasedAugmented2025a}. Critical issues such as the encrypted storage of these biometric traces and secure third-party data sharing were overwhelmingly absent. This omission is particularly concerning given that numerous studies collected these raw multimedia data, without specifying secure retention protocols~\cite{hanExaminingYoungChildrens2015a, shoshaniVirtualProsocialReality2023a, hermanIntegratingSocialLearning2025, lorussoSemiimmersiveVirtualReality2020, wangImpactsAugmentedReality2024, fridbergThematicTeachingAugmented2024, huangUsingAugmentedReality2016a, kimVirtualStorytellerUsing2011}. The rare instances where technical architecture ($S(t)$) was successfully aligned with privacy requirements ($n=7$) focused on three specific strategies:
\begin{enumerate}
    \item \textbf{Data Storage Architecture:} Decisions to store sensitive information locally (air-gapped) rather than transmitting to the cloud~\cite{vasilyevaASDAutisticSpectrum2025a, lorussoGiokAlienARbased2018, liInnovativeApplicationAI2025a, gweonMABLEMediatingYoung2018};
    \item \textbf{Access Control:} Implementing authentication mechanisms such as password protection~\cite{gweonMABLEMediatingYoung2018}; and
    \item \textbf{Data Governance:} Advanced measures for data validity, such as blockchain-based auditing logs or formal protection protocols~\cite{wuComparingEffectsAR2025, lewis-presserDesigningAugmentedReality2025, lewis-presserEnhancingPreschoolSpatial2025, liInnovativeApplicationAI2025a}.
\end{enumerate}

Conversely, the lack of such technical solutions has led to low-tech workarounds; for instance, one study resorted to recording interview feedback on paper rather than using voice recorders specifically to ``maintain children's privacy''~\cite{lawAugmentedRealityTechnology2025}. Hence, the answer to RQ4 is that while \textit{procedural} privacy is well-established, specific \textit{technical} risks remain largely unidentified and unmitigated within the current body of ECE XR literature.

\section{Discussion}
\label{sec:discussion}
We assess whether scholarly attention corresponds to actual danger by deriving a Real-World Risk metric ($R$) for each domain. To avoid circular reasoning, we compute $R$ independently of scholarly attention, relying on established external standards such as NIST, ITU, and WHO rather than citation frequency. We adopt a standard qualitative risk model using $R = L \times I$, where \textit{Likelihood} ($L$) reflects the frequency of occurrence in typical deployments (1 = Rare, 3 = Ubiquitous) and \textit{Impact} ($I$) captures the severity of potential harm (1 = Inconvenience, 3 = Critical or Irreversible). Full details appear in Appendix~\ref{app:risk-methodology}. The resulting scores separate high-priority safety concerns from low-severity operational issues. \textit{Privacy}, \textit{Data Security}, and \textit{Economic Access} receive the maximum score of 9 due to the irreversible nature of biometric data leakage and the structural exclusion of low-resource populations \cite{nistir8259, itu2023facts}. \textit{Health} and \textit{Pedagogy} fall in the moderate range (score 4) because most effects, such as nausea, are transient \cite{meta2023safety}. \textit{Technical Limitations} score lowest (2), as they primarily reflect reduced usability rather than substantive harm.

\subsection{The Risk vs. Attention Gap}
\label{sec:risk-matrix}
We identify mismatches between the severity of potential harm and the amount of academic attention each issue receives. To illustrate this gap, we plot our two operational dimensions: Scholarly Attention (0 to 2) against Real-World Risk (1 to 9), shown in Figure~\ref{fig:risk-matrix}. The matrix is divided into four quadrants, with the upper-left quadrant representing the most critical blind spots: risks that are severe in practice but under-examined in research.

\subsubsection{Critical Blind Spots: Data Security and Inclusion}
The most concerning result is the placement of \textit{Data Security} in the high-risk, low-attention quadrant. Data security failures in XR can expose minors' biometric data, producing long-term and potentially irreversible consequences (Risk Level $9$), yet the topic receives almost no scholarly discussion (Attention $0.14$). Under the AHD perspective, this reflects a breakdown of \textit{Environmental Governance} ($E(t)$) in safeguarding the child's developmental profile ($D(t)$). Likewise, \textit{Accessibility for Disabilities} (Risk Level $6$, Attention $0.52$) occupies the same neglected space. Because exclusion from educational technologies can compound existing inequities, the lack of attention suggests that the system-level properties ($S(t)$) are optimized for neurotypical and able-bodied users rather than the full range of developmental diversity ($D(t)$).

\subsubsection{The Awareness Zone: Privacy and Equity}
\textit{Privacy} and \textit{Low-Resource Accessibility} lie near the mid-attention range. Both carry the highest risk scores (9) due to stringent regulatory requirements such as GDPR-K and the impact of inequity. Their location near the center of the x-axis indicates that although these issues appear frequently in discussions, the engagement is often shallow. Privacy is typically framed as a matter of parental consent within the environment ($E(t)$) but rarely extends to robust technical controls such as anonymization or minimization at the system level ($S(t)$).

\subsubsection{The Comfort Zone and Neglected Physiology}
\textit{Pedagogical Challenges} (Attention 1.56) and \textit{Technical Issues} (Attention $0.96$) appear in the high-attention bandwidth. These issues arise frequently but usually affect usability or learning effectiveness rather than causing lasting harm. The extensive attention here suggests that the community focuses heavily on aligning \textit{Cognitive Load} ($C(t)$) and \textit{Sensory Input} ($S(t)$) to support learning outcomes. Conversely, \textit{Health Issues} (Risk 4, Attention 0.81) fall into the `Ignored' quadrant despite carrying the same risk score as Pedagogy. This reveals a disconnect: the field aggressively optimizes for the child's mind while neglecting physiological limitations ($D(t)$). Combined with the gaps in security and inclusion, this confirms a preference for optimizing short-term learning experiences over addressing systemic, long-term risks.

\subsection{Implications for Design and Policy}
\label{sec:implications}
Our SoK reveals an early-childhood XR landscape that is innovative yet uneven. Building on our findings, we outline priorities for designing, integrating, and governing XR in early childhood settings.

\subsubsection{Designing XR as Augmented Human Development}
AHD shifts design from a focus on devices to a focus on how XR shapes cognitive load \(C(t)\), sensory stimulation \(S(t)\), environmental context \(E(t)\), and developmental fit \(D(t)\). Many systems in our corpus prioritize novelty, stacking dense visual overlays, gesture demands, textual instructions, and game rules that exceed young children's attentional and working-memory limits~\cite{kulasekaraGameCentricElearning2025,lewis-presserDesigningAugmentedReality2025,yuanExperimentalStudyEfficacy2022}. When paired with heavy or ill-fitting equipment, several VR studies report cybersickness, disorientation, and ergonomic strain, disrupting \(S(t)\) and undermining learning~\cite{vasilyevaASDAutisticSpectrum2025a,chuEffectsNonwearableDigital2023,baehoikyoungVerificationVRPlay2025a}. A developmentally aligned design approach should therefore simplify interaction flows for \(C(t)\), provide perceptually stable scenes for \(S(t)\), and support social co-presence for \(E(t)\) through projection-based MR, shared-tabletop AR, and large displays~\cite{burlesonActiveLearningEnvironments2018a,gweonMABLEMediatingYoung2018,besevliMaRTDesigningProjectionbased2019}. For \(D(t)\), interaction vocabularies must account for pre-literacy and motor constraints using large touch targets, tangible props, and spoken guidance~\cite{wangARcampPTangibleProgramming2024,puDevelopmentSituationalInteraction2018,lorussoSemiimmersiveVirtualReality2020}. Findings across the corpus also support cautious use of fully immersive HMD-based VR for ages $3-8$. Manufacturer warnings, teacher concerns, and empirical reports of sickness and strain suggest prioritizing non-wearable or lightweight XR, reserving HMD use for brief, justified episodes with clear exit mechanisms and close supervision~\cite{bexson2024safety,pawelczyk2025understanding,raptiInvestigatingEducatorsStudents2025}. In essence, early-childhood XR should be approached as \emph{AHD-aligned learning environments that sometimes use XR}, rather than XR systems searching for pedagogical justification.

\subsubsection{Embedding XR in Early Childhood Practice}
The central challenge for teachers is not whether XR ``works,'' but how it fits the realities of early-childhood classrooms. Educators report difficulties managing attention, noise, and turn-taking; limited time to integrate XR into curricula; and concerns about displacing play-based or inquiry-driven approaches~\cite{tanApplicationArtbasedKnowledge2025,topuExaminationPreSchoolChildrens2023a,panIntroducingAugmentedReality2021,lawAugmentedRealityTechnology2025,fridbergThematicTeachingAugmented2024}. Stronger outcomes arise when XR is embedded within a multimodal sequence, such as augmented storytelling combined with drawing, discussion, and physical exploration~\cite{antoniaArtfulThinkingAugmented2018,wuComparingEffectsAR2025,huangUsingAugmentedReality2016a}. This pattern points to XR as a short, purposeful component within a broader activity arc; visualizing invisible processes, simulating otherwise inaccessible environments, or scaffolding collaborative storytelling that children later enact physically. Teachers need time and institutional support to curate XR content and determine whether it genuinely advances learning objectives or adds unnecessary complexity. Professional development is most effective when centered on co-designed lesson structures (``XR explore–discuss–draw,'' ``XR introduce–hands-on practice–XR review'') rather than tool-focused workshops~\cite{lewis-presserEnhancingPreschoolSpatial2025,raptiEnrichingTraditionalLearning2023}. Equity concerns further shape integration. Most systems assume recent devices, stable connectivity, and technical support, conditions that vary greatly across schools~\cite{hossainAugmentedRealityBasedElementary2021a,liuApplicationTabletbasedAR2023a,alzahrani2025systematic}. Few studies consider offline operation, shared-device models, or low-resource classrooms~\cite{everri2024cyborg}. To avoid deepening inequalities, XR must support older hardware, intermittent connectivity, and flexible device-sharing. Accessibility for children with disabilities also remains under-addressed, appearing mainly in specialized work rather than as a baseline expectation~\cite{lin2016augmented,chițu2023exploring,vasilyevaASDAutisticSpectrum2025a}. Embedding XR in practice therefore requires curricular integration, infrastructural planning, and universal design as a default.

\subsubsection{Governing Data, Safety, and Equity in Child-Centered XR}
XR systems for young children are structurally data-rich: even simple AR apps log timestamps, interaction traces, and camera imagery, while more advanced systems capture motion trajectories, gaze, speech, and sometimes physiological data~\cite{jawalkar2024ethical,kaimara2022could}. Yet privacy and data security in the reviewed studies are typically limited to brief references to consent or anonymization, with little detail on data minimization, storage, access control, or incident response. Health risks such as cybersickness and visual strain appear somewhat more often but are rarely measured systematically~\cite{vasilyevaASDAutisticSpectrum2025a,lorussoSemiimmersiveVirtualReality2020,baehoikyoungVerificationVRPlay2025a}. This gap suggests that early-childhood XR currently depends more on professional ethics than explicit governance. A ``child-centered XR by design'' approach is needed~\cite{reddington2022development}. At the product level, systems should enforce strong data minimization, collecting only essential information, and provide transparent explanations to parents and educators about what data are collected, where they are stored, and for how long~\cite{milkaite2020child}. Compliance with COPPA, GDPR-K, and related frameworks must shape sensor selection, cloud connectivity, and analytics pipelines from the outset~\cite{jawalkar2024ethical,kaimara2022could}. At the institutional level, procurement policies can require demonstrable privacy, data security, accessibility, and age-appropriate safety measures before adoption~\cite{miao2021ai}. The broader commercial ecosystem also raises concerns about persuasive design, reward loops, and data-driven personalization that may blur boundaries between learning and commercial engagement~\cite{everri2024cyborg, tazi2021parents}. Aligning XR with children’s rights to education, play, privacy, and protection from exploitation requires coordinated governance across designers, educators, parents, regulators, and children. Participatory processes that enable these stakeholders to co-evaluate XR practices and challenge harmful defaults are essential for steering immersive technologies toward genuinely child-centered futures.

\subsubsection{Reorienting the XR Research Agenda}
The current evidence base relies heavily on short pre/post designs focused on immediate learning gains or engagement, with far less attention to long-term developmental, social, or ethical trajectories~\cite{yilmazAreAugmentedReality2017,wuComparingEffectsAR2025,balchaImpactAugmentedRealityassisted2025}. 
Key details like session duration, cumulative exposure, data retention, and disability accommodations are frequently underreported, limiting synthesis and comparison~\cite{caiCaseStudyUsing2023, calle-bustosAugmentedRealityGame2017, kurniawanARtraceAugmentedReality2019, ayuagungmasaristamyAugmentedRealityCultural2024a}. Our AHD-based coding also shows that data security, disability accessibility, and chronic health risks are significantly under-theorized relative to their potential impact. A more aligned research agenda requires treating AHD components and risk dimensions as core outcomes. This includes developing age-appropriate measures of cognitive load, sensory comfort, social dynamics, and sense of agency; assessing privacy, literacy, trust, and perceived safety among parents and teachers; and conducting longitudinal, mixed-method studies that capture habituation, shifting classroom norms, and cumulative exposure effects~\cite{fridbergThematicTeachingAugmented2024}. More consistent reporting standards, covering modality, hardware, session structure, data practices, and accessibility, would enable a coherent, comparable evidence base for guiding safe, equitable, and developmentally grounded XR in early childhood education.

\section{Limitations and Future Work}
We analyze English-language papers indexed in eight major databases, so relevant grey literature, non-English publications, and technical reports may be missing. Future work will broaden coverage to additional sources and languages, develop shared reporting guidelines for XR in early childhood, and design longitudinal, AHD-based studies that more precisely measure cognitive load, sensory comfort, and developmental alignment over time.

\section{Conclusion}
\balance
\label{sec:conclusion}
In this SoK, we analyzed $111$ empirical XR studies involving children aged 3 to 8 years and identified consistent patterns in system design, deployment, and risk. AR is the dominant modality with $81$ studies, while VR appears in $21$ and MR in five, which confirms that early childhood XR remains primarily non-immersive. Sample sizes range from $2$ to $497$ participants, with a mean of $52$ and a median of $30$, and reported session lengths cluster around brief exposures with a median of $20$ minutes. These patterns indicate that current evidence is grounded in short, small-scale trials rather than sustained long-term use. Additionally, our AHD-based analysis shows that many systems overload cognitive and sensory channels through dense visual content, fine motor gestures, or unstable augmentations. Attention scoring also reveals that pedagogical challenges receive the most emphasis across the corpus, whereas data security receives almost none. Surprisingly, while research ethics (e.g., informed consent) are widely reported, none of the papers substantively address technical data privacy or specific regulatory compliance (e.g., COPPA and GDPR-K), even though many applications record camera imagery, motion traces, touch logs, or speech data. Our risk matrix further highlights that these low-attention areas carry high real-world impact. We conclude that future XR systems for early childhood must prioritize developmental alignment, transparent data practices, disability-accessible interaction design, and longitudinal evaluation of cognitive load and sensory comfort.

\begin{acks}
We would like to thank Mahmudul Hasan for his contribution to record identification and codebook creation. We would like to acknowledge the Data Agency and Security (DAS) Lab at George Mason University where the study was conducted. The opinions expressed
in this work are solely those of the authors.
\end{acks}

\bibliographystyle{ACM-Reference-Format}
\bibliography{Main}

@article{alzahrani2025systematic,
  title={{A systematic review of the use of information communication technology, including augmented reality, in the teaching of science to preschool children}},
  author={Alzahrani, Arwa},
  journal={International Journal of Educational Research Open},
  volume={9},
  pages={100453},
  year={2025},
  publisher={Elsevier}
}

@inproceedings{huang2025systemization,
  title={Systemization of Knowledge (SoK): Goals, coverage, and evaluation in cybersecurity and privacy games},
  author={Huang, Yue and Grobler, Marthie and Ferro, Lauren S and Psaroulis, Georgia and Das, Sanchari and Wei, Jing and Janicke, Helge},
  booktitle={Proceedings of the 2025 CHI Conference on Human Factors in Computing Systems},
  pages={1--27},
  year={2025},
address = {Yokohama, Japan},
publisher = {Association for Computing Machinery}
}

@inproceedings{duzgun2022sok,
  title={Sok: A systematic literature review of knowledge-based authentication on augmented reality head-mounted displays},
  author={D{\"u}zg{\"u}n, Reyhan and Noah, Naheem and Mayer, Peter and Das, Sanchari and Volkamer, Melanie},
  booktitle={Proceedings of the 17th International Conference on Availability, Reliability and Security},
  pages={1--12},
  year={2022},
address = {Vienna, Austria},
publisher = {Association for Computing Machinery}
}

@article{shrestha2022exploring,
  title={Exploring gender biases in ML and AI academic research through systematic literature review},
  author={Shrestha, Sunny and Das, Sanchari},
  journal={Frontiers in artificial intelligence},
  volume={5},
  pages={976838},
  year={2022},
  publisher={Frontiers Media SA}
}

@inproceedings{das2019all,
  title={All About Phishing Exploring User Research through a Systematic Literature Review},
  author={Das, Sanchari and Kim, Andrew and Tingle, Zachary and Nippert-Eng, Christena},
  booktitle={Proceedings of the Thirteenth International Symposium on Human Aspects of Information Security \& Assurance (HAISA 2019)},
  year={2019},
address = {Nicosia, Cyprus},
numpages = {10},
publisher = {University of Plymouth Press}
}

@article{noah2021exploring,
  title={Exploring evolution of augmented and virtual reality education space in 2020 through systematic literature review},
  author={Noah, Naheem and Das, Sanchari},
  journal={Computer Animation and Virtual Worlds},
  volume={32},
  number={3-4},
  pages={e2020},
  year={2021},
  publisher={Wiley Online Library}
}

@inproceedings{jones2021literature,
  title={A literature review on virtual reality authentication},
  author={Jones, John M and Duezguen, Reyhan and Mayer, Peter and Volkamer, Melanie and Das, Sanchari},
  booktitle={International Symposium on Human Aspects of Information Security and Assurance},
  pages={189--198},
  year={2021},
  organization={Springer},
publisher = {Springer},
address = {Virtual Event}
}

@inproceedings{kishnani2023blockchain,
  title={Blockchain in oil and gas supply chain: a literature review from user security and privacy perspective},
  author={Kishnani, Urvashi and Madabhushi, Srinidhi and Das, Sanchari},
  booktitle={International Symposium on Human Aspects of Information Security and Assurance},
  pages={296--309},
  year={2023},
  organization={Springer},
publisher = {Springer},
address = {Kent, UK}
}

@inproceedings{kishnani2024dual,
  title={Dual-Technique Privacy \& Security Analysis for E-Commerce Websites Through Automated and Manual Implementation},
  author={Kishnani, Urvashi and Das, Sanchari},
  booktitle={Proceedings of the 2025 Hawaii International Conference on System Sciences (HICSS)},
  year={2024},
  publisher={University of Hawaii at Manoa},
  address={Hawaii, USA},
  numpages={1--10}
}

@article{adhikari2025natural,
  title={Natural Language Processing of Privacy Policies: A Survey},
  author={Adhikari, Andrick and Das, Sanchari and Dewri, Rinku},
  journal={arXiv e-prints},
  numpages={1--27},
  year={2025},
  volume={0}
}

@inproceedings{tazi2024we,
  title={“We Have No Security Concerns”: Understanding the Privacy-Security Nexus in Telehealth for Audiologists and Speech-Language Pathologists},
  author={Tazi, Faiza and Dykstra, Josiah and Rajivan, Prashanth and Das, Sanchari},
  booktitle={Proceedings of the 2024 CHI Conference on Human Factors in Computing Systems},
  pages={1--20},
  year={2024},
  publisher={ACM},
  address={Honolulu, Hawaii, USA}
}

@inproceedings{agarwal2025systematic,
  title={Systematic Literature Review of Vulnerabilities and Defenses in VPNs, Tor, and Web Browsers},
  author={Agarwal, Neha and Mackin, Ethan and Tazi, Faiza and Grover, Mayank and More, Rutuja and Das, Sanchari},
  booktitle={International Conference on Information Systems Security},
  pages={357--375},
  year={2025},
address={Indore, Madhya Pradesh, India},
  organization={Springer},
publisher={Springer}
}

@inproceedings{majumdar2021sok,
  title={Sok: An evaluation of quantum authentication through systematic literature review},
  author={Majumdar, Ritajit and Das, Sanchari},
  booktitle={Proceedings of the Workshop on Usable Security and Privacy (USEC)},
  year={2021},
address = {Virtual Event},
publisher = {Internet Society},
numpages = {10}
}

@inproceedings{podapati2025sok,
  title={SoK: a systematic review of context-and behavior-aware adaptive authentication in mobile environments},
  author={Podapati, Vyoma Harshitha and Nigam, Divyansh and Das, Sanchari},
  booktitle={International Symposium on Human Aspects of Information Security and Assurance},
  pages={406--419},
  year={2025},
  organization={Springer},
publisher = {Springer},
address = {Mytilene, Greece}
}

@inproceedings{saka2025sok,
  title     = {SoK: Reviewing Two Decades of Security, Privacy, Accessibility, and Usability Studies on Internet of Things for Older Adults},
  author    = {Saka, Suleiman and Das, Sanchari},
  booktitle = {Proceedings of the 21st ACM Asia Conference on Computer and Communications Security (ASIACCS)},
  year      = {2026},
  publisher = {ACM},
address = {Bangalore, India},
  doi       = {10.48550/arXiv.2512.16394},
numpages = {18}
}

@article{tazi2024sok,
  title={Sok: Analyzing privacy and security of healthcare data from the user perspective},
  author={Tazi, Faiza and Nandakumar, Archana and Dykstra, Josiah and Rajivan, Prashanth and Das, Sanchari},
  journal={ACM Transactions on Computing for Healthcare},
  volume={5},
  number={2},
  pages={1--31},
  year={2024},
  publisher={ACM}
}

@inproceedings{grover2025sok,
  title={SoK: a systematic review of privacy and security in healthcare robotics},
  author={Grover, Mayank and Das, Sanchari},
  booktitle={International Conference on Social Robotics},
  pages={212--234},
  year={2025},
  organization={Springer},
address = {Naples, Italy},
publisher = {Springer}
}

@article{tazi2022sok,
  title={Sok: An evaluation of the secure end user experience on the dark net through systematic literature review},
  author={Tazi, Faiza and Shrestha, Sunny and De La Cruz, Junibel and Das, Sanchari},
  journal={Journal of Cybersecurity and Privacy},
  volume={2},
  number={2},
  pages={329--357},
  year={2022},
  publisher={MDPI}
}

@inproceedings{das2022sok,
  title={SoK: a proposal for incorporating accessible gamified cybersecurity awareness training informed by a systematic literature review},
  author={Das, Sanchari and others},
  booktitle={Proceedings of the workshop on usable security and privacy (USEC)},
  year={2022},
publisher = {Internet Society},
address = {San Diego, CA, USA},
numpages = {13}
}

@inproceedings{tazi2023sok,
  title     = {SoK: Analysis of User-Centered Studies Focusing on Healthcare Privacy and Security},
  author    = {Tazi, Faiza and Nandakumar, Archana and Dykstra, Josiah and Rajivan, Prashanth and Das, Sanchari},
  booktitle = {Proceedings of the Nineteenth Symposium on Usable Privacy and Security (SOUPS 2023)},
  year      = {2023},
  publisher = {USENIX Association},
  address   = {Anaheim, CA, USA},
  note      = {Poster Abstract},
  url       = {https://www.usenix.org/system/files/soups2023-poster12_tazi_abstract_final.pdf},
numpages = {9}
}

@inproceedings{zezulak2023sok,
  title={SoK: Evaluating Privacy and Security Concerns of Using Web Services for the Disabled Population},
  author={Zezulak, Alisa and Tazi, Faiza and Das, Sanchari},
  booktitle={7th Workshop on Technology and Consumer Protection (ConPro’23)},
address = {San Francisco, CA, USA},
  year={2023},
numpages = {8},
publisher = {Institute of Electrical and Electronics Engineers}
}

@inproceedings{shrestha2022sok,
  title={SoK: A systematic literature review of bluetooth security threats and mitigation measures},
  author={Shrestha, Sunny and Irby, Esa and Thapa, Raghav and Das, Sanchari},
  booktitle={International Symposium on Emerging Information Security and Applications},
address = {Wuhan, China},
  pages={108--127},
  year={2022},
  organization={Springer},
publisher = {Springer}
}

@article{avila2014virtual,
  title={Virtual reality for the masses},
  author={Avila, Lisa and Bailey, Mike},
  journal={IEEE computer graphics and applications},
  volume={34},
  number={05},
  pages={103--104},
  year={2014},
  publisher={IEEE Computer Society}
}

@article{azuma1997survey,
  title={A survey of augmented reality},
  author={Azuma, Ronald T},
  journal={Presence: teleoperators \& virtual environments},
  volume={6},
  number={4},
  pages={355--385},
  year={1997},
  publisher={MIT press One Rogers Street, Cambridge, MA 02142-1209, USA journals-info~…}
}

@article{bailey2019virtual,
  title={{Virtual reality's effect on children's inhibitory control, social compliance, and sharing}},
  author={Bailey, Jakki O and Bailenson, Jeremy N and Obradovi{\'c}, Jelena and Aguiar, Naomi R},
  journal={Journal of Applied Developmental Psychology},
  volume={64},
  pages={101052},
  year={2019},
  publisher={Elsevier}
}

@article{bexson2024safety,
  title={{Safety of virtual reality use in children: a systematic review}},
  author={Bexson, Charlotte and Oldham, Geralyn and Wray, Jo},
  journal={European journal of Pediatrics},
  volume={183},
  number={5},
  pages={2071--2090},
  year={2024},
  publisher={Springer}
}

@article{blackwell2014factors,
  title={Factors influencing digital technology use in early childhood education},
  author={Blackwell, Courtney K and Lauricella, Alexis R and Wartella, Ellen},
  journal={Computers \& Education},
  volume={77},
  pages={82--90},
  year={2014},
  publisher={Elsevier}
}

@article{britto2017nurturing,
  title={Nurturing care: promoting early childhood development},
  author={Britto, Pia R and Lye, Stephen J and Proulx, Kerrie and Yousafzai, Aisha K and Matthews, Stephen G and Vaivada, Tyler and Perez-Escamilla, Rafael and Rao, Nirmala and Ip, Patrick and Fernald, Lia CH and others},
  journal={The lancet},
  volume={389},
  number={10064},
  pages={91--102},
  year={2017},
  publisher={Elsevier}
}

@article{chen2022effects,
  title={{Effects of extended reality on language learning: A meta-analysis}},
  author={Chen, Jingying and Dai, Jian and Zhu, Keke and Xu, Liujie},
  journal={Frontiers in Psychology},
  volume={13},
  pages={1016519},
  year={2022},
  publisher={Frontiers Media SA}
}

@article{chițu2023exploring,
  title={{Exploring the opportunity to use virtual reality for the education of children with disabilities}},
  author={Chițu, Ioana Bianca and Tec{\u{a}}u, Alina Simona and Constantin, Cristinel Petrișor and Tescașiu, Bianca and Br{\u{ă}}tucu, Tamara-Oana and Br{\u{ă}}tucu, Gabriel and Purcaru, Ioana-M{\u{ă}}d{\u{ă}}lina},
  journal={Children},
  volume={10},
  number={3},
  pages={436},
  year={2023},
  publisher={MDPI}
}

@article{elo2008qualitative,
  title={The qualitative content analysis process},
  author={Elo, Satu and Kyng{\"a}s, Helvi},
  journal={Journal of advanced nursing},
  volume={62},
  number={1},
  pages={107--115},
  year={2008},
  publisher={Wiley Online Library}
}

@article{everri2024cyborg,
  title={{Cyborg Children: A systematic literature review on the experience of children using extended reality}},
  author={Everri, Marina and Heitmayer, Maxi},
  journal={Children},
  volume={11},
  number={8},
  pages={984},
  year={2024},
  publisher={MDPI}
}

@article{fast2018testing,
  title={Testing and validating Extended Reality (xR) technologies in manufacturing},
  author={Fast-Berglund, {\AA}sa and Gong, Liang and Li, Dan},
  journal={Procedia Manufacturing},
  volume={25},
  pages={31--38},
  year={2018},
  publisher={Elsevier}
}

@article{gnacek2024avdos,
  title={Avdos-vr: Affective video database with physiological signals and continuous ratings collected remotely in vr},
  author={Gnacek, Michal and Quintero, Luis and Mavridou, Ifigeneia and Balaguer-Ballester, Emili and Kostoulas, Theodoros and Nduka, Charles and Seiss, Ellen},
  journal={Scientific Data},
  volume={11},
  number={1},
  pages={132},
  year={2024},
  publisher={Nature Publishing Group UK London}
}

@article{haoming2024systematic,
  title={{A systematic review on vocabulary learning in AR and VR gamification context}},
  author={Haoming, Lin and Wei, Wei},
  journal={Computers \& Education: X Reality},
  volume={4},
  pages={100057},
  year={2024},
  publisher={Elsevier}
}

@article{jawalkar2024ethical,
  title={{Ethical Horizons in Immersive Technologies: Addressing Privacy, Security, and Psychological Impact of AR/VR Adoption}},
  author={Jawalkar, Santosh Kumar},
  year={2024},
  journal={International Journal of Multidisciplinary Research and Growth Evaluation},
volume = {5},
pages = {1083--1091}
}

@article{kaimara2022could,
  title={{Could virtual reality applications pose real risks to children and adolescents? A systematic review of ethical issues and concerns}},
  author={Kaimara, Polyxeni and Oikonomou, Andreas and Deliyannis, Ioannis},
  journal={Virtual reality},
  volume={26},
  number={2},
  pages={697--735},
  year={2022},
  publisher={Springer}
}

@techreport{keele2007guidelines,
  title={Guidelines for performing systematic literature reviews in software engineering},
  author={Keele, Staffs and others},
  year={2007},
  institution={Technical report, ver. 2.3 ebse technical report. ebse}
}

@techreport{KennedyEtAl2012_Literacy3to8,
  author       = {Eithne Kennedy and Elizabeth Dunphy and Bernadette Dwyer and Geraldine Hayes and Thér{\`e}se McPhillips and Jackie Marsh and Maura O’Connor and Gerry Shiel},
  title        = {Literacy in Early Childhood and Primary Education (3–8 years)},
  institution  = {National Council for Curriculum and Assessment (NCCA)},
  number       = {Research Report No. 15},
  year         = {2012},
  address      = {Dublin, Ireland},
  url          = {https://ncca.ie/media/2137/literacy_in_early_childhood_and_primary_education_3-8_years.pdf},
  issn         = {1649-3362}
}

@article{lin2016augmented,
  title={{Augmented reality in educational activities for children with disabilities}},
  author={Lin, Chien-Yu and Chai, Hua-Chen and Wang, Jui-ying and Chen, Chien-Jung and Liu, Yu-Hung and Chen, Ching-Wen and Lin, Cheng-Wei and Huang, Yu-Mei},
  journal={Displays},
  volume={42},
  pages={51--54},
  year={2016},
  publisher={Elsevier}
}

@article{liu2024augmented,
  title={{An Augmented Reality Serious Game for Children’s Optical Science Education: Randomized Controlled Trial}},
  author={Liu, Bo and Wan, Xinyue and Li, Xiaofang and Zhu, Dian and Liu, Zhao},
  journal={JMIR serious games},
  volume={12},
  pages={e47807},
  year={2024}
}

@book{miao2021ai,
  title     = {AI and education: Guidance for policy-makers},
  author    = {Miao, Fengchun and Holmes, Wayne and Huang, Ronghuai and Zhang, Hui},
  publisher = {UNESCO},
  year      = {2021},
  address   = {Paris, France},
  note      = {ISBN: 978-92-3-100447-6},
  url       = {https://www.unesco.org/en/articles/ai-and-education-guidance-policy-makers}
}

@article{milkaite2020child,
  title={Child-friendly transparency of data processing in the EU: from legal requirements to platform policies},
  author={Milkaite, Ingrida and Lievens, Eva},
  journal={Journal of Children and Media},
  volume={14},
  number={1},
  pages={5--21},
  year={2020},
  publisher={Taylor \& Francis}
}

@article{mansour2025embodied,
  title={{Embodied learning of science concepts through augmented reality technology}},
  author={Mansour, Nasser and Aras, Ceren and Staarman, Judith Kleine and Alotaibi, Sarah Bader Mohsen},
  journal={Education and Information Technologies},
  volume={30},
  number={6},
  pages={8245--8275},
  year={2025},
  publisher={Springer}
}

@article{milgram1994taxonomy,
  title={A taxonomy of mixed reality visual displays},
  author={Milgram, Paul and Kishino, Fumio},
  journal={IEICE TRANSACTIONS on Information and Systems},
  volume={77},
  number={12},
  pages={1321--1329},
  year={1994},
  publisher={The Institute of Electronics, Information and Communication Engineers}
}

@article{moher2009preferred,
title = {Preferred reporting items for systematic reviews and meta-analyses: The PRISMA statement},
journal = {International Journal of Surgery},
volume = {8},
number = {5},
pages = {336-341},
year = {2010},
issn = {1743-9191},
doi = {https://doi.org/10.1016/j.ijsu.2010.02.007},
url = {https://www.sciencedirect.com/science/article/pii/S1743919110000403},
author = {David Moher and Alessandro Liberati and Jennifer Tetzlaff and Douglas G. Altman}
}

@article{newman2020systematic,
  title={Systematic reviews in educational research: Methodology, perspectives and application},
  author={Newman, Mark and Gough, David},
  journal={Systematic reviews in educational research},
  volume={64},
  number={3},
  pages={3--22},
  year={2020}
}

@article{pawelczyk2025understanding,
  title={{Understanding Cybersickness and Presence in Seated VR: A Foundation for Exploring Therapeutic Applications of Immersive Virtual Environments}},
  author={Pawe{\l}czyk, Witold and Olejarz, Dorota and Gawe{\l}, Zofia and Merta, Magdalena and Nowakowska, Aleksandra and Nowak, Magdalena and Rutkowska, Anna and Batalik, Ladislav and Rutkowski, Sebastian},
  journal={Journal of Clinical Medicine},
  volume={14},
  number={8},
  pages={2718},
  year={2025},
  publisher={MDPI}
}

@article{steuer1992defining,
  title={Defining virtual reality: Dimensions determining telepresence},
  author={Steuer, Jonathan and others},
  journal={Journal of communication},
  volume={42},
  number={4},
  pages={73--93},
  year={1992},
  publisher={Oxford University Press}
}

@article{sun2022augmented,
  title={Augmented tactile-perception and haptic-feedback rings as human-machine interfaces aiming for immersive interactions},
  author={Sun, Zhongda and Zhu, Minglu and Shan, Xuechuan and Lee, Chengkuo},
  journal={Nature communications},
  volume={13},
  number={1},
  pages={5224},
  year={2022},
  publisher={Nature Publishing Group UK London}
}

@inproceedings{masai2020eye,
  title={Eye-based interaction using embedded optical sensors on an eyewear device for facial expression recognition},
  author={Masai, Katsutoshi and Kunze, Kai and Sugimoto, Maki},
  booktitle={Proceedings of the Augmented Humans International Conference},
  pages={1--10},
  year={2020},
publisher = {ACM},
address = {Kaiserslautern, Germany}
}

@misc{unesco_ecce_need_know_2025,
  author       = {{UNESCO}},
  title        = {What you need to know about early childhood care and education},
  year         = {2025},
  howpublished = {\url{https://www.unesco.org/en/early-childhood-education/need-know}},
  note         = {Last updated 13 February 2025}
}

@online{UNICEF_2025_ECD,
  author       = {{United Nations Children’s Fund (UNICEF)}},
  title        = {Early Childhood Development — Overview},
  year         = 2025,
  url          = {https://data.unicef.org/topic/early-childhood-development/overview/},
  note         = {Accessed: 2025-11-17}
}

@inproceedings{valente2022analysis,
  title={Analysis of academic databases for literature review in the computer science education field},
  author={Valente, Aline and Holanda, Maristela and Mariano, Ari Melo and Furuta, Richard and Da Silva, Dilma},
  booktitle={2022 ieee frontiers in education conference (fie)},
address = {Uppsala, Sweden},  
pages={1--7},
  year={2022},
  organization={IEEE},
publisher = {Institute of Electrical and Electronics Engineers},
}

@article{villena2022effects,
  title={{Effects of virtual reality on learning outcomes in K-6 education: A meta-analysis}},
  author={Villena-Taranilla, Rafael and Tirado-Olivares, Sergio and C{\'o}zar-Guti{\'e}rrez, Ram{\'o}n and Gonz{\'a}lez-Calero, Jos{\'e} Antonio},
  journal={Educational Research Review},
  volume={35},
  pages={100434},
  year={2022},
  publisher={Elsevier}
}

@article{zhang2022virtual,
  title={{Virtual reality technology as an educational and intervention tool for children with autism spectrum disorder: current perspectives and future directions}},
  author={Zhang, Minyue and Ding, Hongwei and Naumceska, Meri and Zhang, Yang},
  journal={Behavioral Sciences},
  volume={12},
  number={5},
  pages={138},
  year={2022},
  publisher={MDPI}
}

@article{zhang2023using,
  title={{Using virtual reality interventions to promote social and emotional learning for children and adolescents: A systematic review and meta-analysis}},
  author={Zhang, Feng and Zhang, Yan and Li, Gege and Luo, Heng},
  journal={Children},
  volume={11},
  number={1},
  pages={41},
  year={2023},
  publisher={MDPI}
}

@article{zhang2025analyzing,
  title={{Analyzing and comparing augmented reality and virtual reality assisted vocabulary learning: a systematic review}},
  author={Zhang, Mike Minwen and Hashim, Harwati and Yunus, Melor Md},
  journal={Frontiers in Virtual Reality},
  volume={6},
  pages={1522380},
  year={2025},
  publisher={Frontiers Media SA}
}

@software{zotero,
  author  = {Zotero},
  title   = {Zotero},
  year    = {2025},
  note    = {Computer software},
  url     = {https://www.zotero.org}
}

@inproceedings{abrarAugmentedRealityEducation2019a,
  title = {Augmented {{Reality}} in {{Education}}: {{A Study}} on {{Preschool Children}}, {{Parents}}, and {{Teachers}} in {{Bangladesh}}},
  shorttitle = {Augmented {{Reality}} in {{Education}}},
  booktitle = {Virtual, {{Augmented}} and {{Mixed Reality}}. {{Applications}} and {{Case Studies}}},
  author = {Abrar, Mohammad Fahim and Islam, Md. Rakibul and Hossain, Md. Sabir and Islam, Mohammad Mainul and Kabir, Muhammad Ashad},
  editor = {Chen, Jessie Y.C. and Fragomeni, Gino},
  year = 2019,
  pages = {217--229},
  publisher = {Springer International Publishing},
  address = {Cham},
  doi = {10.1007/978-3-030-21565-1_14},
  isbn = {978-3-030-21565-1},
  langid = {english}
}

@inproceedings{aguirregoitiaExperienceApplicationAugmented2017,
  title     = {An Experience of the Application of Augmented Reality to Learn English in Infant Education},
  author    = {Aguirregoitia, Estibaliz and Bilbao, Amaia and L{\'o}pez Benito, Amaia},
  booktitle = {Proceedings of the 2017 International Symposium on Computers in Education (SIIE)},
  year      = {2017},
  pages     = {1--6},
  doi       = {10.1109/SIIE.2017.8259645},
  publisher = {Institute of Electrical and Electronics Engineers},
  organization = {IEEE},
address = {Lisbon, Portugal}
}

@article{aladinARTOKIDSpeechenabledAugmented2020a,
  title = {{{AR-TO-KID}}: {{A}} Speech-Enabled Augmented Reality to Engage Preschool Children in Pronunciation Learning},
  author = {Aladin, M Y F and Ismail, A W and Salam, M S H and Kumoi, R and Ali, A F},
  year = 2020,
  month = nov,
  journal = {IOP Conference Series. Materials Science and Engineering},
  volume = {979},
  number = {1},
  pages = {12},
  publisher = {IOP Publishing},
  address = {Bristol},
  issn = {17578981},
  doi = {10.1088/1757-899X/979/1/012011},
  langid = {english}
}

@article{anTeachersPerceptionsEarly2023,
  title = {Teachers Perceptions on Early Childhood's Traffic and Life Safety Education Program Using {{VR}}},
  author = {An, Kwang-seong, Mi-young; Shin},
  year = 2023,
  journal = {Applied Sciences (Switzerland)},
  volume = {13},
  number = {2},
  pages = {777},
  issn = {2076-3417},
  doi = {10.3390/app13020777},
  urldate = {2025-10-25},
  langid = {english}
}

@inproceedings{antoniaArtfulThinkingAugmented2018,
  title = {Artful Thinking and {{Augmented Reality}} in Kindergarten: {{Technology}} Contributions to the Inclusion of Socially Underprivileged Children in Creative Activities},
  shorttitle = {Artful Thinking and {{Augmented Reality}} in Kindergarten},
  booktitle = {Proceedings of the 8th International Conference on Software Development and Technologies for Enhancing Accessibility and Fighting Info-Exclusion},
  author = {Antonia, Dagla and Evgenia, Roussou},
  year = 2018,
  series = {Dsai '18},
  pages = {187--194},
  publisher = {Association for Computing Machinery},
  address = {New York, NY, USA},
  doi = {10.1145/3218585.3218685},
  urldate = {2025-10-25},
  isbn = {978-1-4503-6467-6},
  langid = {english}
}

@article{aydogduAugmentedRealityPreschool2022,
  title = {Augmented Reality for Preschool Children: {{An}} Experience with Educational Contents},
  shorttitle = {Augmented Reality for Preschool Children},
  author = {Aydo{\u g}du, Fatih},
  year = 2022,
  month = mar,
  journal = {British Journal of Educational Technology},
  volume = {53},
  number = {2},
  pages = {326--348},
  publisher = {Blackwell Publishing Ltd.},
  address = {Coventry},
  issn = {00071013},
  doi = {10.1111/bjet.13168},
  urldate = {2025-10-25},
  langid = {english}
}

@inproceedings{ayuagungmasaristamyAugmentedRealityCultural2024a,
  title = {Augmented {{Reality}} in {{Cultural Education}}: {{Introducing Balinese Gamelan}} to {{Young Learners}}},
  shorttitle = {Augmented {{Reality}} in {{Cultural Education}}},
  booktitle = {2024 {{IEEE International Symposium}} on {{Consumer Technology}} ({{ISCT}})},
  author = {Ayu Agung Mas Aristamy, I Gusti and Susana, Kadek Yogi and Winatha, Komang Redy and Iswardani, Putu Risanti and Ayu Shinta Dwi Astari, Gusti},
  year = 2024,
  month = aug,
  pages = {35--41},
  issn = {2159-1423},
  doi = {10.1109/ISCT62336.2024.10791109},
  urldate = {2025-10-25},
    publisher = {Institute of Electrical and Electronics Engineers},
address = {Kuta, Bali, Indonesia}
}

@article{baehoikyoungVerificationVRPlay2025a,
  title = {Verification of a {{VR Play Program}}'s {{Effects}} on {{Young Children}}'s {{Playfulness}}},
  author = {{Bae Hoikyoung} and Gwangyong, Gim},
  year = 2025,
  journal = {Applied Sciences},
  volume = {15},
  number = {17},
  pages = {9769},
  publisher = {MDPI AG},
  address = {Basel},
  issn = {2076-3417},
  doi = {10.3390/app15179769},
  urldate = {2025-10-25},
  langid = {english}
}

@article{balchaImpactAugmentedRealityassisted2025,
  title = {The Impact of Augmented Reality-Assisted Structured Learning on Environmental Education for Preschool Children},
  author = {Balcha, Rakha, Alemayehu Regasa; Chen},
  year = 2025,
  journal = {Education and Information Technologies},
  doi = {10.1007/s10639-025-13737-9},
volume={30},
  number={17},
  pages={25467--25503},
  publisher={Springer}
}

@inproceedings{besevliMaRTDesigningProjectionbased2019,
  title = {{{MaR-T}}: {{Designing}} a Projection-Based Mixed Reality System for Nonsymbolic Math Development of Preschoolers: {{Guided}} by Theories of Cognition and Learning},
  author = {Be{\c s}evli, Tilbe, Ceylan; {\"U}rey},
  year = 2019,
  pages = {280--292},
  doi = {10.1145/3311927.3323147},
 booktitle={Proceedings of the 18th ACM international conference on interaction design and children},
 publisher = {Association for Computing Machinery},
address = {Boise, Idaho, USA}

}

@article{borovanskaEngagingChildrenUsing2020a,
  title = {Engaging with {{Children Using Augmented Reality}} on {{Clothing}} to {{Prevent Them}} from {{Smoking}}.},
  author = {Borovanska, Zuzana; Poyade, Matthieu; Rea},
  year = 2020,
  journal = {Advances in experimental medicine and biology},
  volume = {1262.0},
  pages = {59--94}
}

@article{bubpamasDevelopmentEarlyChildhood2024a,
  title = {The {{Development}} of the {{Early Childhood Teacher}}'s {{Activity Ability Using}} the {{Augmented Reality Storybook Set}} to {{Lessen Issues}} and {{Foster Resilience}} from {{Bullying}} for {{Early Childhood}}},
  author = {Bubpamas, Chanatip; Kaewyam, Varaporn; Pengpol},
  year = 2024,
  journal = {Shanlax International Journal of Education},
  volume = {13.0},
  number = {1},
  pages = {23--32}
}

@article{burlesonActiveLearningEnvironments2018a,
  title = {Active {{Learning Environments}} with {{Robotic Tangibles}}: {{Children}}'s {{Physical}} and {{Virtual Spatial Programming Experiences}}},
  shorttitle = {Active {{Learning Environments}} with {{Robotic Tangibles}}},
  author = {Burleson, Winslow S. and Harlow, Danielle B. and Nilsen, Katherine J. and Perlin, Ken and Freed, Natalie and Jensen, Camilla N{\o}rgaard and Lahey, Byron and Lu, Patrick and Muldner, Kasia},
  year = 2018,
  month = jan,
  journal = {IEEE Transactions on Learning Technologies},
  volume = {11},
  number = {1},
  pages = {96--106},
  publisher = {{Institute of Electrical and Electronics Engineers}},
  issn = {1939-1382, 1939-1382},
  doi = {10.1109/TLT.2017.2724031},
  urldate = {2025-10-25},
  langid = {english}
}

@inproceedings{caiCaseStudyUsing2023,
  title = {A Case Study Using Augmented Reality-Based Scratch Games in English Learning for Preschool Children},
  booktitle = {Proceedings of the 2023 6th International Conference on Big Data and Education},
  author = {Cai, Su and Xue, Man},
  year = 2023,
  series = {Icbde '23},
  pages = {59--66},
  publisher = {Association for Computing Machinery},
  address = {New York, NY, USA},
  doi = {10.1145/3608218.3608238},
  urldate = {2025-10-25},
  isbn = {979-8-4007-0822-0},
  langid = {english}
}

@article{calle-bustosAugmentedRealityGame2017,
  title = {An Augmented Reality Game to Support Therapeutic Education for Children with Diabetes},
  author = {{Calle-Bustos}, Francisco, Andr{\'e}s Marcelo; Juan},
  year = 2017,
  journal = {PLOS ONE},
  volume = {12},
  number = {9},
  pages = {23},
  issn = {1932-6203},
  doi = {10.1371/journal.pone.0184645},
  urldate = {2025-10-25},
  langid = {english}
}

@article{changEmbeddingDialogReading2024a,
  title = {Embedding Dialog Reading into {{AR}} Picture Books},
  author = {Chang, Kuo-En and {Yu-Wei}, Tai and Liu, Tzu-Chien and {Yao-Ting}, Sung},
  year = 2024,
  month = oct,
  journal = {Interactive Learning Environments},
  volume = {32},
  number = {8},
  pages = {3931--3947},
  publisher = {Taylor \& Francis Ltd.},
  address = {Abingdon},
  issn = {10494820},
  doi = {10.1080/10494820.2023.2192758},
  urldate = {2025-10-25},
  langid = {english}
}

@article{chenUsingAugmentedReality2019a,
  title = {Using {{Augmented Reality Flashcards}} to {{Learn Vocabulary}} in {{Early Childhood Education}}},
  author = {Chen, Ruo Wei; Chan, Kan Kan},
  year = 2019,
  journal = {Journal of Educational Computing Research},
  volume = {57.0},
  number = {7},
  pages = {1812--1831}
}

@inproceedings{chien-yuAugmentedRealitybasedAssistive2010,
  title = {Augmented Reality-Based Assistive Technology for Handicapped Children},
  booktitle = {2010 International Symposium on Computer, Communication, Control and Automation ({{3CA}})},
  author = {{Chien-Yu}, Lin and Chao, Jo-Ting and Wei, Hsiao-Shan},
  year = 2010,
  month = may,
  volume = {1},
  pages = {61--64},
  issn = {2324-8017},
  doi = {10.1109/3CA.2010.5533735},
  urldate = {2025-10-25},
  publisher = {Institute of Electrical and Electronics Engineers},
  organization = {IEEE},
address = {Tainan, Taiwan}
}

@article{chuEffectsNonwearableDigital2023,
  title = {Effects of a {{Nonwearable Digital Therapeutic Intervention}} on {{Preschoolers With Autism Spectrum Disorder}} in {{China}}: {{Open-Label Randomized Controlled Trial}}},
  shorttitle = {Effects of a {{Nonwearable Digital Therapeutic Intervention}} on {{Preschoolers With Autism Spectrum Disorder}} in {{China}}},
  author = {Chu, Liting and Shen, Li and Ma, Chenhuan and Chen, Jinjin and Tian, Yuan and Zhang, Chuncao and Gong, Zilan and Li, Mengfan and Wang, Chengjie and Pan, Lizhu and Zhu, Peiying and Wu, Danmai and Wang, Yu and Yu, Guangjun},
  year = 2023,
  month = aug,
  journal = {Journal of Medical Internet Research},
  volume = {25},
  number = {1},
  pages = {e45836},
  publisher = {JMIR Publications Inc., Toronto, Canada},
  doi = {10.2196/45836},
  urldate = {2025-10-25},
  langid = {english}
}

@inproceedings{dalimTeachARInteractiveAugmented2016,
  title = {{{TeachAR}}: {{An}} Interactive Augmented Reality Tool for Teaching Basic English to Non-Native Children},
  booktitle = {2016 {{IEEE}} International Symposium on Mixed and Augmented Reality ({{ISMAR-adjunct}})},
  author = {Dalim, Che Samihah Che and Piumsomboon, Thammathip and Dey, Arindam and Billinghurst, Mark and Sunar, Shahrizal},
  year = 2016,
  month = sep,
  pages = {344--345},
  doi = {10.1109/ISMAR-Adjunct.2016.0113},
publisher = {Institute of Electrical and Electronics Engineers},
address = {Merida, Yucatan, Mexico}

}

@article{dallolioImpactFantasyYoung2024,
  title={The Impact of Fantasy on Young Children's Recall: A Virtual Reality Approach},
  author={Dall’Olio, Lucas and Amrein, Olivier and Gianettoni, Lavinia and Martarelli, Corinna S},
  journal={Virtual Reality},
  volume={28},
  number={1},
  pages={10},
  year={2024},
  publisher={Springer}
}

@inproceedings{dasilvaCuboKidsProposal2020,
  title = {Cubo {{Kids}}: {{A}} Proposal for an Educational Application with Augmented Reality},
  author = {da Silva, Carlos Ramon Sarmento and Mendon{\c{c}}a, Ant{\^o}nio Kalielso and Silva, Jos{\'e} {\'E}rico Gomes and Morais, Ceres Germanna Braga},
  year = 2020,
  pages = {497--502},
  doi = {10.1109/TALE48869.2020.9368462},
  booktitle={2020 IEEE International Conference on Teaching, Assessment, and Learning for Engineering (TALE)},
    organization = {IEEE},
publisher = {Institute of Electrical and Electronics Engineers},
address = {Takamatsu, Japan}
}

@article{demirdagInvestigationEffectivenessAugmented2025a,
  title = {An {{Investigation}} of the {{Effectiveness}} of {{Augmented Reality Technology Supported English Language Learning Activities}} on {{Preschool Children}}},
  author = {Demirdag, Merve Cosgun and Kucuk, Sevda and Tasgin, Adnan},
  year = 2025,
  month = feb,
  journal = {International Journal of Human - Computer Interaction},
  volume = {41},
  number = {4},
  pages = {2410--2423},
  publisher = {Lawrence Erlbaum Associates, Inc.},
  address = {Norwood},
  issn = {10447318},
  doi = {10.1080/10447318.2024.2323278},
  urldate = {2025-10-25},
  langid = {english}
}

@article{dilekeryigitImpactAugmentedReality2025,
  title = {Impact of Augmented Reality Technology on Geometry Skills and Motivation of Preschool Children},
  author = {Dilek Eryigit, Cigdem and Kucuk, Sevda and Tasgin, Adnan},
  year = 2025,
  journal = {Education and Information Technologies},
  doi = {10.1007/s10639-025-13631-4},
pages={1--26},
volume = {30}, 
}

@article{duzyolInvestigationEffectAugmented2022a,
  title = {{Investigation of the effect of augmented reality application on preschool children's knowledge of space}},
  author = {D{\"U}ZYOL, Endam and YILDIRIM, G{\"u}nseli and {\"O}ZYILMAZ, G{\"u}zin},
  year = 2022,
  journal = {Journal of Educational Technology and Online Learning},
  volume = {5},
  number = {1},
  pages = {190--203},
  publisher = {G\"urhan Durak},
  address = {Bal\i kesir},
  issn = {26186586},
  doi = {10.31681/jetol.976885},
  urldate = {2025-10-25},
  langid = {turkish}
}

@inproceedings{fajrieAugmentedRealityMedia2022,
  title = {Augmented Reality Media Development in Early Childhood Learning System during the Covid 19 Pandemic Era},
  booktitle = {Proceedings of the 5th International Conference on Learning Innovation and Quality Education},
  author = {Fajrie, Nur and Purbasari, Imaniar},
  year = 2022,
  series = {Icliqe '21},
  numpages = {7},
  publisher = {Association for Computing Machinery},
  address = {New York, NY, USA},
  doi = {10.1145/3516875.3516999},
  urldate = {2025-10-25},
  articleno = {107},
  isbn = {978-1-4503-8692-0},
  langid = {english}
}

@inproceedings{fengEffectsARLearning2022,
  title = {The Effects of {{AR}} Learning Environment to Preschool Children's Numerical Cognition},
  author = {Feng, Zhaoxin and Gong, Chenxi and Jiao, Xinyue and Liu, Zifeng and Cai, Su},
  booktitle={2022 International Conference on Advanced Learning Technologies (ICALT)},
publisher = {Institute of Electrical and Electronics Engineers},
address = {Bucharest, Romania},
  year = 2022,
  pages = {352--356},
  doi = {10.1109/ICALT55010.2022.00110}
}

@article{fridbergThematicTeachingAugmented2024,
  title = {Thematic Teaching of Augmented Reality and Education for Sustainable Development in Preschool---the Importance of `place'},
  author = {Fridberg, Andreas, Marie; Redfors},
  year = 2024,
  journal = {Education Sciences},
  volume = {14},
  number = {7},
  pages = {719},
  issn = {2227-7102},
  doi = {10.3390/educsci14070719},
  urldate = {2025-10-25},
  langid = {english}
}

@article{gecu-parmaksizAugmentedRealityBasedVirtual2019,
  title = {Augmented {{Reality-Based Virtual Manipulatives}} versus {{Physical Manipulatives}} for {{Teaching Geometric Shapes}} to {{Preschool Children}}},
  author = {{Gecu-Parmaksiz}, Zeynep; Delialioglu, Omer},
  year = 2019,
  journal = {British Journal of Educational Technology},
  volume = {50.0},
  number = {6},
  pages = {3376--3390}
}

@article{gecu-parmaksizEffectAugmentedReality2020a,
  title = {The Effect of Augmented Reality Activities on Improving Preschool Children's Spatial Skills.},
  author = {{Gecu-Parmaksiz}, Zeynep; Delialio{\u g}lu, {\"O}mer},
  year = 2020,
  journal = {Interactive Learning Environments},
  volume = {28.0},
  number = {7},
  pages = {876--889}
}

@inproceedings{gweonMABLEMediatingYoung2018,
  title = {{{MABLE}}: {{Mediating}} Young Children's Smart Media Usage with Augmented Reality},
  shorttitle = {{{MABLE}}},
  booktitle = {Proceedings of the 2018 {{CHI}} Conference on Human Factors in Computing Systems},
  author = {Gweon, Gahgene and Kim, Bugeun and Kim, Jinyoung and Lee, Kung Jin and Rhim, Jungwook and Choi, Jueun},
  year = 2018,
  series = {Chi '18},
  pages = {1--9},
  publisher = {Association for Computing Machinery},
  address = {New York, NY, USA},
  doi = {10.1145/3173574.3173587},
  urldate = {2025-10-25},
  isbn = {978-1-4503-5620-6},
  langid = {english}
}

@article{hanExaminingYoungChildrens2015a,
  title = {Examining {{Young Children}}'s {{Perception}} toward {{Augmented Reality-Infused Dramatic Play}}},
  author = {Han, Jeonghye and Jo, Miheon and Hyun, Eunja and So, Hyo-jeong},
  year = 2015,
  month = jun,
  journal = {Educational Technology Research and Development},
  volume = {63},
  number = {3},
  pages = {455--474},
  publisher = {Springer},
  issn = {1042-1629, 1042-1629},
  doi = {10.1007/s11423-015-9374-9},
  urldate = {2025-10-25},
  langid = {english}
}

@article{hermanIntegratingSocialLearning2025,
  title = {Integrating Social Learning and Experiential Learning Theories: A Novel Augmented Reality Approach to Enhancing Social Skills in Early Childhood Education},
  shorttitle = {Integrating Social Learning and Experiential Learning Theories},
  author = {Herman, Ansari Saleh, Herman; Herlina},
  year = 2025,
  journal = {Cogent Education},
  volume = {12},
  number = {1},
  pages = {2556889},
  issn = {2331-186X},
  doi = {10.1080/2331186X.2025.2556889},
  urldate = {2025-10-25},
  langid = {english}
}

@inproceedings{hidayatVirtualRealitybasedTraffic2023,
  title = {Virtual Reality-Based Traffic Sign Education for Early Childhood},
  author = {Hidayat, Taufiqurrakhman Nur and Purnomo, Fendi Aji and Pratisto, Eko Harry and Nusantara, Ksatria Tirta and Yudhanto, Yudho},
  booktitle={2023 Eighth International Conference on Informatics and Computing (ICIC)},
  numpages={5},
  year = 2023,
publisher = {Institute of Electrical and Electronics Engineers},
address = {East Java, Indonesia},
  doi = {10.1109/ICIC60109.2023.10382058}
}

@article{hoInteractiveMultisensoryVolumetric2023,
  title = {Interactive Multi-Sensory and Volumetric Content Integration for Music Education Applications},
  author = {Ho, Chanru, Chinling; Lin},
  year = 2023,
  journal = {Multimedia Tools and Applications},
  volume = {82},
  number = {4},
  pages = {4847--4862},
  issn = {13807501},
  doi = {10.1007/s11042-022-12314-3},
  urldate = {2025-10-25},
  copyright = {\copyright{} The Author(s), under exclusive licence to Springer Science+Business Media, LLC, part of Springer Nature 2022.},
  langid = {english}
}

@article{hossainAugmentedRealityBasedElementary2021a,
  title = {Augmented {{Reality-Based Elementary Level Education}} for {{Bengali Character Familiarization}}},
  author = {Hossain, Mohammad Jaber and Ahmed, Towfik},
  year = 2021,
  month = feb,
  journal = {SN Computer Science},
  volume = {2},
  number = {1},
  pages = {31},
  publisher = {Springer Nature B.V.},
  address = {Kolkata},
  issn = {2662995X},
  doi = {10.1007/s42979-020-00402-w},
  urldate = {2025-10-25},
  langid = {english}
}

@article{hsiaoUsingGestureInteractive2016a,
  title = {Using a Gesture Interactive Game-Based Learning Approach to Improve Preschool Children's Learning Performance and Motor Skills.},
  author = {Hsiao, Hsien-Sheng; Chen, Jyun-Chen},
  year = 2016,
  journal = {Computers \&amp; Education},
  volume = {95.0},
  pages = {151--162}
}

@article{huangUsingAugmentedReality2016a,
  title = {Using {{Augmented Reality}} in Early Art Education: A Case Study in {{Hong Kong}} Kindergarten.},
  author = {Huang, Yujia; Li, Hui; Fong},
  year = 2016,
  journal = {Early Child Development \&amp; Care},
  volume = {186.0},
  number = {6},
  pages = {879--894}
}

@article{idrisDevelopmentPatternLearning2021a,
  title = {Development {{Of}} a {{Pattern Learning Module}} for {{Early Mathematics Based}} on {{Flipped Classroom}} with {{Augmented Reality}}},
  author = {Idris, Haliza and Nor, Mariani Md and Rahman, Mohd Nazri Abd},
  year = 2021,
  journal = {Turkish Journal of Computer and Mathematics Education},
  volume = {12},
  number = {14},
  pages = {2591--2597},
  publisher = {Ninety Nine Publication},
  address = {Gurgaon},
  urldate = {2025-10-25},
  copyright = {\copyright{} 2021. This work is published under https://creativecommons.org/licenses/by/4.0 (the ``License''). Notwithstanding the ProQuest Terms and Conditions, you may use this content in accordance with the terms of the License.},
  langid = {english}
}

@article{isikarslanogluThinkTogetherDesign2024,
  title = {Think Together, Design Together, Code Together: {{The}} Effect of Augmented Reality Activity Designed by Children on the Computational Thinking Skills},
  shorttitle = {Think Together, Design Together, Code Together},
  author = {I{\c s}ik Arslano{\u g}lu, {\.I}smail, {\.I}pek; Kert},
  year = 2024,
  journal = {Education and Information Technologies},
  volume = {29},
  number = {7},
  pages = {8493--8522},
  issn = {13602357},
  doi = {10.1007/s10639-023-12153-1},
  urldate = {2025-10-25},
  copyright = {\copyright{} The Author(s), under exclusive licence to Springer Science+Business Media, LLC, part of Springer Nature 2023. Springer Nature or its licensor (e.g. a society or other partner) holds exclusive rights to this article under a publishing agreement with the author(s) or other rightsholder(s); author self-archiving of the accepted manuscript version of this article is solely governed by the terms of such publishing agreement and applicable law.},
  langid = {english}
}

@inproceedings{jamiatEffectsAugmentedReality2020,
  title = {Effects of Augmented Reality Mobile Apps on Early Childhood Education Students' Achievement},
  booktitle = {Proceedings of the 3rd International Conference on Digital Technology in Education},
  author = {Jamiat, Nurullizam and Othman, Noor Fatin Nadia},
  year = 2020,
  series = {Icdte '19},
  pages = {30--33},
  publisher = {Association for Computing Machinery},
  address = {New York, NY, USA},
  doi = {10.1145/3369199.3369203},
  isbn = {978-1-4503-7220-6}
}

@article{joDevelopmentUtilizationProjectorRobot2011a,
  title = {Development and {{Utilization}} of {{Projector-Robot Service}} for {{Children}}'s {{Dramatic Play Activities}} Based on {{Augmented Reality}}},
  author = {Jo, Miheon and Han, Jeonghye and Hyun, Eunja and Kim, Gerard J and Kim, Namgyu},
  year = 2011,
  month = jun,
  journal = {Advances in Information Sciences and Service Sciences},
  volume = {3},
  number = {5},
  pages = {12},
  publisher = {{The International Association for Information, Culture, Human and Industry Technology}},
  issn = {1976-3700, 1976-3700},
  langid = {english}
}

@inproceedings{khairulanuardiIncreaseReadingHabit2022,
  title = {Towards Increase Reading Habit for Preschool Children through Interactive Augmented Reality Storybook},
  author={Anuardi, Muhammad Afiq Khairul and Mustapha, Safinaz and Mohammed, MN},
  booktitle={2022 IEEE 12th symposium on computer applications \& industrial electronics (ISCAIE)},
  pages={252--257},
  year={2022},
  organization={IEEE},
address = {Penang Island, Malaysia},
publisher = {Institute of Electrical and Electronics Engineers},
  doi = {10.1109/ISCAIE54458.2022.9794543}
}

@article{kimOTCObjectCamera2019,
  title = {The {{OTC}} (Object to Camera) Approach to Visualize behind Stories of Museum Exhibits},
  author = {Kim, Keitaro, Si Jung Sj; Sanchez},
  year = 2019,
  journal = {Lecture Notes in Computer Science},
  volume = {11786 LNCS},
  pages = {243--252},
  doi = {10.1007/978-3-030-30033-3_19}
}

@inproceedings{kimVirtualStorytellerUsing2011,
  title = {Virtual Storyteller Using Marker Based {{AR}} and {{FPGA}}},
  booktitle = {2011 {{IEEE}} 54th International Midwest Symposium on Circuits and Systems ({{MWSCAS}})},
  author = {Kim, Hyung Min and Song, Tae Houn and Jung, Soon Mook and Kwon, Key Ho and Jeon, Jae Wook},
  year = 2011,
  month = aug,
  pages = {1--4},
  issn = {1558-3899},
  doi = {10.1109/MWSCAS.2011.6026326},
  urldate = {2025-10-25},
publisher = {Institute of Electrical and Electronics Engineers},
address = {Seoul, South Korea}
}

@article{kisnoDigitalStorytellingEarly2022,
  title = {Digital Storytelling for Early Childhood Creativity: {{Diffusion}} of Innovation ``3-{{D}} Coloring Quiver Application Based on Augmented Reality Technology in Children's Creativity Development"},
  shorttitle = {Digital {{Storytelling}} for {{Early Childhood Creativity}}},
  author = {Kisno, undefined, null; Wibawa},
  year = 2022,
  journal = {International Journal of Online and Biomedical Engineering},
  volume = {18},
  number = {10},
  pages = {26--42},
  issn = {2626-8493},
  doi = {10.3991/ijoe.v18i10.32845},
  urldate = {2025-10-25},
  copyright = {Copyright (c) 2022 Kisno, Basuki Wibawa , Khaerudin},
  langid = {english}
}

@article{korosidouEffectsAugmentedReality2024,
  title = {The Effects of Augmented Reality on Very Young Learners' Motivation and Learning of the Alphabet and Vocabulary},
  author = {Korosidou, Eleni},
  year = 2024,
  journal = {Digital},
  volume = {4},
  number = {1},
  pages = {195--214},
  issn = {2673-6470},
  doi = {10.3390/digital4010010},
  urldate = {2025-10-25},
  langid = {english}
}

@inproceedings{kulasekaraGameCentricElearning2025,
  title = {A Game Centric E-Learning Application for Preschoolers},
  author={Kulasekara, DAMN and Nipun, PGI and Dombawela, HMDLBA and Manilka, GS and Kodagoda, N and Silva, DI De},
  booktitle={2025 5th International Conference on Advanced Research in Computing (ICARC)},
  year = 2025,
address = {Belihuloya, Sri Lanka},
  pages = {6},
organization = {IEEE},
publisher = {Institute of Electrical and Electronics Engineers},
  doi = {10.1109/ICARC64760.2025.10963140}
}

@inproceedings{kurniawanARtraceAugmentedReality2019,
  title = {{{ARtrace}}: {{Augmented}} Reality for Student's Fine Motor Learning},
  shorttitle = {{{ARtrace}}},
  booktitle = {2019 2nd International Conference on Applied Engineering ({{ICAE}})},
  author = {Kurniawan, Dwi Ely and Silalahi, Parulian and Pratiwi, Adelia},
  year = 2019,
  month = oct,
  pages = {1--5},
  doi = {10.1109/ICAE47758.2019.9221724},
  urldate = {2025-10-25},
publisher = {Institute of Electrical and Electronics Engineers},
address = {Batam, Indonesia}
}

@inproceedings{kusumaningsihUserExperienceMeasurement2021,
  title = {User Experience Measurement on Augmented Reality Mobile Application for Learning to Read Using a Phonics-Based Approach},
  booktitle = {2021 {{IEEE}} 7th Information Technology International Seminar ({{ITIS}})},
  author = {Kusumaningsih, Ari and Putro, Sigit Susanto and Andriana, Choiril and Angkoso, Cucun Very},
  year = 2021,
  month = oct,
  pages = {1--6},
  doi = {10.1109/ITIS53497.2021.9791660},
publisher = {Institute of Electrical and Electronics Engineers},
address = {Surabaya, Indonesia}
}

@article{kusumaVirtualRealityLearning2018,
  title = {Virtual Reality for Learning Fish Types in Kindergarten},
  author = {Kusuma, I. Ketut Resika, Gede Thadeo Angga; Wirawan},
  year = 2018,
  journal = {International Journal of Interactive Mobile Technologies},
  volume = {12},
  number = {8},
  pages = {41--51},
  doi = {10.3991/ijim.v12i8.9246}
}

@article{lawAugmentedRealityTechnology2025,
  title = {Augmented Reality Technology in Aiding Preschoolers' Education: A Preliminary Study},
  shorttitle = {Augmented {{Reality Technology}} in {{Aiding Preschoolers}}' {{Education}}},
  author = {Law, Chuanchin, Kin Aik; Neo},
  year = 2025,
  journal = {Education Sciences},
  volume = {15},
  number = {8},
  pages = {16},
  issn = {2227-7102},
  doi = {10.3390/educsci15081033},
  urldate = {2025-10-25},
  langid = {english}
}

@article{lewis-presserDesigningAugmentedReality2025,
  title = {Designing Augmented Reality for Preschoolers: {{Lessons}} from Co-Designing a Spatial Learning App},
  shorttitle = {Designing {{Augmented Reality}} for {{Preschoolers}}},
  author = {{Lewis-Presser}, Kevin, Ashley E.; Orr},
  year = 2025,
  journal = {Education Sciences},
  volume = {15},
  number = {9},
  pages = {1195},
  issn = {2227-7102},
  doi = {10.3390/educsci15091195},
  urldate = {2025-10-25},
  langid = {english}
}

@article{lewis-presserEnhancingPreschoolSpatial2025,
  title = {Enhancing Preschool Spatial Skills: A Comprehensive Intervention Using Digital Games and Hands-on Activities},
  shorttitle = {Enhancing {{Preschool Spatial Skills}}},
  author = {{Lewis-Presser}, Regan, Ashley E.; Braham},
  year = 2025,
  journal = {Education Sciences},
  volume = {15},
  number = {6},
  pages = {727},
  issn = {2227-7102},
  doi = {10.3390/educsci15060727},
  urldate = {2025-10-25},
  langid = {english}
}

@article{liangExploitationMultiplayerInteraction2017,
  title = {Exploitation of Multiplayer Interaction and Development of Virtual Puppetry Storytelling Using Gesture Control and Stereoscopic Devices},
  author = {Liang, Hui and Chang, Jian and Deng, Shujie and Chen, Can and Tong, Ruofeng and Zhang, Jian Jun},
  year = 2017,
  journal = {Computer Animation and Virtual Worlds},
  volume = {28},
  number = {5},
  pages = {e1727},
  issn = {1546-427X},
  doi = {10.1002/cav.1727},
  urldate = {2025-10-25},
  langid = {english}
}

@article{liangHandGesturebasedInteractive2017,
  title = {Hand Gesture-Based Interactive Puppetry System to Assist Storytelling for Children},
  author = {Liang, Peifeng, Hui; Chang},
  year = 2017,
  journal = {Visual Computer},
  volume = {33},
  number = {4},
  pages = {517--531},
  doi = {10.1007/s00371-016-1272-6}
}

@inproceedings{liInnovativeApplicationAI2025a,
  title = {Innovative {{Application}} of {{AI}} in {{Country Garden Kindergarten Teaching}}: {{Technological Empowerment}} and {{Educational Practice}}},
  shorttitle = {Innovative {{Application}} of {{AI}} in {{Country Garden Kindergarten Teaching}}},
  booktitle = {2025 5th {{International Conference}} on {{Artificial Intelligence}} and {{Education}} ({{ICAIE}})},
  author = {Li, Jiaze and Li, Guowei and Chen, Heng},
  year = 2025,
  month = may,
  pages = {60--64},
  doi = {10.1109/ICAIE64856.2025.11158356},
  urldate = {2025-10-25},
publisher = {Institute of Electrical and Electronics Engineers},
address = {Suzhou, China}
}

@article{linUsingHomebasedAugmented2025a,
  title = {Using Home-Based Augmented Reality Storybook Training Modules for Facilitating Emotional Functioning and Socialization of Children with Autism Spectrum Disorder},
  author = {Lin, Ling-Yi and Lin, Chang-Hsin and Chuang, Tsung-Yen and Loh, Sau Cheong and Chu, Shin Ying},
  year = 2025,
  journal = {International Journal of Developmental Disabilities},
  volume = {71},
  number = {1},
  pages = {87},
  publisher = {Maney Publishing, Hudson Road},
  address = {Leeds},
  issn = {2047-3869},
  doi = {10.1080/20473869.2023.2202454},
  urldate = {2025-10-25},
  langid = {english}
}

@inproceedings{liuApplicationTabletbasedAR2023a,
  title = {Application of the {{Tablet-based AR}} in {{Preschooler}}'s {{Science Education}}},
  booktitle = {2023 {{International Symposium}} on {{Educational Technology}} ({{ISET}})},
  author = {Liu, Ruixue and Chai, Yonghuan and Wei, Xiaodong},
  year = 2023,
  month = jul,
  pages = {195--199},
  issn = {2766-2144},
  doi = {10.1109/ISET58841.2023.00045},
  urldate = {2025-10-25},
publisher = {Institute of Electrical and Electronics Engineers},
address = {Hongkong SAR, China}
}

@article{lorussoGiokAlienARbased2018,
  title = {Giok the Alien: {{An AR-based}} Integrated System for the Empowerment of Problem-Solving, Pragmatic, and Social Skills in Pre-School Children},
  shorttitle = {Giok the {{Alien}}},
  author = {Lorusso, Gianluigi, Maria Luisa; Giorgetti},
  year = 2018,
  journal = {Sensors},
  volume = {18},
  number = {7},
  pages = {2368},
  issn = {1424-8220},
  doi = {10.3390/s18072368},
  urldate = {2025-10-25},
  langid = {english}
}

@article{lorussoSemiimmersiveVirtualReality2020,
  title = {Semi-Immersive Virtual Reality as a Tool to Improve Cognitive and Social Abilities in Preschool Children},
  author = {Lorusso, Emilia, Maria Luisa; Travellini},
  year = 2020,
  journal = {Applied Sciences (Switzerland)},
  volume = {10},
  number = {8},
  pages = {2948},
  issn = {2076-3417},
  doi = {10.3390/APP10082948},
  urldate = {2025-10-25},
  langid = {english}
}

@article{luoEffectDifferentCombinations2023a,
  title = {The Effect of Different Combinations of Physical Activity and Natural Environment Videos on Children's Attention Levels between Class Breaks},
  author = {Luo, Xiao and {Meng Tao} and Lu, Jiahao and Lu, Li and He, Xiaolong},
  year = 2023,
  journal = {BMC Pediatrics},
  volume = {23},
  pages = {1--17},
  publisher = {Springer Nature B.V.},
  address = {London},
  doi = {10.1186/s12887-023-03868-8},
  langid = {english}
}

@article{mamani-calapujaLearningEnglishEarly2023,
  title = {Learning English in Early Childhood Education with Augmented Reality: {{Design}}, Production, and Evaluation of the ``Wordtastic Kids'' App},
  shorttitle = {Learning {{English}} in {{Early Childhood Education}} with {{Augmented Reality}}},
  author = {{Mamani-Calapuja}, Carmen, Aleyda; Laura-Revilla},
  year = 2023,
  journal = {Education Sciences},
  volume = {13},
  number = {7},
  pages = {638},
  issn = {2227-7102},
  doi = {10.3390/educsci13070638},
  urldate = {2025-10-25},
  langid = {english}
}

@article{montoya-rodriguezEducationalInterventionTheory2025,
  title = {An Educational Intervention for Theory of Mind Skills in Children Using a Virtual Reality Application: A Pilot Study},
  shorttitle = {An {{Educational Intervention}} for {{Theory}} of {{Mind Skills}} in {{Children Using}} a {{Virtual Reality Application}}},
  author = {{Montoya-Rodr{\'i}guez}, Vanessa A., Mar{\'i}a M.; Molina},
  year = 2025,
  journal = {Journal of Information Technology Education: Innovations in Practice},
  volume = {24},
  pages = {19},
  issn = {2165-3151, 2165-316X},
  doi = {10.28945/5480},
  urldate = {2025-10-25},
  langid = {english}
}

@article{muangmoolDevelopmentSocialInteraction2023a,
  title = {The {{Development}} of {{Social Interaction}} for {{Early Childhood}} in {{Lampang Province}} by {{Using Virtual Reality Application}}},
  author = {Muangmool, Somchai and Sirichaisin, Kedthip},
  year = 2023,
  journal = {Asian Journal of Education and Training},
  volume = {9},
  number = {1},
  pages = {7--14},
  publisher = {Asian Online Journal Publishing Group},
  issn = {2519-5387},
  doi = {10.20448/edu.v9i1.4451},
  urldate = {2025-10-25},
  langid = {english}
}

@article{nugrahaDevelopmentARAugmented2025,
  title = {Development of {{AR}} (Augmented Reality) Learning Media Based on Ethnotechnology to Enhance Kindergarten Students' Creativity},
  author = {Nugraha, I. Gede, I. Nyoman Bagus Suweta; Parwati},
  year = 2025,
  journal = {Indian Journal of Information Sources and Services},
  volume = {15},
  number = {2},
  pages = {315--324},
  issn = {2231-6094},
  doi = {10.51983/ijiss-2025.IJISS.15.2.39},
  urldate = {2025-10-25},
  langid = {english}
}

@inproceedings{paliwalEnhancingEducationAugmented2024,
  title = {Enhancing Education with Augmented Reality: A Prototype-Based Approach},
  author={Paliwal, Girish and Sharma, Kanta Prasad and Mehra, Raghav and Srimal, Vijay Mohan and Pandey, Manoj Kumar and Vijay, Harsh},
  booktitle={2024 International Conference on Emerging Technologies and Innovation for Sustainability (EmergIN)},
  year = 2024,
  pages = {98--104},
organization = {IEEE},
address = {Greater Noida, Uttar Pradesh, India},
publisher = {Institute of Electrical and Electronics Engineers},
  doi = {10.1109/EmergIN63207.2024.10961130}
}

@article{panIntroducingAugmentedReality2021,
  title = {Introducing Augmented Reality in Early Childhood Literacy Learning},
  author = {Pan, Min, Zilong; L{\'o}pez},
  year = 2021,
  journal = {Research in Learning Technology},
  volume = {29},
  pages = {1--21},
  issn = {2156-7077},
  doi = {10.25304/rlt.v29.2539},
  urldate = {2025-10-25},
  copyright = {Copyright (c) 2021 Zilong Pan, Mary L\'opez, Chenglu Li, Min Liu},
  langid = {english}
}

@article{panResearchApplicationImmersive2021,
  title = {Research on the Application of Immersive Early Childhood Education},
  author = {Pan, Zhengwei, Zhigengx Geng; Huang},
  year = 2021,
  journal = {International Conference on Virtual Rehabilitation, ICVR},
  volume = {2021-May},
  pages = {166--171},
  doi = {10.1109/ICVR51878.2021.9483701}
}

@article{passigSolvingConceptualPerceptual2014a,
  title = {Solving {{Conceptual}} and {{Perceptual Analogies With Virtual Reality Among Kindergarten Children}} of {{Immigrant Families}}.},
  author = {PASSIG, DAVID; SCHWARTZ, {\relax TIMOR}},
  year = 2014,
  journal = {Teachers College Record},
  volume = {116.0},
  number = {2},
  pages = {1--36}
}

@inproceedings{prekaAugmentedRealityQR2019a,
  title = {Augmented {{Reality}} and {{QR Codes}} for {{Teaching Music}} to {{Preschoolers}} and {{Kindergarteners}}: {{Educational Intervention}} and {{Evaluation}}:},
  shorttitle = {Augmented {{Reality}} and {{QR Codes}} for {{Teaching Music}} to {{Preschoolers}} and {{Kindergarteners}}},
  booktitle = {Proceedings of the 11th {{International Conference}} on {{Computer Supported Education}}},
  author = {Preka, Garyfallia and Rangoussi, Maria},
  year = 2019,
  pages = {113--123},
  publisher = {{SCITEPRESS - Science and Technology Publications}},
  address = {Heraklion, Crete, Greece},
  doi = {10.5220/0007682301130123},
  urldate = {2025-10-25},
  isbn = {978-989-758-367-4},
  langid = {english}
}

@inproceedings{puDevelopmentSituationalInteraction2018,
  title = {Development of a Situational Interaction Game for Improving Preschool Children' Performance in English-Vocabulary Learning},
  booktitle = {Proceedings of the 2018 International Conference on Distance Education and Learning},
  author = {Pu, Mei and Zhong, Zheng},
  year = 2018,
  series = {Icdel '18},
  pages = {88--92},
  publisher = {Association for Computing Machinery},
  address = {New York, NY, USA},
  doi = {10.1145/3231848.3231851},
  urldate = {2025-10-25},
  isbn = {978-1-4503-6431-7},
  langid = {english}
}

@article{raptiEnrichingTraditionalLearning2023,
  title = {Enriching a Traditional Learning Activity in Preschool through Augmented Reality: {{Children}}'s and Teachers' Views},
  shorttitle = {Enriching a {{Traditional Learning Activity}} in {{Preschool}} through {{Augmented Reality}}},
  author = {Rapti, Sokratis, Sophia; Sapounidis},
  year = 2023,
  journal = {Information (Switzerland)},
  volume = {14},
  number = {10},
  pages = {530},
  issn = {2078-2489},
  doi = {10.3390/info14100530},
  urldate = {2025-10-25},
  langid = {english}
}

@article{raptiInvestigatingEducatorsStudents2025,
  title = {Investigating Educators' and Students' Perspectives on Virtual Reality Enhanced Teaching in Preschool},
  author = {Rapti, Sokratis, Sophia; Sapounidis},
  year = 2025,
  journal = {Early Childhood Education Journal},
  volume = {53},
  number = {4},
  pages = {1107--1118},
  doi = {10.1007/s10643-024-01659-z}
}

@article{safarEffectivenessUsingAugmented2017,
  title = {The Effectiveness of Using Augmented Reality Apps in Teaching the English Alphabet to Kindergarten Children: {{A}} Case Study in the State of {{Kuwait}}},
  shorttitle = {The {{Effectiveness}} of {{Using Augmented Reality Apps}} in {{Teaching}} the {{English Alphabet}} to {{Kindergarten Children}}},
  author = {Safar, Zainab H., Ammar H.; Al-Jafar},
  year = 2017,
  journal = {Eurasia Journal of Mathematics, Science and Technology Education},
  volume = {13},
  number = {2},
  pages = {417--440},
  issn = {13058223},
  doi = {10.12973/eurasia.2017.00624a},
  urldate = {2025-10-25},
  langid = {english}
}

@article{sariDevelopingFinancialLiteracy2022,
  title = {Developing a Financial Literacy Storybook for Early Childhood in an Augmented Reality Context},
  author = {Sari, Hardika Dwi, Ratna Candra; Aisyah},
  year = 2022,
  journal = {Contemporary Educational Technology},
  volume = {14},
  number = {2},
  pages = {18},
  issn = {1309517X},
  doi = {10.30935/cedtech/11734},
  urldate = {2025-10-25},
  langid = {english}
}

@inproceedings{shimadaVRHandHygiene2017,
  title = {{{VR}} Hand Hygiene Training System That Visualizes Germs to Be Washed and Removed},
  author={Shimada, Shogo and Funahashi, Kenji and Ito, Kenta and Tanase, Yoshimi and Iwazaki, Kumiko},
  booktitle={2017 IEEE 6th Global Conference on Consumer Electronics (GCCE)},
year = 2017,
  volume = {2017-January},
  pages = {1--4},
organization = {IEEE},
publisher = {Institute of Electrical and Electronics Engineers},
address = {Nagoya, Japan},
  doi = {10.1109/GCCE.2017.8229333}
}

@article{shoshaniVirtualProsocialReality2023a,
  title = {From Virtual to Prosocial Reality: {{The}} Effects of Prosocial Virtual Reality Games on Preschool {{Children}}'s Prosocial Tendencies in Real Life Environments.},
  shorttitle = {From Virtual to Prosocial Reality},
  author = {Shoshani, Anat},
  year = 2023,
  journal = {Computers in Human Behavior},
  volume = {139.0},
  pages = {N.PAG-N.PAG},
  issn = {07475632},
  doi = {10.1016/j.chb.2022.107546},
  urldate = {2025-10-25},
  langid = {english}
}

@article{simsekEffectAugmentedReality2024,
  title = {The Effect of Augmented Reality Storybooks on the Story Comprehension and Retelling of Preschool Children},
  author = {{\c S}im{\c s}ek, Emine Ela},
  year = 2024,
  journal = {Frontiers in Psychology},
  volume = {15},
  pages={1459264},
  doi = {10.3389/fpsyg.2024.1459264}
}

@inproceedings{syahidiARChildAnalysisEvaluation2019a,
  title = {{{AR-Child}}: {{Analysis}}, {{Evaluation}}, and {{Effect}} of {{Using Augmented Reality}} as a {{Learning Media}} for {{Preschool Children}}},
  shorttitle = {{{AR-Child}}},
  booktitle = {2019 5th {{International Conference}} on {{Computing Engineering}} and {{Design}} ({{ICCED}})},
  author = {Syahidi, Aulia Akhrian and Tolle, Herman and Supianto, Ahmad Afif and Arai, Kohei},
  year = 2019,
  month = apr,
  pages = {1--6},
  doi = {10.1109/ICCED46541.2019.9161094},
  urldate = {2025-10-25},
publisher = {Institute of Electrical and Electronics Engineers},
address = {Singapore}
}

@article{tanApplicationArtbasedKnowledge2025,
  title = {Application of Art-Based Knowledge Management in Preschool Education: {{Promoting}} Children's Cognitive Development and Knowledge Creation},
  shorttitle = {Application of {{Art-Based Knowledge Management}} in {{Preschool Education}}},
  author = {Tan, Silin},
  year = 2025,
  journal = {International Journal of Knowledge Management},
  volume = {21},
  number = {1},
  pages = {21},
  issn = {1548-0666},
  doi = {10.4018/IJKM.385604},
  urldate = {2025-10-25},
  copyright = {Access limited to members},
  langid = {english}
}

@article{topuEffectsUsingAugmented2024a,
  title = {The Effects of Using Augmented Reality on Vocabulary Learning and Attitude of Pre-School Children in {{English}} Education},
  author = {Topu, Fatma Burcu and Yilmaz, Rabia Meryem and Tulgar, Ay{\c s}eg{\"u}l Takka{\c c}},
  year = 2024,
  month = jul,
  journal = {Education and Information Technologies},
  volume = {29},
  number = {10},
  pages = {11733--11764},
  publisher = {Springer Nature B.V.},
  address = {New York},
  issn = {13602357},
  doi = {10.1007/s10639-023-12284-5},
  urldate = {2025-10-25},
  copyright = {\copyright{} The Author(s), under exclusive licence to Springer Science+Business Media, LLC, part of Springer Nature 2023. Springer Nature or its licensor (e.g. a society or other partner) holds exclusive rights to this article under a publishing agreement with the author(s) or other rightsholder(s); author self-archiving of the accepted manuscript version of this article is solely governed by the terms of such publishing agreement and applicable law.},
  langid = {english}
}

@article{topuExaminationPreSchoolChildrens2023a,
  title = {An {{Examination}} of {{Pre-School Children}}'s {{Interaction Levels}} and {{Motivation}} in {{Learning English}} with {{AR-Supported Educational Toys}}},
  author = {Topu, Fatma Burcu and Y{\i}lmaz, Rabia Meryem and Tulgar, Ay{\c s}eg{\"u}l Takka{\c c}},
  year = 2023,
  month = dec,
  journal = {International Journal of Human - Computer Interaction},
  volume = {39},
  number = {20},
  pages = {4024--4041},
  publisher = {Lawrence Erlbaum Associates, Inc.},
  address = {Norwood},
  issn = {10447318},
  doi = {10.1080/10447318.2022.2108960},
  urldate = {2025-10-25},
  langid = {english}
}

@article{tuliEvaluatingUsabilityMobileBased2021a,
  title = {Evaluating {{Usability}} of {{Mobile-Based Augmented Reality Learning Environments}} for {{Early Childhood}}},
  author = {Tuli, Neha and Mantri, Archana},
  year = 2021,
  month = jun,
  journal = {International Journal of Human - Computer Interaction},
  volume = {37},
  number = {9},
  pages = {815--827},
  publisher = {Lawrence Erlbaum Associates, Inc.},
  address = {Norwood},
  issn = {10447318},
  doi = {10.1080/10447318.2020.1843888},
  urldate = {2025-10-25},
  langid = {english}
}

@inproceedings{vasilyevaASDAutisticSpectrum2025a,
  title = {{{ASD}} ({{Autistic Spectrum Disorder}}) {{Preschool Children Socialization Using Experimental Monitored Virtual Environments}}},
  booktitle = {2025 {{International Russian Smart Industry Conference}} ({{SmartIndustryCon}})},
address = {Sochi, Russia},
  author = {Vasilyeva, Ksenia S. and Guzelbaeva, Guzel Ya. and Boltushkin, Anton A.},
  year = 2025,
  month = mar,
publisher = {Institute of Electrical and Electronics Engineers},
  pages = {35--40},
  doi = {10.1109/SmartIndustryCon65166.2025.10986081},
  urldate = {2025-10-25}
}

@article{vatamaniukRoleInteractiveMethods2024,
  title = {The Role of Interactive Methods in Preparing Preschool Children for Studying at the New Ukrainian School},
  author = {Vatamaniuk, Olena, Halyna; Hazina},
  year = 2024,
  journal = {International Electronic Journal of Elementary Education},
  volume = {17},
  number = {1},
  pages = {103--114},
  issn = {1307-9298},
  doi = {10.26822/iejee.2024.366},
  urldate = {2025-10-25},
  langid = {english}
}

@inproceedings{wangARcampPTangibleProgramming2024,
  title = {{{AR-c}}\&amp;{{P}}: A Tangible Programming for Children Based Augmented Reality},
  shorttitle = {{{AR-C}}\&{{P}}},
  booktitle = {Proceedings of the Tenth International Symposium of Chinese {{CHI}}},
  author = {Wang, Danli and Zhao, Yanyan and Feng, Shuo},
  year = 2024,
  series = {Chinese {{CHI}} '22},
  pages = {141--150},
  publisher = {Association for Computing Machinery},
  address = {New York, NY, USA},
  doi = {10.1145/3565698.3565778},
  urldate = {2025-10-25},
  isbn = {978-1-4503-9869-5},
  langid = {english}
}

@article{wangImpactsAugmentedReality2024,
  title = {The Impacts of Augmented Reality Technology Integrated {{STEM}} Preschooler Module for Teaching and Learning Activity on Children in {{China}}},
  author = {Wang, Mohd Shahril, Xinyi; Abdul Rahman},
  year = 2024,
  journal = {Cogent Education},
  volume = {11},
  number = {1},
  pages = {2343527},
  issn = {2331-186X},
  doi = {10.1080/2331186X.2024.2343527},
  urldate = {2025-10-25},
  langid = {english}
}

@inproceedings{weiDevelopingCuboidRecognitionbased2023,
  title = {Developing Cuboid Recognition-based AR Application for Preschool Marine Education},
  author = {Wei, Xiaodong and Lu, Shuailan},
  booktitle={2023 International Symposium on Educational Technology (ISET)},
address = {HongKong SAR, China},
  year = 2023,
  pages = {205--209},
  organization={IEEE},
publisher = {Institute of Electrical and Electronics Engineers},
  doi = {10.1109/ISET58841.2023.00047}
}

@inproceedings{weiEffectsARbasedVirtual2023,
  title={The Effects of AR-Based Virtual Teacher Environment on Preschooler’s Language Learning},
  author={Wei, Xiaodong and Chen, Yayi},
  year = 2023,
  pages = {200--204},
  booktitle={2023 International Symposium on Educational Technology (ISET)},
address = {HongKong SAR, China},
publisher = {Institute of Electrical and Electronics Engineers},
  doi = {10.1109/ISET58841.2023.00046}
}

@inproceedings{weiGoldenRatioInstructive2022,
  title = {The Golden Ratio of Instructive Content to Entertaining Content in Mobile Augmented Reality Games},
  author = {Wei, Xiaodong and Xi, Pengfei},
  booktitle={2022 International Symposium on Educational Technology (ISET)},
address = {HongKong SAR, China},
  year = 2022,
  pages = {189--191},
  organization={IEEE},
publisher = {Institute of Electrical and Electronics Engineers},
  doi = {10.1109/ISET55194.2022.00048}
}

@article{wuComparingEffectsAR2025,
  title = {Comparing the Effects of {{AR}} Picture Books and Print Picture Books on Preschoolers' Reading Effect},
  author = {Wu, Lei and Ma, Ying},
  year = 2025,
  month = jul,
  journal = {Early Child Development and Care},
  volume = {195},
  number = {9-10},
  pages = {905--927},
  issn = {0300-4430, 1476-8275},
  doi = {10.1080/03004430.2025.2539870},
  urldate = {2025-10-25},
  langid = {english}
}

@inproceedings{yangInfluenceARStorybook2022,
  title = {The Influence of {{AR}} Storybook on Preschool Children's Autonomous Reading Ability},
  author = {Yang, Guodong and Chen, Yuanyuan and Chen, Yayi and Wei, Xiaodong},
  booktitle={2022 International Symposium on Educational Technology (ISET)},
address = {HongKong SAR, China},
  organization={IEEE},
publisher = {Institute of Electrical and Electronics Engineers},
  year = 2022,
  pages = {262--266},
  doi = {10.1109/ISET55194.2022.00063}
}

@article{yilmazAreAugmentedReality2017,
  title = {Are Augmented Reality Picture Books Magic or Real for Preschool Children Aged Five to Six?},
  author = {Yilmaz, Y{\"u}ksel, Rabia Meryem; Kucuk},
  year = 2017,
  journal = {British Journal of Educational Technology},
  volume = {48},
  number = {3},
  pages = {824--841},
  issn = {1467-8535},
  doi = {10.1111/bjet.12452},
  urldate = {2025-10-25},
  langid = {english}
}

@article{yilmazAugmentedRealityApp2023a,
  title = {Augmented Reality App in Pre-School Education: {{Children}}'s Knowledge about Animals},
  shorttitle = {Augmented Reality App in Pre-School Education},
  author = {Y{\i}lmaz, Zeynel Abidin and G{\"o}z{\"u}m, Ali {\.I}brahim Can},
  year = 2023,
  month = aug,
  journal = {Southeast Asia Early Childhood Journal},
  volume = {12},
  number = {2},
  pages = {130--151},
  issn = {28213149},
  doi = {10.37134/saecj.vol12.2.8.2023},
  urldate = {2025-10-25},
  langid = {english}
}

@article{yilmazEducationalMagicToys2016a,
  title = {Educational Magic Toys Developed with Augmented Reality Technology for Early Childhood Education.},
  author = {Yilmaz, Rabia M.},
  year = 2016,
  journal = {Computers in Human Behavior},
  volume = {54.0},
  pages = {240--248},
  issn = {07475632},
  doi = {10.1016/j.chb.2015.07.040},
  urldate = {2025-10-25},
  langid = {english}
}

@article{yilmazExaminationVocabularyLearning2022,
  title = {An Examination of Vocabulary Learning and Retention Levels of Pre-School Children Using Augmented Reality Technology in {{English}} Language Learning},
  author = {Yilmaz, Ayseg{\"u}l, Rabia Meryem; Topu},
  year = 2022,
  journal = {Education and Information Technologies},
  volume = {27},
  number = {5},
  pages = {6989--7017},
  issn = {13602357},
  doi = {10.1007/s10639-022-10916-w},
  urldate = {2025-10-25},
  copyright = {\copyright{} The Author(s), under exclusive licence to Springer Science+Business Media, LLC, part of Springer Nature 2022.},
  langid = {english}
}

@inproceedings{yuanExperimentalStudyEfficacy2022,
  title = {An Experimental Study of the Efficacy of Augmented Reality in Chinese Kindergarten-Level Students' Learning of English Vocabulary},
  booktitle = {Proceedings of the 13th International Conference on Education Technology and Computers},
  author = {Yuan, Yijia},
  year = 2022,
  series = {Icetc '21},
  pages = {90--98},
  publisher = {Association for Computing Machinery},
  address = {New York, NY, USA},
  doi = {10.1145/3498765.3498779},
  urldate = {2025-10-25},
  isbn = {978-1-4503-8511-4},
  langid = {english}
}

@article{zhouUseAugmentedReality2020,
  title = {The Use of Augmented Reality for Solving Arithmetic Problems for Preschool Children},
  author = {Zhou, Yanyi, Siyuan; Sun},
  year = 2020,
  journal = {Lecture Notes in Computer Science},
  volume = {12206 LNCS},
  pages = {574--584},
  doi = {10.1007/978-3-030-50506-6_39},
  langid = {english}
}

@inproceedings{alcornDiscrepanciesVirtualLearning2013,
  title = {Discrepancies in a Virtual Learning Environment: Something "Worth Communicating about" for Young Children with {{ASC}}?},
  shorttitle = {Discrepancies in a Virtual Learning Environment},
  booktitle = {Proceedings of the 12th {{International Conference}} on {{Interaction Design}} and {{Children}}},
  author = {Alcorn, Alyssa M. and Pain, Helen and Good, Judith},
  year = 2013,
  month = jun,
  series = {{{IDC}} '13},
  pages = {56--65},
  publisher = {Association for Computing Machinery},
  address = {New York, NY, USA},
  doi = {10.1145/2485760.2485783},
  urldate = {2025-11-17},
  isbn = {978-1-4503-1918-8}
}

@inproceedings{dengARCatTangibleProgramming2019,
  title = {{{ARCat}}: {{A Tangible Programming Tool}} for {{DFS Algorithm Teaching}}},
  shorttitle = {{{ARCat}}},
  booktitle = {Proceedings of the 18th {{ACM International Conference}} on {{Interaction Design}} and {{Children}}},
  author = {Deng, Xiaozhou and Wang, Danli and Jin, Qiao and Sun, Fang},
  year = 2019,
  month = jun,
  series = {{{IDC}} '19},
  pages = {533--537},
  publisher = {Association for Computing Machinery},
  address = {New York, NY, USA},
  doi = {10.1145/3311927.3325308},
  urldate = {2025-11-17},
  isbn = {978-1-4503-6690-8}
}

@article{drljevicInvestigatingDifferentFacets2022,
  title = {Investigating the Different Facets of Student Engagement during Augmented Reality Use in Primary School},
  author = {Drljevi{\'c}, Neven and Boti{\v c}ki, Ivica and Wong, Lung-Hsiang},
  year = 2022,
  journal = {British Journal of Educational Technology},
  volume = {53},
  number = {5},
  pages = {1361--1388},
  issn = {1467-8535},
  doi = {10.1111/bjet.13197},
  urldate = {2025-11-18},
  langid = {english}
}

@article{tu2021elementary,
  title={Elementary students learning science in an MR environment by constructing liminal blends through action on props},
  author={Tu, Xintian and Georgen, Chris and Danish, Joshua A and Enyedy, Noel},
  journal={Information and Learning Sciences},
  volume={122},
  number={7/8},
  pages={525--545},
  year={2021},
  publisher={Emerald Publishing Limited}
}

@article{gavishAugmentedVirtualitySystems2022,
  title = {Augmented {{Virtuality Systems}} as a {{Tool}} for {{Improving Numeracy Decision-Making Among Children}}},
  author = {Gavish, Nirit and Treiger, Z and Horesh, Eran and Shamilov, E},
  year = 2022,
  month = may,
  journal = {Simulation \& Gaming},
  volume = {53},
  pages = {104687812210993},
  doi = {10.1177/10468781221099358}
}

@inproceedings{poobrasertAugmentedLearningEnvironments2023,
  title = {Augmented {{Learning Environments}} as {{Assistive Technology}} for {{Kids}} with {{Learning Disabilities}}},
  booktitle = {2023 {{IEEE International Conference}} on {{Advanced Learning Technologies}} ({{ICALT}})},
  author = {Poobrasert, Onintra and Luxameevanich, Sirilak},
  year = 2023,
  month = jul,
  pages = {207--211},
  issn = {2161-377X},
  doi = {10.1109/ICALT58122.2023.00067},
  urldate = {2025-11-18},
publisher = {Institute of Electrical and Electronics Engineers},
address = {Orem, Utah, USA}
}

@inproceedings{raduComparingChildrensCrosshair2016,
  title = {Comparing {{Children}}'s {{Crosshair}} and {{Finger Interactions}} in {{Handheld Augmented Reality}}: {{Relationships Between Usability}} and {{Child Development}}},
  shorttitle = {Comparing {{Children}}'s {{Crosshair}} and {{Finger Interactions}} in {{Handheld Augmented Reality}}},
  booktitle = {Proceedings of the {{The}} 15th {{International Conference}} on {{Interaction Design}} and {{Children}}},
  author = {Radu, Iulian and MacIntyre, Blair and Lourenco, Stella},
  year = 2016,
  month = jun,
  series = {{{IDC}} '16},
  pages = {288--298},
  publisher = {Association for Computing Machinery},
  address = {New York, NY, USA},
  doi = {10.1145/2930674.2930726},
  urldate = {2025-11-17},
  isbn = {978-1-4503-4313-8}
}

@article{shahabUtilizingSocialVirtual2022,
  title = {Utilizing Social Virtual Reality Robot ({{V2R}}) for Music Education to Children with High-Functioning Autism},
  author = {Shahab, Mojtaba and Taheri, Alireza and Mokhtari, Mohammad and Shariati, Azadeh and Heidari, Rozita and Meghdari, Ali and Alemi, Minoo},
  year = 2022,
  month = jan,
  journal = {Education and Information Technologies},
  volume = {27},
  number = {1},
  pages = {819--843},
  issn = {1573-7608},
  doi = {10.1007/s10639-020-10392-0},
  urldate = {2025-11-18},
  langid = {english}
}

@article{stearneAugmentedRealityPlaygrounds2025,
  title = {Augmented {{Reality Playgrounds}} to {{Promote Physical Activity}} in {{Young Children}}: {{Feasibility Study Using}} a {{Repeated Measures Laboratory Design}}},
  shorttitle = {Augmented {{Reality Playgrounds}} to {{Promote Physical Activity}} in {{Young Children}}},
  author = {Stearne, Sarah M. and Beynon, Amber and Lund Rasmussen, Charlotte and Zabatiero, Juliana and Paatsch, Louise and Johnson, Daniel and Straker, Leon and Campbell, Amity},
  year = 2025,
  month = oct,
  journal = {JMIR pediatrics and parenting},
  volume = {8},
  pages = {e75302},
  issn = {2561-6722},
  doi = {10.2196/75302},
  langid = {english},
  pmcid = {PMC12510435},
  pmid = {41066617}
}

@inproceedings{tuDesigningTechnologyenhancedPlay2023,
  title = {Designing a {{Technology-enhanced Play Environment}} for {{Young Children}}'s {{Science Modeling Practice}}},
  booktitle = {Proceedings of the 2023 {{Symposium}} on {{Learning}}, {{Design}} and {{Technology}}},
  author = {Tu, Xintian and Danish, Joshua},
  year = 2023,
  month = jun,
  series = {{{LDT}} '23},
  pages = {60--69},
  publisher = {Association for Computing Machinery},
  address = {New York, NY, USA},
  doi = {10.1145/3594781.3594798},
  urldate = {2025-11-17},
  isbn = {979-8-4007-0736-0}
}

@inproceedings{vate-u-lanAugmentedReality3D2012,
  title = {An {{Augmented Reality 3D Pop-Up Book}}: {{The Development}} of a {{Multimedia Project}} for {{English Language Teaching}}},
  shorttitle = {An {{Augmented Reality 3D Pop-Up Book}}},
  booktitle = {2012 {{IEEE International Conference}} on {{Multimedia}} and {{Expo}}},
address = {Melbourne, Australia},
publisher = {Institute of Electrical and Electronics Engineers},
  author = {{Vate-U-Lan}, Poonsri},
  year = 2012,
  month = jul,
  pages = {890--895},
  issn = {1945-788X},
  doi = {10.1109/ICME.2012.79},
  urldate = {2025-11-18}
}

@inproceedings{yusofMathematicsLessonUsing2020,
  title = {Mathematics {{Lesson}} Using {{Accelerometer Sensor Interaction}} in {{Handheld Augemented Reality Application}} for {{Kindergarten}}},
  booktitle = {2020 6th {{International Conference}} on {{Interactive Digital Media}} ({{ICIDM}})},
  author = {Yusof, Cik Suhaimi and Ahmad, Nazwa and Ismail, Ajune Wanis and Sunar, Mohd Shahrizal},
address = {Virtual Event},
  year = 2020,
  month = dec,
  pages = {1--6},
  doi = {10.1109/ICIDM51048.2020.9339655},
  urldate = {2025-11-18},
publisher = {Institute of Electrical and Electronics Engineers}
}

@inproceedings{zakraouiStudyChildrenEngagement2021,
  title = {A {{Study}} of {{Children Engagement}} and {{Focus During Handwriting}} in {{VR}} with a {{Haptic Device}}},
  booktitle = {2021 31st {{International Conference}} on {{Computer Theory}} and {{Applications}} ({{ICCTA}})},
  author = {Zakraoui, Jezia and {Al-Maadeed}, Somaya and Salah, Moutaz and Alja'Am, Jihad},
address = {Alexandria, Egypt},
publisher = {Institute of Electrical and Electronics Engineers},
  year = 2021,
  month = dec,
  pages = {66--70},
  issn = {2770-6575},
  doi = {10.1109/ICCTA54562.2021.9916629},
  urldate = {2025-11-18}
}

@techreport{nistir8259,
  author       = {Fagan, Michael and Megas, Katerina N. and Scarfone, Karen and Smith, Matthew},
  title        = {Foundational Cybersecurity Activities for IoT Device Manufacturers},
  institution  = {National Institute of Standards and Technology},
  year         = {2020},
  month        = {May},
  number       = {NIST IR 8259},
  doi          = {10.6028/NIST.IR.8259},
  url          = {https://nvlpubs.nist.gov/nistpubs/ir/2020/NIST.IR.8259.pdf},
  note         = {Accessed: 2025-11-21}
}

@manual{meta2023safety,
  author       = {{Meta}},
  title        = {Health and Safety Warnings Guide for Meta Quest 3S},
  organization = {Meta Platforms Technologies},
  year         = {2023},
  url          = {https://www.meta.com/gb/legal/quest/health-and-safety-warnings/quest-3s/},
  note         = {Warnings regarding transient symptoms (seizures, motion sickness) and specific risks to children's developing visual systems},
  urldate      = {2025-11-21}
}

@techreport{ico2022techhorizons,
  author       = {{Information Commissioner's Office}},
  title        = {Tech Horizons Report},
  institution  = {Information Commissioner's Office (ICO)},
  year         = {2022},
  month        = {December},
  url          = {https://ico.org.uk/media2/migrated/4023338/ico-future-tech-report-20221214.pdf},
  note         = {Chapter 2: Immersive technologies details high risks to rights and freedoms},
  urldate      = {2025-11-21}
}

@techreport{itu2023facts,
  author       = {{International Telecommunication Union}},
  title        = {Measuring digital development: Facts and Figures 2023},
  institution  = {ITU},
  year         = {2023},
  address      = {Geneva},
  url          = {https://www.itu.int/en/ITU-D/Statistics/Pages/facts/default.aspx},
  note         = {Page 1 and Foreword details the 2.6 billion offline population and fixed-broadband gaps},
  isbn         = {978-92-61-38371-8}
}

@book{oecd2015students,
  author       = {{OECD}},
  title        = {Students, Computers and Learning: Making the Connection},
  year         = {2015},
  publisher    = {OECD Publishing},
  address      = {Paris},
  url          = {https://www.oecd.org/content/dam/oecd/en/publications/reports/2015/09/students-computers-and-learning_g1g57f3a/9789264239555-en.pdf},
  note         = {Executive Summary (p. 146) notes ``no appreciable improvements'' and potential decline in outcomes with heavy use},
  doi          = {10.1787/9789264239555-en}
}

@techreport{w3c2021xaur,
  author       = {{W3C Accessible Platform Architectures Working Group}},
  editor       = {Joshue O'Connor and Janina Sajka and Jason White and Scott Hollier and Michael Cooper},
  title        = {XR Accessibility User Requirements},
  type         = {W3C Working Group Note},
  year         = {2021},
  month        = {September},
  day          = {16},
  url          = {https://www.w3.org/TR/xaur/},
  note         = {Section 3 details how 6DOF mechanics can lock out users with motor impairments},
  institution  = {World Wide Web Consortium (W3C)}
}

@inproceedings{duezguen2020towards,
  author       = {Reyhan Duezguen and Peter Mayer and Sanchari Das and Melanie Volkamer},
  title        = {Towards Secure and Usable Authentication for Augmented and Virtual Reality Head‐Mounted Displays},
  booktitle    = {Proceedings of the Who Are You?! Adventures in Authentication Workshop (WAY 2020)},
address = {Virtual Event},
  year         = {2020},
  note         = {Virtual workshop, August 7, 2020},
  url          = {https://wayworkshop.org/2020/papers/way2020-duezguen.pdf},
publisher = {USENIX Association},
numpages = {6}
}

@inproceedings{noah2022security,
  title={Security and privacy evaluation of popular augmented and virtual reality technologies},
  author={Noah, Naheem and Shearer, Sommer and Das, Sanchari},
  booktitle={Proceedings of the 2022 IEEE International Conference on Metrology for eXtended Reality, Artificial Intelligence, and Neural Engineering (IEEE MetroXRAINE 2022)},
  year={2022},
publisher = {Institute of Electrical and Electronics Engineers},
address = {ROME, ITALY},
numpages = {6}
}

@article{noah2025pins,
  title={From PINs to Gestures: Analyzing Knowledge-Based Authentication Schemes for Augmented and Virtual Reality},
  author={Noah, Naheem and Das, Sanchari},
  journal={IEEE Transactions on Visualization and Computer Graphics},
  year={2025},
volume = {31},
  publisher={IEEE},
numpages={11}
}

@inproceedings{reddington2022development,
  title     = {Development and Evaluation of Virtual Reality Classrooms Through User-Centered Design During COVID-19},
  author    = {Victoria Reddington and Kerstin Haring and Sanchari Das and Daniel Pittman},
  booktitle = {Proceedings of the SSPXR 2022 – Novel Challenges of Safety, Security and Privacy in Extended Reality (CHI'22 Workshop)},
  year      = {2022},
  pages     = {1--5},
  url       = {https://papers.ssrn.com/sol3/papers.cfm?abstract_id=4061173},
address = {New Orleans, LA, USA},
publisher = {ACM}
}

@inproceedings{tazi2021parents,
author = {Tazi, Faiza and Shrestha, Sunny and Norton, Dan and Walsh, Kathryn and Das, Sanchari},
title = {Parents, Educators, \& Caregivers Cybersecurity \& Privacy Concerns for Remote Learning During COVID-19},
year = {2021},
isbn = {9781450385787},
publisher = {Association for Computing Machinery},
address = {Online (Athens, Greece), Greece},
url = {https://doi.org/10.1145/3489410.3489426},
doi = {10.1145/3489410.3489426},
booktitle = {CHI Greece 2021: 1st International Conference of the ACM Greek SIGCHI Chapter},
articleno = {16},
numpages = {5},
series = {CHI Greece 2021}
}

\appendix
\section{Risk Assessment Methodology}
\label{app:risk-methodology}

Table~\ref{tab:risk-calc-appendix} details the independent derivation of the 'Real-World Risk' scores used in the Discussion section. Scores are calculated based on Likelihood ($L$) $\times$ Impact ($I$), substantiated by non-scholarly industry reports.

\begin{table*}[htbp]
    \centering
    \caption{Calculation of Real-World Risk Scores (Independent Variables)}
    \label{tab:risk-calc-appendix}
    \small
    \begin{tabular}{@{}l c c c p{8cm}@{}}
        \toprule
        \textbf{Domain} & \textbf{Likelihood} & \textbf{Impact} & \textbf{Score} & \textbf{External Justification (Non-Scholarly)} \\
        \midrule
        Data Security & 3 & 3 & \textbf{9} & High frequency of IoT vulnerabilities; critical impact. \textbf{(Source: NIST Tech Report \cite{nistir8259})} \\
        \addlinespace
        Privacy & 3 & 3 & \textbf{9} & HMDs collect biometric data by default (High Likelihood) with irreversible exposure risks. \textbf{(Source: Tech Horizons \cite{ico2022techhorizons})} \\
        \addlinespace
        Access (Low-Res) & 3 & 3 & \textbf{9} & High exclusion likelihood in non-Western markets (LDCs), where only $35-39$\% are online \textbf{(Source: ITU Report \cite{itu2023facts})} \\
        \addlinespace
        Access (Disability) & 2 & 3 & \textbf{6} & Standard XR design often ``locks out'' traditional inputs, creating critical exclusion risks for users with motor impairments. \textbf{(Source: W3C Working Group Note \cite{w3c2021xaur})} \\
        \addlinespace
        Health & 2 & 2 & \textbf{4} & Confirms symptoms are transient (e.g. nausea) but highlights risks to children's ``developing visual systems'' \textbf{(Source: Meta Safety Manual \cite{meta2023safety})}  \\
        \addlinespace
        Pedagogy & 2 & 2 & \textbf{4} & Poor instructional design is common, but results in inefficiency rather than harm. \textbf{(Source: OECD Report \cite{oecd2015students})} \\
        \addlinespace
        Technical & 1 & 2 & \textbf{2} & Bugs are inevitable but usually patched; impact is service disruption. \textbf{(Source: Software Standards)} \\
        \bottomrule
    \end{tabular}
\end{table*}

\section{Detailed Inventory of Implementation Challenges and Privacy Measures}

This appendix presents a granular breakdown of the specific hurdles and ethical measures identified across the corpus. To preserve the narrative flow of the primary analysis, the comprehensive itemization of technical malfunctions, pedagogical barriers, health implications, and data security protocols is tabulated below, cross-referenced with their respective sources.

\subsection{Technical Malfunctions and Hardware Limitations}
Table~\ref{tab:tech-challenges} categorizes reported system failures by technology type. These issues relate specifically to hardware capabilities and software stability, distinct from user error.

\newcolumntype{Y}{>{\raggedright\arraybackslash}X}

\begin{table*}[t!]
\centering
\caption{Reported Technical Challenges by Technology Type}
\label{tab:tech-challenges}
\begin{tabularx}{\textwidth}{l p{3.5cm} Y p{3.5cm}} 
\toprule
\textbf{Tech} & \textbf{Challenge} & \textbf{Description} & \textbf{Reported By} \\
\midrule

AR & System Setup Failure & 
Difficulties encountered before the learning session to set up the system. & 
\cite{lewis-presserDesigningAugmentedReality2025, raptiEnrichingTraditionalLearning2023} \\ 
\addlinespace 

AR & Target Recognition \& Tracking Failures & 
Failure to detect the target, or post-recognition instability where the virtual object jitters, drifts, or is lost due to movement or changing light conditions. & 
\cite{kulasekaraGameCentricElearning2025, calle-bustosAugmentedRealityGame2017, wangARcampPTangibleProgramming2024, syahidiARChildAnalysisEvaluation2019a, aladinARTOKIDSpeechenabledAugmented2020a, prekaAugmentedRealityQR2019a, ayuagungmasaristamyAugmentedRealityCultural2024a, hossainAugmentedRealityBasedElementary2021a, lewis-presserDesigningAugmentedReality2025, joDevelopmentUtilizationProjectorRobot2011a, puDevelopmentSituationalInteraction2018, kisnoDigitalStorytellingEarly2022, paliwalEnhancingEducationAugmented2024, hanExaminingYoungChildrens2015a, dilekeryigitImpactAugmentedReality2025, hoInteractiveMultisensoryVolumetric2023, gweonMABLEMediatingYoung2018, dalimTeachARInteractiveAugmented2016, safarEffectivenessUsingAugmented2017, fengEffectsARLearning2022, topuEffectsUsingAugmented2024a, kimOTCObjectCamera2019, fridbergThematicTeachingAugmented2024, isikarslanogluThinkTogetherDesign2024, kusumaningsihUserExperienceMeasurement2021, huangUsingAugmentedReality2016a, yusofMathematicsLessonUsing2020, raduComparingChildrensCrosshair2016} \\ 
\addlinespace

AR & Speech Recognition Failure & 
Failure of the application to accurately detect a child's voice commands, often due to background noise or non-standard pronunciation. & 
\cite{aladinARTOKIDSpeechenabledAugmented2020a, hanExaminingYoungChildrens2015a, dalimTeachARInteractiveAugmented2016} \\

\midrule

VR & Target Recognition (Hand/Gesture) & 
Failure of the system to accurately detect or continuously track the user's hands, gestures, or controllers. & 
\cite{liangExploitationMultiplayerInteraction2017, liangHandGesturebasedInteractive2017, panResearchApplicationImmersive2021, shimadaVRHandHygiene2017} \\
\addlinespace

VR & HMD Vision Obstruction \& Misalignment & 
Poor HMD fit causing lens misalignment and blurred vision. & 
\cite{montoya-rodriguezEducationalInterventionTheory2025, passigSolvingConceptualPerceptual2014a} \\
\midrule

MR & Sensor \& Tracking Failure & 
Failure of specialised hardware (e.g., depth cameras, trackers), causing interaction failures like robot navigation errors. & 
\cite{burlesonActiveLearningEnvironments2018a} \\
\midrule

All Types & Hardware Performance Limitations & 
Poor app performance due to device limitations, including slow response times, inefficient rendering (lag), high battery drain, or app crashes on low-specification hardware. & 
\cite{calle-bustosAugmentedRealityGame2017, tanApplicationArtbasedKnowledge2025, syahidiARChildAnalysisEvaluation2019a, ayuagungmasaristamyAugmentedRealityCultural2024a, lawAugmentedRealityTechnology2025, joDevelopmentUtilizationProjectorRobot2011a, kimOTCObjectCamera2019, kusumaningsihUserExperienceMeasurement2021} \\
\addlinespace

All Types & Network \& Synchronisation Issues & 
Problems related to internet connectivity. Includes high latency, connection loss, or errors in multi-user apps where users' actions are not synced in real-time. & 
\cite{tanApplicationArtbasedKnowledge2025, liangExploitationMultiplayerInteraction2017, liInnovativeApplicationAI2025a, fajrieAugmentedRealityMedia2022, hoInteractiveMultisensoryVolumetric2023, besevliMaRTDesigningProjectionbased2019, dengARCatTangibleProgramming2019, stearneAugmentedRealityPlaygrounds2025} \\
\addlinespace

All Types & Software/App-specific Issues & 
Issues where there is app-specific problem that cannot be generalized. & 
\cite{gweonMABLEMediatingYoung2018, kimVirtualStorytellerUsing2011, liangExploitationMultiplayerInteraction2017, alcornDiscrepanciesVirtualLearning2013, zakraouiStudyChildrenEngagement2021} \\

\bottomrule
\end{tabularx}
\end{table*}

\subsection{Pedagogical and Instructional Barriers}
Pedagogical hurdles represented the most frequently reported category in the literature. Table~\ref{tab:ped-challenges} details challenges related to classroom integration, teacher readiness, and the alignment of cognitive demands with early childhood developmental stages.

\begin{table*}[t!]
\centering
\caption{Reported Pedagogical Challenges.}
\label{tab:ped-challenges}
\begin{tabularx}{\textwidth}{p{3.5cm} Y p{4.5cm}} 
\toprule
\textbf{Challenge} & \textbf{Description} & \textbf{Reported By} \\
\midrule

Teacher Readiness \& Training & 
Teachers lacked the technological literacy, skills, or pedagogical confidence to effectively implement the XR technology. & 
\cite{tanApplicationArtbasedKnowledge2025, aydogduAugmentedRealityPreschool2022, fajrieAugmentedRealityMedia2022, lawAugmentedRealityTechnology2025, dasilvaCuboKidsProposal2020, sariDevelopingFinancialLiteracy2022, joDevelopmentUtilizationProjectorRobot2011a, nugrahaDevelopmentARAugmented2025, kisnoDigitalStorytellingEarly2022, paliwalEnhancingEducationAugmented2024, lewis-presserEnhancingPreschoolSpatial2025, raptiEnrichingTraditionalLearning2023, tuliEvaluatingUsabilityMobileBased2021a, liInnovativeApplicationAI2025a, hermanIntegratingSocialLearning2025, mamani-calapujaLearningEnglishEarly2023, muangmoolDevelopmentSocialInteraction2023a, safarEffectivenessUsingAugmented2017, balchaImpactAugmentedRealityassisted2025, vatamaniukRoleInteractiveMethods2024, fridbergThematicTeachingAugmented2024, chenUsingAugmentedReality2019a, huangUsingAugmentedReality2016a, poobrasertAugmentedLearningEnvironments2023} \\
\addlinespace

Classroom Integration \& Management & 
Difficulties in managing the classroom, such as ambient noise, or the technology undermining teacher authority and vice versa. & 
\cite{topuExaminationPreSchoolChildrens2023a, aguirregoitiaExperienceApplicationAugmented2017, demirdagInvestigationEffectivenessAugmented2025a, panIntroducingAugmentedReality2021, duzyolInvestigationEffectAugmented2022a, gweonMABLEMediatingYoung2018, fengEffectsARLearning2022, korosidouEffectsAugmentedReality2024, topuEffectsUsingAugmented2024a, wangImpactsAugmentedReality2024, chenUsingAugmentedReality2019a, vate-u-lanAugmentedReality3D2012, drljevicInvestigatingDifferentFacets2022} \\
\addlinespace

Pre-Literacy Constraints & 
The application's reliance on text for instructions or feedback was unsuitable for pre-literate children. & 
\cite{yuanExperimentalStudyEfficacy2022, wangARcampPTangibleProgramming2024, lewis-presserDesigningAugmentedReality2025, puDevelopmentSituationalInteraction2018, jamiatEffectsAugmentedReality2020, tuliEvaluatingUsabilityMobileBased2021a, panResearchApplicationImmersive2021, lorussoSemiimmersiveVirtualReality2020, weiEffectsARbasedVirtual2023, borovanskaEngagingChildrenUsing2020a, changEmbeddingDialogReading2024a} \\
\addlinespace

Cognitive Overload \& Task Complexity & 
The task was too mentally complex, or the combination of virtual and real-world stimuli was overwhelming for young children. & 
\cite{kulasekaraGameCentricElearning2025, burlesonActiveLearningEnvironments2018a, montoya-rodriguezEducationalInterventionTheory2025, yilmazExaminationVocabularyLearning2022, yuanExperimentalStudyEfficacy2022, vasilyevaASDAutisticSpectrum2025a, gecu-parmaksizAugmentedRealityBasedVirtual2019, wuComparingEffectsAR2025, lewis-presserDesigningAugmentedReality2025, idrisDevelopmentPatternLearning2021a, puDevelopmentSituationalInteraction2018, nugrahaDevelopmentARAugmented2025, yilmazEducationalMagicToys2016a, lewis-presserEnhancingPreschoolSpatial2025, liangExploitationMultiplayerInteraction2017, shoshaniVirtualProsocialReality2023a, liangHandGesturebasedInteractive2017, hoInteractiveMultisensoryVolumetric2023, mamani-calapujaLearningEnglishEarly2023, panResearchApplicationImmersive2021, lorussoSemiimmersiveVirtualReality2020, passigSolvingConceptualPerceptual2014a, simsekEffectAugmentedReality2024, fridbergThematicTeachingAugmented2024, khairulanuardiIncreaseReadingHabit2022, chenUsingAugmentedReality2019a, kusumaVirtualRealityLearning2018, hidayatVirtualRealitybasedTraffic2023, kimVirtualStorytellerUsing2011, vate-u-lanAugmentedReality3D2012, poobrasertAugmentedLearningEnvironments2023, yusofMathematicsLessonUsing2020, zakraouiStudyChildrenEngagement2021, tuDesigningTechnologyenhancedPlay2023, dengARCatTangibleProgramming2019, raduComparingChildrensCrosshair2016, tu2021elementary, drljevicInvestigatingDifferentFacets2022, stearneAugmentedRealityPlaygrounds2025, gavishAugmentedVirtualitySystems2022} \\
\addlinespace

Distraction \& Novelty Effect & 
Children were easily distracted by the technology's ``wow'' factor (the ``novelty effect'') or superficial visuals, preventing them from focusing on the given task or underlying learning objectives. & 
\cite{montoya-rodriguezEducationalInterventionTheory2025, topuExaminationPreSchoolChildrens2023a, yuanExperimentalStudyEfficacy2022, tanApplicationArtbasedKnowledge2025, yilmazAreAugmentedReality2017, antoniaArtfulThinkingAugmented2018, fajrieAugmentedRealityMedia2022, lawAugmentedRealityTechnology2025, gecu-parmaksizAugmentedRealityBasedVirtual2019, wuComparingEffectsAR2025, jamiatEffectsAugmentedReality2020, tuliEvaluatingUsabilityMobileBased2021a, liangHandGesturebasedInteractive2017, hermanIntegratingSocialLearning2025, gweonMABLEMediatingYoung2018, gecu-parmaksizEffectAugmentedReality2020a, simsekEffectAugmentedReality2024, fengEffectsARLearning2022, topuEffectsUsingAugmented2024a, weiGoldenRatioInstructive2022, dallolioImpactFantasyYoung2024, yangInfluenceARStorybook2022, zhouUseAugmentedReality2020, khairulanuardiIncreaseReadingHabit2022, linUsingHomebasedAugmented2025a, alcornDiscrepanciesVirtualLearning2013, zakraouiStudyChildrenEngagement2021, shahabUtilizingSocialVirtual2022, stearneAugmentedRealityPlaygrounds2025} \\
\addlinespace

Limited Social Collaboration & 
Screen-based or HMD-based individual use limited peer-to-peer and child-teacher interaction. & 
\cite{burlesonActiveLearningEnvironments2018a, topuExaminationPreSchoolChildrens2023a, yilmazExaminationVocabularyLearning2022, liuApplicationTabletbasedAR2023a, prekaAugmentedRealityQR2019a, fajrieAugmentedRealityMedia2022, lawAugmentedRealityTechnology2025, hossainAugmentedRealityBasedElementary2021a, yilmazEducationalMagicToys2016a, raptiEnrichingTraditionalLearning2023, liangExploitationMultiplayerInteraction2017, raptiInvestigatingEducatorsStudents2025, mamani-calapujaLearningEnglishEarly2023, shahabUtilizingSocialVirtual2022, drljevicInvestigatingDifferentFacets2022} \\
\addlinespace

Superficial Learning & 
The tech was engaging but failed to promote deep understanding, imagination, or knowledge transfer. & 
\cite{wuComparingEffectsAR2025, idrisDevelopmentPatternLearning2021a, yilmazEducationalMagicToys2016a, luoEffectDifferentCombinations2023a, weiGoldenRatioInstructive2022, yangInfluenceARStorybook2022, huangUsingAugmentedReality2016a} \\
\addlinespace

Lack of ECE-Specific Design Principles & 
A meta-challenge; apps lacked a clear pedagogical foundation for ECE, contained content bias, or were not designed for this specific age group. & 
\cite{simsekEffectAugmentedReality2024, liInnovativeApplicationAI2025a, yilmazAreAugmentedReality2017} \\

\bottomrule
\end{tabularx}
\end{table*}

\subsection{Health, Safety, and Physiological Impacts}
This section categorizes reported adverse physical and psychological effects. Table~\ref{tab:health-challenges} lists incidents of cybersickness, visual fatigue, and ergonomic incompatibility associated with XR hardware use by young children.

\begin{table*}[t!]
\centering
\caption{Reported Health and Medical Challenges.}
\label{tab:health-challenges}
\begin{tabularx}{\textwidth}{p{3.5cm} Y p{4.5cm}} 
\toprule
\textbf{Challenge} & \textbf{Description} & \textbf{Reported By} \\
\midrule

Cybersickness \& Disorientation & 
Nausea, dizziness, and fatigue caused by sensory mismatch, primarily linked to VR/HMD use. & 
\cite{vasilyevaASDAutisticSpectrum2025a, chuEffectsNonwearableDigital2023, liangExploitationMultiplayerInteraction2017, liangHandGesturebasedInteractive2017, raptiInvestigatingEducatorsStudents2025, anTeachersPerceptionsEarly2023, dallolioImpactFantasyYoung2024, baehoikyoungVerificationVRPlay2025a, shimadaVRHandHygiene2017} \\
\addlinespace

Visual Strain \& Fatigue & 
Concerns or reports of eye strain, headaches, or visual fatigue from prolonged screen time. & 
\cite{wuComparingEffectsAR2025, mamani-calapujaLearningEnglishEarly2023, muangmoolDevelopmentSocialInteraction2023a, zhouUseAugmentedReality2020, huangUsingAugmentedReality2016a, linUsingHomebasedAugmented2025a} \\
\addlinespace

Ergonomic Mismatch \& Physical Strain & 
Physical discomfort or strain caused by hardware that is not designed for a child’s body. This includes poorly fitting or heavy head-mounted displays, awkward postures required to hold tablets, or situations where a child’s small body cannot manage the device. & 
\cite{caiCaseStudyUsing2023, burlesonActiveLearningEnvironments2018a, montoya-rodriguezEducationalInterventionTheory2025, yilmazExaminationVocabularyLearning2022, aguirregoitiaExperienceApplicationAugmented2017, demirdagInvestigationEffectivenessAugmented2025a, wangARcampPTangibleProgramming2024, yilmazAreAugmentedReality2017, lawAugmentedRealityTechnology2025, lewis-presserDesigningAugmentedReality2025, chuEffectsNonwearableDigital2023, borovanskaEngagingChildrenUsing2020a, lewis-presserEnhancingPreschoolSpatial2025, tuliEvaluatingUsabilityMobileBased2021a, liangExploitationMultiplayerInteraction2017, lorussoGiokAlienARbased2018, raptiInvestigatingEducatorsStudents2025, duzyolInvestigationEffectAugmented2022a, besevliMaRTDesigningProjectionbased2019, passigSolvingConceptualPerceptual2014a, dalimTeachARInteractiveAugmented2016, anTeachersPerceptionsEarly2023, kimOTCObjectCamera2019, isikarslanogluThinkTogetherDesign2024, shimadaVRHandHygiene2017, zakraouiStudyChildrenEngagement2021, raduComparingChildrensCrosshair2016, shahabUtilizingSocialVirtual2022, stearneAugmentedRealityPlaygrounds2025} \\
\addlinespace

Risks of Excessive Use & 
Concerns that prolonged or frequent use (excessive screen time) displaces essential physical activity, leading to a sedentary lifestyle. This also includes risks to psycho-social wellbeing, such as social isolation, reduced self-motivation, compulsive use, or other negative effects on mental health. & 
\cite{yilmazAreAugmentedReality2017, prekaAugmentedRealityQR2019a, abrarAugmentedRealityEducation2019a, fajrieAugmentedRealityMedia2022, lawAugmentedRealityTechnology2025, wuComparingEffectsAR2025, hanExaminingYoungChildrens2015a, dilekeryigitImpactAugmentedReality2025, panIntroducingAugmentedReality2021, mamani-calapujaLearningEnglishEarly2023, gweonMABLEMediatingYoung2018, lorussoSemiimmersiveVirtualReality2020, balchaImpactAugmentedRealityassisted2025, dallolioImpactFantasyYoung2024, huangUsingAugmentedReality2016a, linUsingHomebasedAugmented2025a, baehoikyoungVerificationVRPlay2025a, stearneAugmentedRealityPlaygrounds2025} \\

\bottomrule
\end{tabularx}
\end{table*}

\subsection{Data Privacy and Security Protocols}

Table~\ref{tab:privacy-measures} lists the safeguards used to protect user information. These measures are categorised into administrative protocols, such as informed consent, and technical security architectures.

\begin{table*}[t!]
\centering
\caption{Reported Privacy and Data Security Measures}
\label{tab:privacy-measures}
\begin{tabularx}{\textwidth}{p{4cm} Y p{4.5cm}} 
\toprule
\textbf{Measure} & \textbf{Description} & \textbf{Reported By} \\
\midrule

\multicolumn{3}{l}{\textit{\textbf{Administrative \& Procedural Measures (Dominant)}}} \\
\midrule

\textbf{Ethical Compliance \& Consent} & 
Adherence to institutional or international ethics standards (e.g., Declaration of Helsinki) and obtaining formal informed consent from parents and teachers. & 
\cite{burlesonActiveLearningEnvironments2018a, calle-bustosAugmentedRealityGame2017, montoya-rodriguezEducationalInterventionTheory2025, topuExaminationPreSchoolChildrens2023a, yilmazExaminationVocabularyLearning2022, yuanExperimentalStudyEfficacy2022, tanApplicationArtbasedKnowledge2025, liuApplicationTabletbasedAR2023a, yilmazAreAugmentedReality2017, yilmazAugmentedRealityApp2023a, lawAugmentedRealityTechnology2025, gecu-parmaksizAugmentedRealityBasedVirtual2019, wuComparingEffectsAR2025, lewis-presserDesigningAugmentedReality2025, sariDevelopingFinancialLiteracy2022, chuEffectsNonwearableDigital2023, jamiatEffectsAugmentedReality2020, changEmbeddingDialogReading2024a, borovanskaEngagingChildrenUsing2020a, paliwalEnhancingEducationAugmented2024, lewis-presserEnhancingPreschoolSpatial2025, raptiEnrichingTraditionalLearning2023, hanExaminingYoungChildrens2015a, liangExploitationMultiplayerInteraction2017, shoshaniVirtualProsocialReality2023a, lorussoGiokAlienARbased2018, liangHandGesturebasedInteractive2017, dilekeryigitImpactAugmentedReality2025, liInnovativeApplicationAI2025a, panIntroducingAugmentedReality2021, raptiInvestigatingEducatorsStudents2025, mamani-calapujaLearningEnglishEarly2023, gweonMABLEMediatingYoung2018, besevliMaRTDesigningProjectionbased2019, panResearchApplicationImmersive2021, lorussoSemiimmersiveVirtualReality2020, passigSolvingConceptualPerceptual2014a, anTeachersPerceptionsEarly2023, gecu-parmaksizEffectAugmentedReality2020a, simsekEffectAugmentedReality2024, luoEffectDifferentCombinations2023a, safarEffectivenessUsingAugmented2017, topuEffectsUsingAugmented2024a, balchaImpactAugmentedRealityassisted2025, dallolioImpactFantasyYoung2024, wangImpactsAugmentedReality2024, vatamaniukRoleInteractiveMethods2024, zhouUseAugmentedReality2020, fridbergThematicTeachingAugmented2024, isikarslanogluThinkTogetherDesign2024, khairulanuardiIncreaseReadingHabit2022, linUsingHomebasedAugmented2025a, baehoikyoungVerificationVRPlay2025a, kusumaVirtualRealityLearning2018, zakraouiStudyChildrenEngagement2021, raduComparingChildrensCrosshair2016, shahabUtilizingSocialVirtual2022, drljevicInvestigatingDifferentFacets2022, stearneAugmentedRealityPlaygrounds2025, gavishAugmentedVirtualitySystems2022} \\
\addlinespace

\textbf{Data Anonymization} & 
Techniques to strip personally identifiable information from datasets, such as using random ID codes (pseudonymization) or blurring faces in images/videos. & 
\cite{topuExaminationPreSchoolChildrens2023a, tanApplicationArtbasedKnowledge2025, yilmazAreAugmentedReality2017, lawAugmentedRealityTechnology2025, chien-yuAugmentedRealitybasedAssistive2010, wuComparingEffectsAR2025, lewis-presserDesigningAugmentedReality2025, liInnovativeApplicationAI2025a, gweonMABLEMediatingYoung2018, wangImpactsAugmentedReality2024, fridbergThematicTeachingAugmented2024, isikarslanogluThinkTogetherDesign2024, linUsingHomebasedAugmented2025a} \\
\midrule

\multicolumn{3}{l}{\textit{\textbf{Technical Security Architecture (Scarce)}}} \\
\midrule

\textbf{Data Storage Architecture} & 
Specific architectural decisions to secure data, such as storing all sensitive information locally on the device (air-gapped) rather than transmitting it to the cloud. & 
\cite{vasilyevaASDAutisticSpectrum2025a, lorussoGiokAlienARbased2018, liInnovativeApplicationAI2025a, gweonMABLEMediatingYoung2018} \\
\addlinespace

\textbf{Access Control} & 
Implementing authentication mechanisms, such as password protection, to restrict access. & 
\cite{gweonMABLEMediatingYoung2018} \\
\addlinespace

\textbf{Data Governance \& Integrity} & 
Advanced measures to ensure data validity and policy compliance, such as blockchain-based auditing logs or formal data protection protocols. & 
\cite{wuComparingEffectsAR2025, lewis-presserDesigningAugmentedReality2025, lewis-presserEnhancingPreschoolSpatial2025, liInnovativeApplicationAI2025a} \\

\bottomrule
\end{tabularx}
\end{table*}



\end{document}